%% file: main.tex
\documentclass[longauth]{aaEC}

\usepackage[utf8]{inputenc}
\usepackage{txfonts}

\usepackage{graphicx}
\graphicspath{{Figures/}}
\usepackage{subcaption}

\usepackage{scalerel}
\usepackage[dvipsnames]{xcolor}

\usepackage{siunitx}            % Units
\usepackage{natbib}
\usepackage[nolist]{acronym}    % Don't include acronym list nor hyperlinks

\usepackage[pdfencoding=auto,psdextra]{hyperref}
\hypersetup{
    colorlinks=true,
    linkcolor=blue,
    filecolor=magenta,
    urlcolor=blue,
    citecolor=blue
}
\makeatletter
\renewcommand*\aa@pageof{, page \thepage{} of \pageref*{LastPage}}
\makeatother

\usepackage[switch, modulo]{lineno} % needed by euclid.sty
\usepackage{euclid}             % Euclid commands

\usepackage{orcidlink}          % This has to come after hyperref
\newcommand{\orcid}[1]{\orcidlink{#1}}

\newcommand{\JMASS}{\ensuremath{J_{\sfont{2MASS}}}\xspace}
\renewcommand{\d}{\ensuremath{\mathrm{d}}} % upright d, e.g. \d x
\newcommand{\defeq}{\triangleq}            % Definition

\begin{document}

% Title for a Key Project paper
\title{Euclid Quick Data Release (Q1)} %
% \subtitle{SIR Spectroscopic Processing and Data Products}
\subtitle{From spectrograms to spectra: the SIR spectroscopic Processing
  Function\thanks{Dedicated to our friend and colleague Bianca Garilli (1959–
    2024), for her central contributions to \Euclid in general, and NISP and
    SIR in particular.}}

%% please do not edit the author list once you copy it from the Publication
%% Portal -- contact ECEB Bureau for changes
%\input{short_authors}  % Long author list is not working on TeX 2019
\maxdeadcycles=500              % helps TeX 2019 to compile pages
\input{authors}                 % affiliations will be mal-formatted w/ TeX 2019

%
% Put your abstract here:
%
\abstract{ % Context
  The \Euclid space mission aims to investigate the nature of dark energy and
  dark matter by mapping the large-scale structure of the Universe.  A key
  component of \Euclid's observational strategy is slitless spectroscopy,
  conducted using the Near Infrared Spectrometer and Photometer (NISP).  This
  technique enables the acquisition of large-scale spectroscopic data without
  the need for targeted apertures, allowing precise redshift measurements for
  millions of galaxies.  These data are essential for \Euclid's core science
  objectives, including the study of cosmic acceleration and the evolution of
  galaxy clustering, as well as enabling many non-cosmological
  investigations. %
%}{ % Aims
  This study presents the SIR processing function (PF), which is responsible
  for processing slitless spectroscopic data from \Euclid's NISP instrument.
  The objective is to generate science-grade fully-calibrated one-dimensional
  spectra, ensuring high-quality spectroscopic data for cosmological or
  astrophysical analyses. %
%}{ % Methods
  The processing function relies on a source catalogue generated from
  photometric data, effectively corrects detector effects, subtracts
  cross-contaminations, minimizes self-contamination, calibrates wavelength
  and flux, and produces reliable spectra for later scientific use. %
%}{ % Results
  The first Quick Data Release (Q1) of \Euclid's spectroscopic data provides
  approximately three million validated spectra for sources observed in the
  red-grism mode from a selected portion of the Euclid Wide Survey.  We find
  that wavelength accuracy and measured resolving power are within
  requirements, thanks to the excellent optical quality of the instrument. %
%}{ % Conclusions
  The SIR~PF represents a significant step in processing slitless spectroscopic
  data for the \Euclid mission.  As the survey progresses, continued
  refinements and additional features will enhance its capabilities, supporting
  high-precision cosmological and astrophysical measurements. %
}

%
% Provide up to five key words from the list in
% https://www.aanda.org/for-authors/latex-issues/information-files#pop
%
\keywords{
  Cosmology: observations --
  Instrumentation: spectrographs --
  Techniques: imaging spectroscopy --
  Methods: data analysis
}
%
% Add short versions of title and author list for page headings
%
\titlerunning{The SIR Processing Function}
\authorrunning{Euclid Collaboration: Y.\ Copin et al.}

\maketitle

% -------------------------------------------------------------------
%
%   Start the main text of your paper here
%

\section{Introduction}
\label{sc:Intro}%

Slitless spectroscopy, also known as dispersed imaging, is one of the two
operating modes of the Near Infrared Spectrometer and Photometer
\citep[NISP,][]{EuclidSkyNISP}, one of the two instruments, along with VIS
\citep{EuclidSkyVIS}, on board \Euclid \citep{EuclidSkyOverview}.  This high
multiplexing spectrographic technique allows for the simultaneous dispersion of
light from all sources within a given field of view, eliminating the need for
traditional targeted apertures or slits, thereby enabling efficient
spectroscopic measurements across vast regions of the sky.

The spectroscopic exposures captured by the NISP spectrometer (hereafter
NISP-S) undergo comprehensive processing to generate scientifically valuable,
decontaminated, wavelength- and flux-calibrated, combined one-dimensional (1D)
spectra.  This processing is handled by the SIR \ac{PF} within the \Euclid
\ac{SGS}.  The SIR \ac{PF} produces spectra for all entries listed in the
source catalogue independently produced by the MER \ac{PF} \citep{Q1-TP004}
from photometric data from \Euclid visible \citep[VIS \ac{PF},][]{Q1-TP002} and
near-infrared \citep[NIR \ac{PF},][]{Q1-TP003} observations, as well as
selected external observations (EXT \ac{PF}).  The calibrated and validated
spectra are subsequently passed to the SPE \ac{PF} \citep{Q1-TP007} for
advanced spectral analyses, such as redshift determination and other spectral
feature extractions.  The SIR \ac{PF} processes exposures from the NISP-S
instrument, covering both wide and deep acquisitions, as well as red (`RGS',
with passband $\RGE \approx \text{\SIrange{1200}{1900}{nm}}$ and resolving
power $\mathcal{R} > 480$) and blue (`BGS', with passband
$\BGE \approx \text{\SIrange{920}{1370}{nm}}$ and resolving power
$\mathcal{R} > 400$) grisms, although the Q1 release focuses only on red-grism
data from the \aclu{EWS} \citep[\acs{EWS},][]{Q1-TP001}.

During the \ac{EWS}, the \ac{ROS} plays a crucial role in structuring the
observations \citep{Scaramella-EP1}.  It is executed at every pointing, and
consists of four dithers, where NISP-S and VIS observe simultaneously.  Each
dither involves a spectroscopic exposure with 549.6~s of integration time,
covering approximately the same sky portion but with a distinct combination of
red grism (RGS000 or RGS180) and \ac{GWA} tilt ($\ang{0}, \pm\ang{4}$),
following the dithering `K' sequence: RGS000\texttt{+}0 $\to$ RGS180\texttt{+}4
$\to$ RGS000\texttt{-}4 $\to$ RGS180\texttt{+}0.  Consequently, each source
will be observed approximately four times (except if they unfortuitously fall
on detector gaps or field edges), providing as many `single-dither' spectra.

In spectroscopy, it is important to distinguish between two key concepts,
`spectrogram' and `spectrum', the counterparts of photometric concepts of
observed image (2D) and inferred integrated flux (scalar).

In slitless spectroscopy, a spectrogram specifically refers to the observed
two-dimensional trace of dispersed light on the detector, representing the
source's spectral content as a function of both spatial position and
wavelength.  By the design of the NISP instrument, SIR focuses on the trace of
the first dispersion order of the grism (hereafter `1st-order spectrogram'),
but such traces also exist for other dispersion orders (e.g., the 0th-order
spectrogram).

In contrast, a spectrum refers to the one-dimensional spectrum of a source,
representing its chromatic flux density independently of the dispersion order.
The SIR \ac{PF} will infer this `intrinsic' spectrum from the 1st-order
spectrogram under the `spectral separability' hypothesis, which posits that the
light distribution in both spatial and spectral directions can be factored into
independent components: \(C(x, y, \lambda) = I(x, y) \, S(\lambda)\), where
\(I(x, y)\) represents the spatial intensity distribution, and \(S(\lambda)\)
represents the spectral flux distribution.  This assumption is valid for
unresolved sources or spatially-resolved uniform sources.  However, this
hypothesis neglects potential spatial gradients in colour, internal flux
distribution, or internal kinematics, which would require full-forward
modelling of the spectrograms \citep{2020A&A...633A..43O}.

While slitless spectroscopy offers significant advantages in efficiency and sky
coverage, it is also susceptible to two major sources of contamination:
`cross-contamination', where the spectrograms of neighbouring sources (in the
first or other dispersion orders) may overlap with the target source's
spectrogram, and `self-contamination', which arises from the degeneracy of the
spatial and spectral dimensions along the dispersion direction and which leads
to an effective resolving power function of source spatial extent.  The SIR
\ac{PF} mitigates these contaminations with sophisticated decontamination and
virtual-slit techniques (see below), but these issues, still active fields of
research, are not discussed further here.

This paper provides an overview of the SIR \ac{PF} at the time of the data
production (November~2024) for the \Euclid Q1 release \citep{Q1cite}.  It is
organised as follows.  Section~\ref{sc:spectro} describes the individual
processing steps that progressively transform raw slitless spectroscopic data
into precise, calibrated spectra, including scientific
(Sect.~\ref{sec:scientific-pipeline}), calibration
(Sect.~\ref{sec:calibration-pipeline}), and validation and data quality control
(Sect.~\ref{sec:dqc-validation}) pipelines.  Section~\ref{sc:perfs} presents
validation of the spectroscopic performance of the SIR \ac{PF} in the light of
the \Euclid's top-level mission requirements, and Sect.~\ref{sc:conclusion}
concludes.  All magnitudes are in the AB~mag system \citep{Schirmer-EP18}.

\section{Spectroscopic calibration and measurements}
\label{sc:spectro}

\subsection{Overview}
\label{sec:overview}

The SIR \ac{PF} is split into three sets of pipelines, each of which contains
individual \acp{PE} that will be described in their respective sections.  In
addition to the main `scientific' and `calibration' pipelines (described
below), a third `validation' pipeline runs independently on a control field to
assess and validate the software quality (Sect.~\ref{sec:dqc-validation}).

\subsubsection{The scientific pipelines}

There are two independent scientific pipelines
(Sect.~\ref{sec:scientific-pipeline}), which are run sequentially for the
processing of all scientific exposures acquired by NISP-S during the survey.
Their objectives is to produce science-grade products, internal to SIR \ac{PF}
or for general publication, based on some pre-computed and validated
calibration products.

\paragraph{The \emph{Spectra Extraction} pipeline} delivers single-dither
calibrated spectra; it runs sequentially on an observation basis during the
processing of the scientific exposures.  It includes the following eight
science-related \acp{PE}.
\begin{description}
\item[Preprocessing:] identification and correction of NISP detector artefacts
  (e.g., dark current, nonlinearity, persistence, etc.).  This step is common
  to the NIR \ac{PF} \citep{Q1-TP003}, since the same detectors are used both
  for photometry and slitless spectroscopy, although in different readout modes
  (Sect.~\ref{sec:sci-pre}).
\item[Spectra location:] full mapping between the sky coordinates of an
  arbitrary source and the precise position of the corresponding spectrograms
  in the \ac{FP} as a function of wavelength and dispersion orders
  (Sect.~\ref{sec:sci-loc}).
\item[Detector scaling:] estimate of the incident spectrum on each pixel (scene
  model) and correction for (potentially chromatic) fluctuations of detector
  response (Sect.~\ref{sec:sci-ff}).
\item[Background subtraction:] subtraction of the zodiacal light and other
  additive backgrounds (Sect.~\ref{sec:sci-sub}).
\item[Spectra decontamination:] correction (or masking) of 1st-order
  spectrogram for additive crosstalk from adjacent sources
  (Sect.~\ref{sec:sci-dec}).
\item[Spectra extraction:] estimate of the (1D) spectrum of a source from a
  single (2D) 1st-order spectrogram (Sect.~\ref{sec:sci-ext}).
\item[Relative flux scaling:] rescaling of all spectra to a common
  instrumental flux scale (internal consistency from different detectors,
  pointings, epochs, instrumental configurations, etc.), up to a chromatic
  external zero-point (Sect.~\ref{sec:sci-rfx}).
\item[Absolute flux scaling:] rescaling of all spectra to an astronomical flux
  scale (external consistency, Sect.~\ref{sec:sci-afx}).
\end{description}

\paragraph{The \emph{Spectra Combination} pipeline} includes a single \ac{PE}
to combine single-dither spectra on a MER tile basis \citep{Q1-TP004}:
\begin{description}
\item[Spectra combination:] merging of all flux-calibrated spectra for a single
  source (from different detectors, dithers, and pointings) into a single
  consolidated estimate (Sect.~\ref{sec:sci-com}).
\end{description}

\subsubsection{The calibration pipelines}

Five calibration-specific \acsp{PE} are needed to provide adequate calibrations
to the scientific \acsp{PE} from dedicated observations -- obtained during the
\ac{PV} phase or monthly monitoring of the self-calibration field --,
processing, and analyses (Sect.~\ref{sec:calibration-pipeline}).
\begin{description}
\item[Preprocessing calibration:] %(Sect.~\ref{sec:cal-pre})
  production of preprocessing calibration maps (e.g., dark current, detector
  bad pixels, etc.), derived from dedicated ground- and space-based detector
  characterisation measurements; this is addressed in \cite{Q1-TP003}.
\item[Spectra location model:] description of the distortion and dispersive
  behaviour of NISP-S, derived from prior knowledge of the instrumental
  properties and dedicated calibration exposures, including wavelength
  calibrators (Sect.~\ref{sec:cal-loc}).
\item[Detector scaling calibration:] detector response to a spatially and
  spectrally uniform illumination, derived from ground-based measurements
  (Sect.~\ref{sec:cal-ff}).
\item[Relative flux calibration:] transmission estimate of the NISP-S
  instrument (end-to-end, including telescope and detectors) as a function of
  position in the \ac{FP}, derived from comparison of repeated observations of
  the same sources (Sect.~\ref{sec:cal-rfx}).
\item[Absolute flux calibration:] conversion factor between instrumental and
  physical flux units (as a function of wavelength), derived from observations
  of flux calibrators (Sect.~\ref{sec:cal-afx}).
\end{description}

\subsection{Interfaces}
\label{sec:interfaces}

In its standard configuration (data processing of the NISP-S exposures from the
\Euclid telescope), the SIR \ac{PF} interfaces with LE1 (a technical \ac{PF} in
charge of crafting and complementing raw exposures received from spacecraft
with operational meta-data), MER \citep{Q1-TP004}, and NIR \citep{Q1-TP003}
\ac{PF}s on the input side, and the SPE \ac{PF} \citep{Q1-TP007} on the output
side.  The format of the files released with Q1 is described in the \Euclid
\ac{SGS} Data Product Description Document\footnote{%
  \url{http://st-dm.pages.euclid-sgs.uk/data-product-doc/dm10/sirdpd/sirindex.html}}
\citep{Q1cite}.

\paragraph{Input.} SIR \ac{PF} relies on the following input data set.
\begin{description}
\item[\path{DpdNispRawFrame} (LE1):] raw NISP-S exposures (signal and 8-bit
  quality factor computed on-board following \citealt{2016PASP..128j4504K}) and
  associated meta-data \citep{EuclidSkyNISP}.
\item[\path{DpdMerFinalCatalog} (MER):] consolidated source catalogue,
  including source identifier (ID), sky coordinates, size and shape
  information, NIR broadband photometry.
\item[\path{DpdMerBksMosaic} and \path{DpdMerSegmentationMap} (MER):]
  astrometrically-registered background-subtracted flux-calibrated image cutout
  of individual sources in each of the NISP photometer (hereafter NISP-P)
  filters, and their associated variance and segmentation maps.
\item[\path{DpdExtTwoMassCutout} (EXT):] elements from the Two Micron All Sky
  Survey catalogue \citep[2MASS,][]{2006AJ....131.1163S} to complement the MER
  catalogue on bright sources (see Sect.~\ref{sec:sci-dec}).  Since the 2MASS
  $J$, $H$, and $K$ bands are not identical to \Euclid \YE, \JE, and \HE bands,
  colour corrections are included even in the overlapping $J$ and $H$ filters
  \citep{Schirmer-EP18}.
\end{description}

\paragraph{Output.} There are two SIR products delivered for the Q1 release.
\begin{description}
\item[\path{DpdSirScienceFrame}:] preprocessed and background-subtracted
  dispersed image with approximate world coordinate system set from commanded
  pointing.
\item[\path{DpdSirCombinedSpectra}:] fully calibrated decontaminated integrated
  (1D) spectrum (both single-dither and combined) of each source identified in
  the input MER source catalogue.  Each spectrum consists of a signal vector,
  an associated estimate of the variance, and a bitmask vector (see
  Table~\ref{tab:bitmask}).  Along each spectrum, an exhaustive list of source
  IDs potentially contaminating the spectrogram, and the standard deviation of
  the effective \ac{LSF} is provided.
  % \item[LE3 \path{DpdSirLE3LocationTable}:] elements of the spectrometric
  %   model (so-called \emph{Location Table}).  This file allow to locate
  %   spectrogram starting and ending wavelength on the background subtracted
  %   frame of each extradted spectra
\end{description}

\subsection{Scientific processing elements}
\label{sec:scientific-pipeline}

\subsubsection{Usage of MER catalogue}
\label{sec:mer-sel}

% \qr{MSco}

The MER (photometric) source catalogue plays an important role in the running
of both the main SIR \ac{PF}, and the various SIR calibration \ac{PF}s.
Because of the very nature of the slitless spectroscopic data, it would be
rather complex and prone to significant uncertainties to carry out the object
detection step, necessary for the spectra extraction stage, directly on the
spectroscopic data.  It was therefore decided to design all of the SIR \acp{PF}
to use as external input the list of detected objects provided by the imaging
data (VIS and NIR, as well as EXT), as constructed and delivered by the MER
catalogue.  From this catalogue, the SIR \acp{PF} extract each object's sky
coordinates, VIS \IE and NIR \YE, \JE, and \HE magnitudes (based on MER
template-fitting photometry, see \citealt{Q1-TP004}), as well as object
isophotal data (semi-major axis size, position angle, and axial ratio).

Since the MER catalogue includes all detections, either on the VIS or NIR
images, irrespective of the \ac{S/N} associated with those detections, it is
likely that at low flux limits a growing fraction of the detections included in
the catalogue are actually false positive and do not correspond to real objects
in the sky (e.g., due to persistence, see \citealt{Q1-TP003}).  Moreover, these
faint sources would not contribute any significant signal on the spectroscopic
data, as a result of the already faint flux being dispersed over approximately
500~pixels (the median counts-per pixel signal from an object of magnitude
$\HE=21.5$ is approximately 10, compared with a median background level of
approximately 800).  It was therefore decided that in the early stages of the
\Euclid spectroscopic data analysis for the Q1 release, only a subset of the
objects listed in the MER catalogue would be included in the SIR \ac{PF} data
analysis, specifically only objects with a measured NIR magnitude
$\HE \leq 22.5$.  For bright objects (mostly point-like) that result in
saturated images in the NISP-P imaging exposures ($\lesssim 16$~mag,
\citealt{EuclidSkyNISP}), MER is not able to produce reliable flux
measurements, and the SIR \ac{PF} then falls back on using 2MASS photometry
instead \citep{2006AJ....131.1163S}.

\subsubsection{Preprocessing} % (SCI-PRE)
\label{sec:sci-pre}

% \qr{Shooby, SCo}

The preprocessing is the general terminology for the processing steps needed to
correct for detector-related artefacts, e.g., identification of cosmetic
defects (bad pixels), nonlinearity corrections, intrinsic signal pollution
(dark current or persistence signal), cosmic ray hits, etc.  It ultimately
generates preprocessed science exposures from LE1 raw exposures.  This
`composite' \ac{PE} includes all preprocessing-related software components
developed in common with NIR \ac{PF}.  For a detailed description of these
steps, we refer to \citet{Q1-TP003}; a summarised list is provided below.
\begin{description}
  % (SCI-PRE-INI)
\item[Initialisation:] initialisation of the SIR frame from LE1 raw data,
  including signal in \ac{ADU}, variance estimate, as well as \aclu{QF}
  (\acsu{QF}, i.e., up-the-ramp $\chi^{2}$, see \citealt{2016PASP..128j4504K})
  and bitmask layers.
  % (SCI-PRE-BAD)
\item[Bad pixel masking:] identification in the bitmask layer of known pixels
  giving unusable or suspicious signal.
  % (SCI-PRE-LIN)
\item[Linearity correction:] correction of the nonlinear detector behaviour,
  saturation flagging, and conversion of signal from \ac{ADU} to electrons.
  % (SCI-PRE-DK)
\item[Dark subtraction:] subtraction of the `dark' (thermal) contribution from
  the detector.
  % (SCI-PRE-CR)
\item[Cosmic ray rejection:] identification and masking of pixels affected by
  cosmic rays from analysis of \ac{QF}\footnote{The \ac{QF} layer is not
    propagated further in the pipeline.}.
  % (SCI-PRE-SIR)
\item[SIR-specific initialisation:] interpolation-free rotation of the frames
  to align spectrograms mostly horizontally -- in so-called SIR-coordinates
  $(X, Y)_{\text{SIR}} \equiv (Z, -Y)_{\text{MOSAIC}}$
  \citep[see][]{EuclidSkyNISP} -- and creation of the SIR-specific bitmask.
\end{description}
Persistence flagging -- identification and masking of pixels affected by
persistent signal \citep{2024SPIE13103E..15K} -- was still in development for
spectroscopic exposures at the time of production, and is therefore not
implemented for the Q1~release.  Accordingly, some high-\ac{S/N} sources may in
fact be spurious.

\subsubsection{Spectra location} % (SCI-LOC)
\label{sec:sci-loc}

% \qr{MFu}

The primary objective of the SIR \ac{PF} is to estimate spectra for all
selected entries in the source catalogue provided by MER \ac{PF}
\citep{Q1-TP004}, independently of their spectral signatures in NISP-S images
(e.g., apparent continuum or noticeable emission lines).  The exact mapping --
hereafter the spectrometric model -- between a source, identified by its sky
coordinates, and the corresponding spectrogram on the detectors (including
wavelength solution), is the goal of the `spectra location' \ac{PE}.

This \ac{PE} is split into two software components.
\begin{description}
\item[Pointing registration:] commanded spacecraft pointing coordinates, as
  stored in the LE1 frames, can significantly differ from the effective values,
  up to \ang{;;7} \citep{EuclidSkyOverview}.  The first step of the pipeline
  calculates the actual pointing of the spacecraft: positions of selected
  bright 0th-order spots are measured in the four central detectors (where the
  0th-order optical quality is better\footnote{Given the dispersive power of
    the prism \emph{and} the blazing function of the grating, the 0th-order
    spectrogram has a distinctive non-trivial double-peaked roughly
    10~pixel-long shape.}), and a roto-translation is computed against known
  sky coordinates of the stars to evaluate the effective spacecraft pointing
  and roll angle.

\item[Spectra location:] three geometric models provide the location and an
  effective description of the spectrograms of all the objects selected in the
  MER catalogue.  For each source, a reference position of the spectrogram in
  the \ac{FP} is first computed using the `astrometric model' (so-called OPT
  model), mapping its sky coordinates (RA, Dec) to the 1st-order position
  $(x_{1}, y_{1})$ of reference wavelength $\lambda_{1}$.  Then, the `curvature
  model' (so-called CRV model) is used to map the cross-dispersion position of
  incident light along the spectral trace for any wavelength and dispersion
  orders (limited to 0th- and 1st-orders for the Q1 release).  Finally, the
  `\aclu{IDS}' (\acs{IDS}) provides a mapping between incident wavelength
  $\lambda$ and position $D$ along the spectral trace.  The full `spectroscopic
  model' is stored for all sources of the input catalogue into a single
  \path{DpdSirLocationTable} product, a precise description of \emph{all} 0th-
  and 1st-order spectrograms in the frame.
\end{description}
The associated calibration \acsp{PE} will provide astrometric % (CAL-LOC-AST)
and spectroscopic % (CAL-LOC-SPE)
models (see Sect.~\ref{sec:cal-loc}), to be used as input for the \ac{PE}.

\begin{figure*}
  % \centering
  % \includegraphics[width=.8\linewidth]{J+RGS000_2684805874647806467}
  \sidecaption
  \includegraphics[width=12cm]{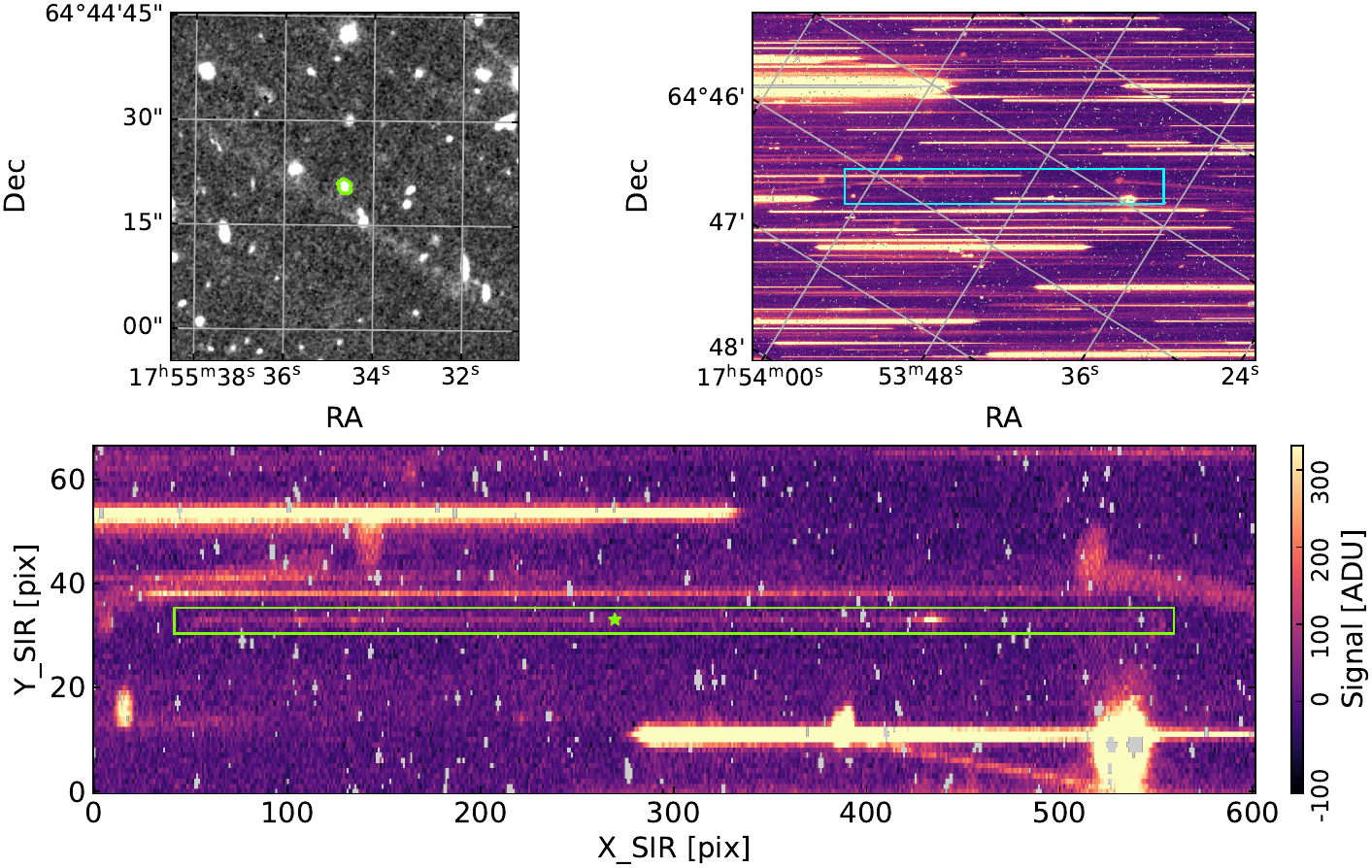}
  \caption{Illustration of the various exposures entering the SIR pipeline for
    object ID 2684805874647806467, a $z=1.63$ galaxy. % from observation ID=2712.
    \emph{Upper left:} \ang{;;50}-cutout from the MER \JE-band stack,
    % (tile ID=102158584),
    centred on the object (green contour).  \emph{Upper right:} close-up on
    DET42 of preprocessed background-subtracted RGS000\texttt{+}0 spectroscopic
    exposure (pointing ID 11953) around the spectrogram of the same object
    (blue box).  \emph{Bottom:} zoom into the blue box in SIR coordinates; the
    effective extraction window is indicated as a green box, and the position
    of the reference wavelength $\lambda_{1}$ is marked with a star, while
    original pixels flagged as unusable are in grey.  We note the bright
    (saturated) 0th-order spectrogram in the lower right, as well as low-level
    persistent traces of previously-observed tilted 1st-order and 0th-order
    spectrograms.}
  \label{fig:J+RGS000}
\end{figure*}

\subsubsection{Detector scaling} % (SCI-FF)
\label{sec:sci-ff}

% \qr{Phil, Ranga}

\begin{figure}
  \centering
  \includegraphics[width=.6\linewidth]{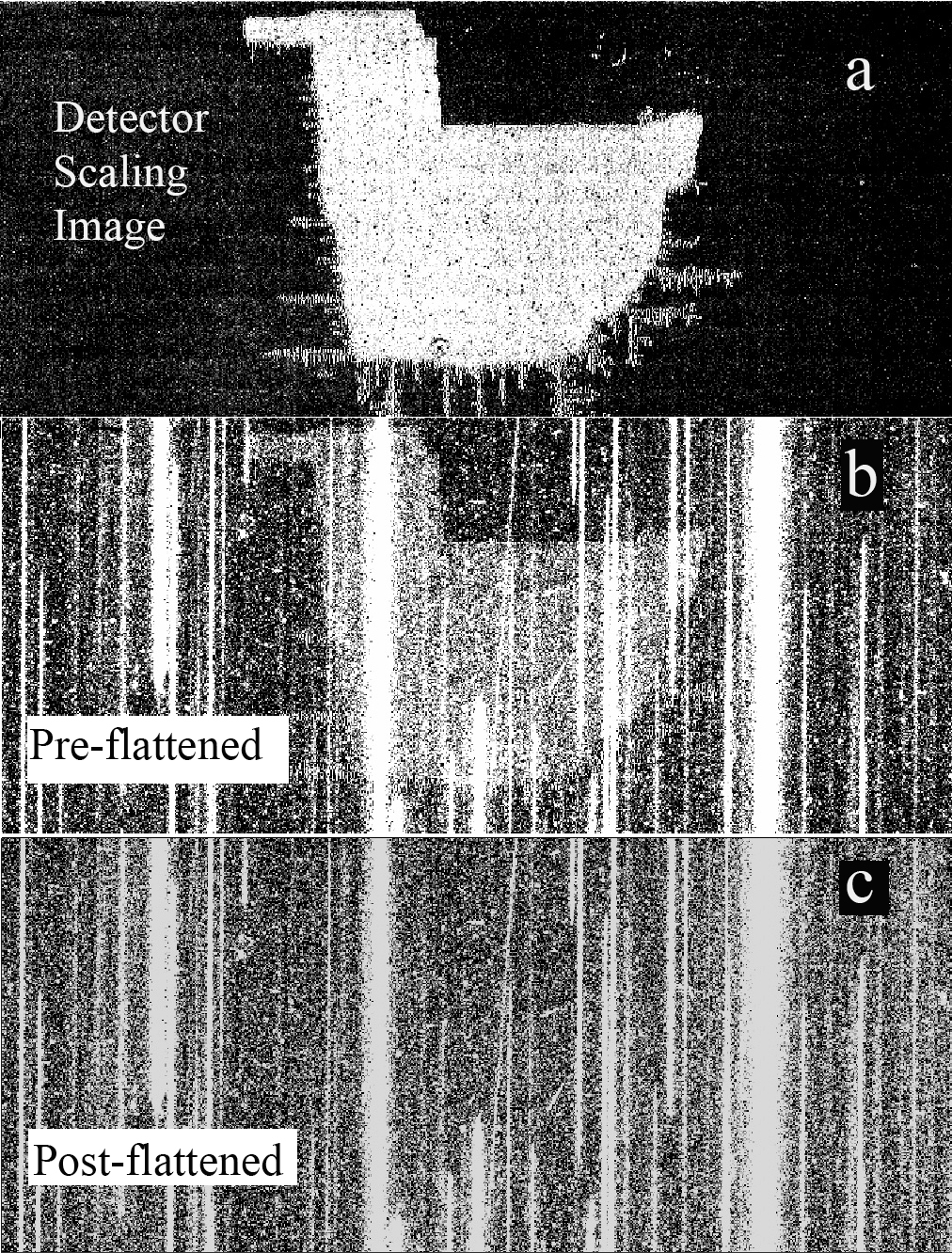}
  \caption{Illustration of the application of the detector scaling product to a
    small section of the DET11 images: (a) the detector scaling image centred
    on the `duck' structure (see Sect.~\ref{sec:cal-ff}); (b) the same section
    of a dispersed image prior to correction, and (c) after the application of
    the detector scaling.  The `duck' structure has been successfully
    mitigated.}
  \label{fig:duckgone}
\end{figure}

The purpose of the `detector scaling' is to correct intensity variations within
a detector, and across the detectors, due to differences in each pixel's
\ac{QE}.  This eliminates not only individual pixel-to-pixel variations on the
small scale, but also larger-scale fluctuations in the detector response due to
various surface properties, sometimes leading to distinctive `islands' of
pixels with lower-than-normal~\ac{QE} \citep[see][Euclid Collaboration: Kubik
et al., in prep.]{EuclidSkyNISP}.  These structures can show spatially abrupt
changes in \ac{QE}, especially at the island boundaries, and need to be
corrected to restore spatial continuity at the detector scale (see
Fig.~\ref{fig:duckgone}).

In order to eliminate intensity variations in the data due to \ac{QE}
variations, the spectroscopic image is divided by a `master flat', one per
detector. As explained in Sect.~\ref{sec:cal-ff}, the master flat is presumed
to be \emph{a}chromatic, and computed assuming a uniform illumination scene
dominated by the zodiacal background.

We note that, unlike standard photometry (where the mapping between sky and
detector positions is bijective), the master flat for dispersed imaging only
corrects for detector-scale effects, but cannot account for relative flux
calibration at all positions and wavelengths, which are degenerate quantities
on the detector; this is therefore specifically adressed by the relative flux
scaling (Sect.~\ref{sec:sci-rfx}).

\subsubsection{Background subtraction} % (SCI-SUB)
\label{sec:sci-sub}

% \qr{MFu}

The `background subtraction' \ac{PE} is aiming at estimating and subtracting
the additional flux component not directly associated with individual
spatially-localised sources, e.g., zodiacal diffuse background, scattered, and
stray light (diffusion), ghosts (reflections), etc.

By lack of elaborate ghost and stray light models for the spectroscopic channel
at time of production, the Q1 version of the pipeline only computes a uniform
background value independently for each detector, estimated from the mode of
the distribution of the `signal-free' pixels, i.e., not covered by 0th- and
1st-order spectrograms of sources with $\JE \leq 23$, and not masked during
preprocessing.

\subsubsection{Spectra decontamination} % (SCI-DEC)
\label{sec:sci-dec}

% \qr{Ranga}

Dispersed imaging -- as obtained with \Euclid NISP-S -- suffers from
\emph{cross}-contamination, i.e., the spectrogram of each source is potentially
contaminated by flux from \emph{other} sources in its vicinity.  Although the
use of the four different dispersion directions in the observation strategy
mitigates the contamination to a certain extent, the sensitivity of \Euclid
implies there is a large number of potentially-contaminating sources
(\SIrange{e4}{e5}{deg^{-2}}) relative to the number of H$\alpha$ emitters (less
than \SI{4000}{deg^{-2}}, see~\citealt{Scaramella-EP1, Gabarra-EP31}).  These
H$\alpha$ emitters are used measuring the imprint of the baryon acoustic
oscillations on galaxy clustering between $0.9 < z < 1.8$ to determine the
redshift evolution of dark energy, one of the primary science goals of \Euclid
\citep{Laureijs11}.

Furthermore, the relatively coarse spatial sampling of NISP (\ang{;;0.3},
\citealt{EuclidSkyNISP}) and significant extent of the NISP-S~\aclu{PSF}
(\acsu{PSF}, 20\% flux outside \ang{;;0.68}) mean that the spatial wings of
bright sources can affect many pixels beyond the typical source extraction
aperture of five pixels used in the pipeline (see Sect.~\ref{sec:sci-ext}).  As
an estimation, there are 10 to 30 sources that overlap the 1st-order
spectrogram of each source of interest; this number is even larger in dense
regions of sky such as galaxy clusters.

Contamination of the 1st-order spectrogram of a source of interest occurs
because the 0th-order and the 1st-order -- and possibly other dispersion orders
for extremely bright sources -- spectrograms of unrelated sources fall on the
same region of the detector.  In all cases, this will result in an extrinsic
dither-dependent flux excess in the extracted spectrum of the target source,
degrading both its continuum and its spectral features.

Due to the volume of data being run through the spectroscopic pipeline, there
are stringent memory and computing time requirements, which result in
limitations on the range of algorithms that can be used for decontaminating the
spectra.
% (SCI-DEC-STD)
For the Q1 release, a `standard' decontamination \ac{PE} was implemented and
tested.  It identifies all contaminating sources, gathers their positions,
brightnesses, and surface brightness profiles from NISP-P imaging data,
estimates their (1D) spectrum, builds (2D) pixel-level spectrogram models at
their specific locations, and subtracts these models from the spectrogram of
each source of interest at each individual roll angle (dither).  If the
contamination appears too large (above requirements, see below), the
contaminated pixels are flagged as unusable, and will not enter the extraction
step (see Fig.~\ref{fig:xtract}).  The procedure is also used to identify and
mask out 0th-order spectrograms in the dispersed images.

\paragraph{Contaminant catalogue.} The first step in this process is to compile
accurate photometry for all the sources in the field of view.  While the NISP-P
photometry is accurate for sources fainter than the limit of 16th~mag, there is
no reliable \Euclid measurement for brighter saturated sources
\citep{EuclidSkyNISP}.  As mentioned earlier, we address this issue by using
the external 2MASS photometry \citep{2006AJ....131.1163S} to estimate their
brightness\footnote{All-sky $Y$-band flux densities from PanSTARRS
  \citep{2016arXiv161205560C} and DECaLS \citep{2019AJ....157..168D} surveys
  are not yet incorporated into the pipeline.}.

The second step is to use the position, size, and brightness of all the sources
in the input source catalogue and corresponding spectra-location table to
define the effective area within which the source spectrogram is located.  For
typical sources, the source size is adopted as the larger of the size in the
photometric data and 5~pix; for the brightest sources ($\JE < 16$), however,
the size of the location table is progressively widened up to 20~pixels (for
sources with $\JE < 12$) % see #29968
to account for the flux of the wings of the \ac{PSF}, as described earlier.  If
this were not done, a fraction of sources of interest would still be
contaminated by the brightest sources in the field of view.

\paragraph{0th-order masking.} We next use the spectrometric model
(Sect.~\ref{sec:sci-loc}) to mask out the 0th-order spectrograms for all
sources.  We have estimated that the magnitude threshold at which the 0th-order
spectrogram of a source is below the Poisson noise threshold from the
background corresponds to $\JE = 19.5$.  Not only is the 0th-order \ac{PSF} not
as sharp as the 1st-order one, but its extent also depends on its radial
position in the \ac{FP}.  Additionally, for bright resolved galaxies, the
spatial extent of the source matters as well.  For the Q1 release, we have been
conservative in the size of the 0th-order masking box by calibrating it on
bright stars; however, for bright sources, particularly galaxies detected by
2MASS, the 0th order may still extend beyond the box and be left unmasked,
polluting distant spectrograms.  An improved modelling and masking of the
0th-order will be included in a subsequent version of the pipeline.

\paragraph{1st-order contaminants.} The location table is used for each source
to identify all 1st-order contaminants, i.e., adjacent sources whose 1st-order
spectrogram overlap the 1st-order spectrogram of interest.  Since the typical
spectrogram extent is 531~pixels long and approximately 5~pixels wide, if any
of those 2500~pixels include flux from even the wings of an adjacent source, it
is classified and listed as a contaminant. In the current pipeline runs, a
catalog magnitude cut of $\HE < 22.5$ is adopted for identifying the sources
and their contaminants.  This is partly because of spurious sources being
present at fainter magnitudes, likely due to persistence.

\paragraph{Spectral (1D) model of contaminants.} We next estimate the
contribution from each identified contaminant to the source of interest.  To
model the continuum of each contaminant, we adopt two approaches.  We first try
to model the continuum by fitting a power law to the measured flux densities in
the spectrograms over the uncontaminated domains.  If a line is strong enough
to be seen at $\gg 5\sigma$ in a single spectrogram, it is masked out when
deriving the power-law fits to the continuum flux model and then added back in
as a Gaussian line to the model.  The contaminating source is then defined as
`bright with detectable continuum' if the derived continuum is consistent with
the \JE and \HE broadband flux densities within 10\%; we find that this
criterion is matched only for a few percent of sources, mostly because of
contamination, and because we have not yet included optimal profile-weighted
extraction in the pipeline (see Sect.~\ref{sec:sci-ext}).  For the bright
sources fulfilling this consistency criterion, the power-law fit from the
spectrogram continuum is used.

For fainter contaminating sources, or sources for which no consistent
measurements of the continuum can be obtained from spectrograms, we directly
fit a power law to its broadband flux densities.  For sources with \emph{all}
\YE, \JE, and \HE measurements available from NISP-P, we adopt one power law
between \YE and \JE, and another between \JE and \HE, with continuity between
the two interpolations.  For sources missing any NISP-P magnitude, the spectral
model falls back to a single power law fit to the 2MASS $J$- and $H$-band flux
densities.  Various tests have shown that the double power-law results in
better residuals than a single power-law.

\paragraph{Spectrogram (2D) model of contaminants.} The next challenge is to
spatially distribute the model flux density of the contaminant in the spatial
(cross-dispersion) direction to build a contaminating spectrogram.  While one
would naively adopt the spatial extent of the source in the imaging data, this
is inaccurate, since the grism has optical power: the imaging and spectroscopic
\acsp{PSF} are different, with the imaging \ac{PSF} being narrower\footnote{The
  spatial resolution of NISP-P is ${\sigma\simeq\ang{;;0.15}}$
  \citep{EuclidSkyNISP} for all bands, while the NISP-S red-grism one is
  ${\sigma\simeq\ang{;;0.18}}$ (see Sect.~\ref{sec:specres-reqs}).}.  We use an
ad hoc wavelength-dependent Gaussian kernel to degrade the source profile
derived from the segmented thumbnail extracted from the \JE stack produced by
MER \citep{Q1-TP004}.  For saturated sources that do not have source profiles
measured by MER, we assume that they are point sources; this is obviously
inaccurate for bright nearby galaxies and will be revised in the future.
Appropriate corrections for the fraction of flux outside the extraction
aperture are also applied.

We then take the Gaussian fits to the imaging profiles of the sources and
degrade them with the imaging-to-spectroscopic cross-kernel to obtain the
estimated wavelength-dependent spatial profile of each source in the
spectroscopic data.  The model flux densities derived are then distributed
chromatically using this spatial profile within its corresponding location
table.

\paragraph{Contaminant subtraction.} These modelled spectrograms of all
contaminants are finally subtracted pixel-by-pixel from the native spectrogram
of the source of interest.  This is done for each dither separately on the
preprocessed, detector-rescaled, background-subtracted dispersed images.
Pixels where the total contaminating flux is larger than 10\% that of the
source of interest are flagged out for excessive contamination, and do not
enter the extraction procedure (Sect.~\ref{sec:sci-ext}).

\begin{figure}
  \centering
  \includegraphics[width=\linewidth]{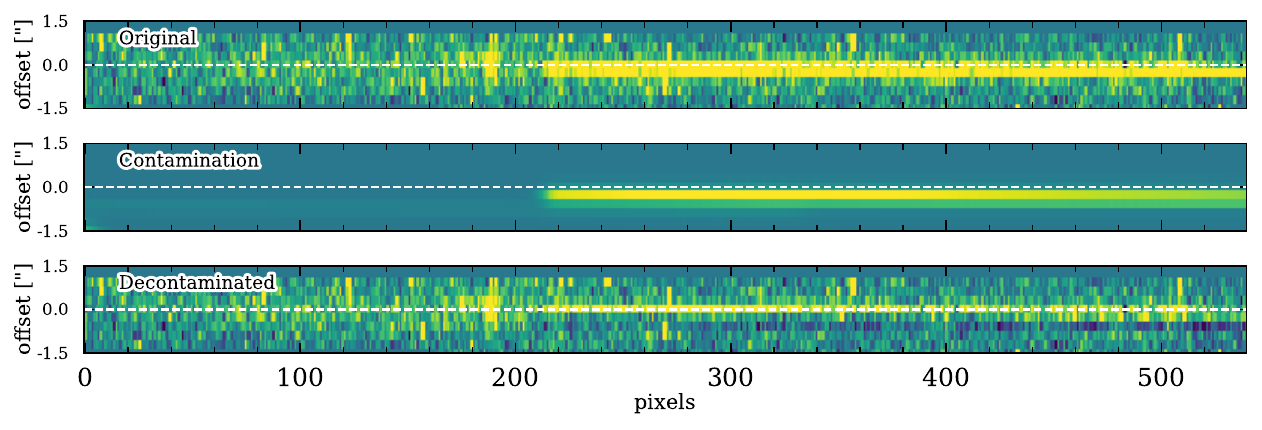}
  \caption{Illustration of the basic decontamination procedure for a
    line-emitting source.  The top panel shows the original RGS000\texttt{+}0
    spectrogram of object ID 2709725257636288279 in SIR coordinates, i.e., the
    $y$-axis being the spatial direction and $x$-axis effectively the
    dispersion direction.  The middle panel shows the model for the bright
    contaminant, in this case created from broadband photometry, which affects
    a part of the target spectrogram.  The bottom panel shows the
    decontaminated spectrogram of the source of interest.  The flux scaling of
    all panels is the same.}
  \label{fig:basdec1}
\end{figure}

The end result from this decontamination process is a decontaminated
spectrogram for each source of interest for each dither, along with
corresponding bitmask and variance layers (see Fig.~\ref{fig:basdec1}).  The
bitmask layers are crucial for identifying which pixels should be ignored
either due to the 0th-order contamination or due to excessive contamination
from a bright source in the subsequent steps in the pipeline (notably spectrum
extraction).

\subsubsection{Spectra extraction} % (SCI-EXT)
\label{sec:sci-ext}

% \qr{MFu, YCo}

Once the 1st-order spectrogram of a given object has been precisely located
within the NISP-S exposure (see Sect.~\ref{sec:sci-loc}) and properly
decontaminated from external sources (see Sect.~\ref{sec:sci-dec}), one needs
to extract and build an estimate of the source spectrum.  This includes proper
handling of optical distortions and application of the wavelength solution to
produce a linear wavelength ramp.  This is the objective of the `spectra
extraction' \ac{PE}, which provides both a `recti-linear' 2D~spectrogram (not
integrated over the cross-dispersion spatial direction) and a 1D~spectrum
(integrated over the source extent in the cross-dispersion direction).

\paragraph{Spectrogram resampling.}
The first step of the extraction is to resample the 2D~spectrogram, in order
to:
\begin{enumerate}
\item align and rectify the spectrogram along the horizontal direction,
  accounting for mean grism tilt and distortion-induced curvature;
\item include the \ac{IDS} to generate a spectrogram linearly sampled in
  wavelength in the dispersion direction;
\item move the virtual slit (see below) perpendicular to the dispersion
  direction to minimise the effective \ac{LSF}.
\end{enumerate}
The decontaminated spectrogram is resampled a single time using a $4\times 4$
hyperbolic-tangent kernel, with proper handling of masked pixels.

\begin{figure}
  \centering
  \includegraphics[width=.9\linewidth]{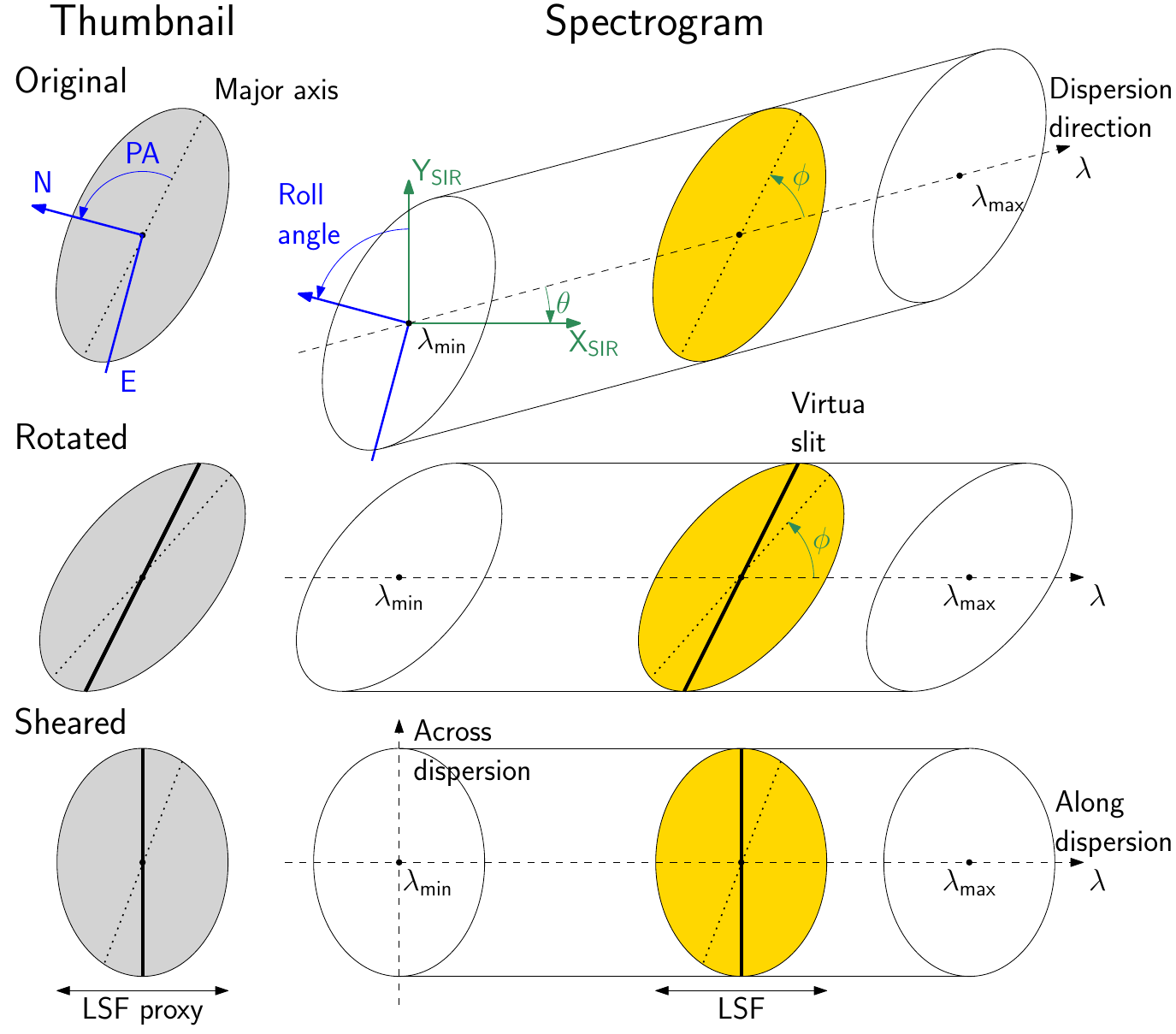}
  \caption{Illustration of the different steps in the extraction of a
    spectrogram (\emph{right}) of an extended source, along its photometry
    thumbnail (\emph{left}).  \emph{Top:} original orientation in the \ac{FP}.
    \emph{Middle:} rotation to bring the dispersion direction to horizontal.
    \emph{Bottom:} shear to bring the virtual slit along the cross-dispersion
    direction and minimise self-contamination.  This illustration is a
    simplified case with no initial tilt or curvature in the spectral trace;
    furthermore, in practice, rotation and shear are performed in a single step
    to minimise correlations between resampled pixels.}
  \label{fig:virtualslit}
\end{figure}

While the wavelength- and distortion-resamplings are classical, the virtual
slit deserves more explanation.  In order to minimise self-contamination --
i.e., the degeneracy between the effective spectral resolution and the spatial
extent of the source in the dispersion direction -- and therefore improve
spectral resolution by minimising the effective \ac{LSF}, the 2D~spectrogram of
a resolved source is resampled to align the source maximal elongation in the
cross-dispersion direction, the so-called `virtual slit', perpendicular to the
dispersion direction (see Fig.~\ref{fig:virtualslit}).

In practice, the resampling includes a transformation locally similar to
\begin{equation}
  \label{eq:resampling}
  T =
  \begin{bmatrix}
    1 & m \\
    0 & 1 \\
  \end{bmatrix}
  \begin{bmatrix}
    \cos\theta & \sin\theta \\
    -\sin\theta & \cos\theta
  \end{bmatrix},
  \quad\text{with}\quad
  m = \frac{1 - q^{2}}{\tan\phi + q^{2}/\tan\phi},
\end{equation}
where $\theta$ is the dispersion direction with respect to horizontal
(potentially wavelength-dependent, due to distortions), $\phi$ the source
position angle with respect to dispersion direction, and $q$ the flattening of
the source (both supposed achromatic).  This transformation guarantees that the
spectrogram is resampled horizontally, and the virtual slit brought to
vertical; as a consequence, the apparent extent of the source along the
dispersion direction, which directly sets the effective \ac{LSF}, is minimised.

For the Q1 release, the actual width of the extraction aperture used by the
spectrogram resampling is defined as follows, depending on the nature of the
source.  For extended objects, the size of the rectified virtual slit is set
from the semi-major axis of the source as quoted in the MER
catalogue\footnote{This actually neglects the $\sim\ang{;;0.1}$ spatial
  resolution (quadratic) difference between NISP-P and NISP-S, a reasonable
  assumption for extended objects.}  \citep{Q1-TP004}; this size is further
limited to five (lower limit) and 31~pixels (upper limit).  For point-like
objects -- defined as objects with a point-like probability $> 0.7$ in the MER
catalogue or cross-matched in the 2MASS catalogue, the virtual slit is five
pixels long.

In addition to the `spatial' components of the resampling (curvature and
virtual slit), resulting in a scale of \ang{;;0.3}\,\si{pix^{-1}} in the
cross-dispersion direction, the resampling transformation also includes the
wavelength solution in the spectral direction, so that the resampled
spectrogram is linearly sampled along the dispersion direction, from
$\lambda_{\min} = \SI{1190.0}{nm}$ to $\lambda_{\max} = \SI{1900.2}{nm}$, with
a step of $\delta\lambda = \SI{1.34}{nm}$ for the red grism (531~pixels).

\begin{figure*}
  % \centering
  % \includegraphics[width=\linewidth]{EUC_SIR_W-XTRACT_2684805874647806467}
  \sidecaption
  \includegraphics[width=12cm]{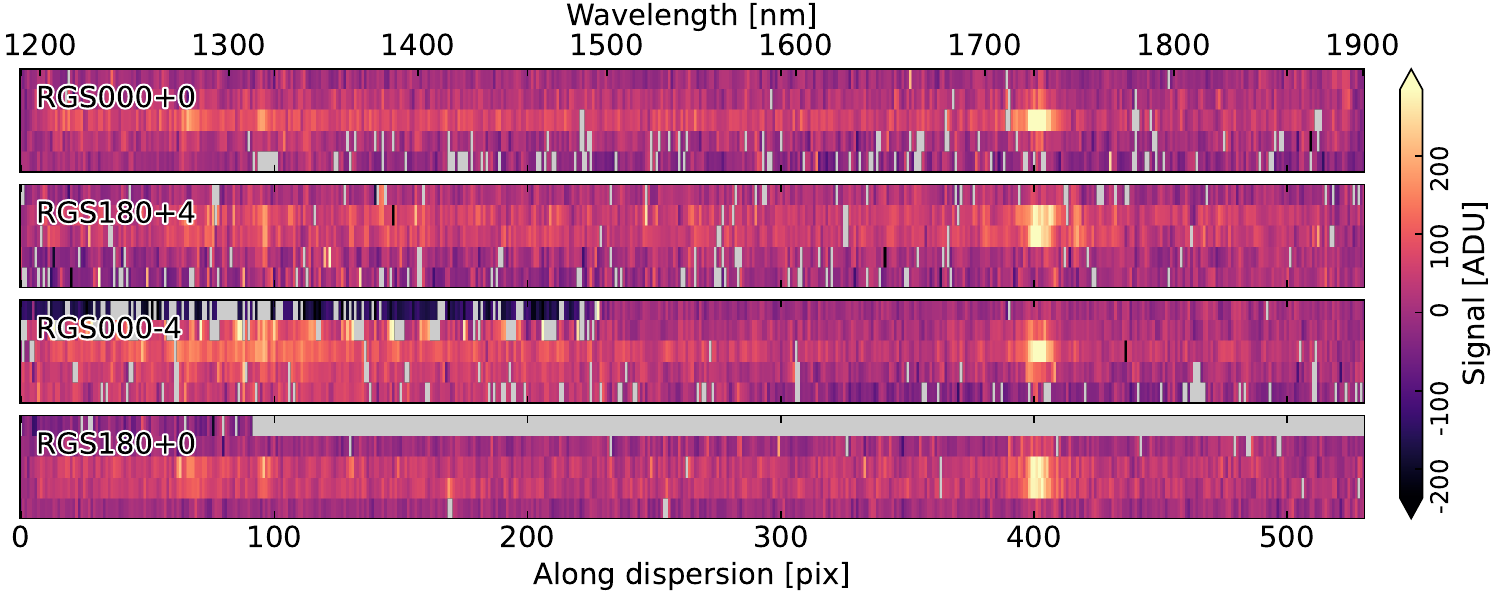}
  \caption{The four $5 \times 531$~pixels (corresponding to
    $\ang{;;1.5} \times \SI{711.54}{nm}$) resampled spectrograms for object ID
    2684805874647806467 (RGS000\texttt{+}0, RGS180\texttt{+}4,
    RGS000\texttt{-}4 and RGS180\texttt{+}0 from top to bottom) from
    observation ID 2712.  Resampled pixels flagged as unusable
    (\protect\path{NOT_USE}) are in grey.}
  \label{fig:xtract}
\end{figure*}

During the resampling process, resampled pixels are a mixture of numerous (up
to 16) original pixels, with no longer direct inheritance: all the original
quality bits of the input pixels cannot be propagated to the output ones.  For
this reason, a new bitmask is computed and stored along with the 2D~spectrogram
(see Table~\ref{tab:bitmask}).  Since the resampling is a weighted average of
the pixels within the kernel extent, a final pixel is flagged as:
\begin{description}
\item[\path{NOT_USE},] if the numeral fraction of unmasked pixels used during
  resampling is lower than 25\%;
\item[\path{LOW_SNR},] if the same fraction is lower than 50\%;
\item[\path{LOW},] if only outer weights of the kernel are used, resulting in a
  suspicious interpolated value.
\end{description}
Figure~\ref{fig:xtract} shows examples of signal spectrograms after resampling.
Following \cite{2000AJ....120.2747C}, the variance layer of the spectrogram is
resampled with the same procedure, rather than being propagated.

\paragraph{Averaged summation.}
For the Q1 release, the spectral extraction, which generates the 1D~spectrum of
the source, is performed by averaging unmasked pixels along the
cross-dispersion direction, and rescaling by the aperture size, i.e., the width
of the 2D~spectrogram along the cross-dispersion direction.
% \footnote{This is formally equivalent to a least-square fit of the
% cross-dispersion profile by a constant.}.
This `averaged summation' attenuates the impact on flux of masked pixels in the
spectrogram.

The 1D~bitmask is computed as:
\begin{description}
\item[\path{NOT_USE},] if the fraction of unmasked pixels used in the average
  is lower than 50\%;
\item[\path{LOW_SNR},] if the same fraction is lower than 75\%.
\end{description}
Figure~\ref{fig:combspec} shows examples of extracted spectra (after relative
and absolute flux calibrations).

\paragraph{Line-spread function.}
As mentioned earlier, the effective \ac{LSF} of a slitless spectrum is an
intricate mixing of instrumental \ac{PSF} and intrinsic source extent.  While
the \ac{PSF} part can be estimated independently from pure point sources
(e.g., stars), the spatial contribution depends on the extended source
properties, and presumably varies with wavelength due to colour gradients and
distribution differences between stellar and gaseous components.

In practice, the effective extent of the source is estimated from the segmented
\ang{;;0.1}\,\si{pix^{-1}} thumbnail extracted from the \JE stack produced by
MER \citep{Q1-TP004}.  The thumbnail at NISP-P resolution
($\sigma \approx \ang{;;0.15}$) is convolved by a 2D~Gaussian to match mean
NISP-S \ac{PSF} ($\sigma \approx \ang{;;0.18}$), rotated, and sheared according
to the spectrogram resampling procedure (see Fig.~\ref{fig:virtualslit}), and
finally rebinned by a factor of three to match the NISP spatial scale of
\ang{;;0.3}\,\si{pix^{-1}}.  The \ac{LSF} standard deviation is estimated from
a 1D~Gaussian fit to the resulting thumbnail marginalised over the
cross-dispersion direction.

\subsubsection{Relative flux scaling} % (SCI-RFX)
\label{sec:sci-rfx}

% \qr{Dan}

Large-scale transmission variations in the instrument, arising from a
combination of optical and detector effects, must be measured and corrected to
ensure consistent flux measurement for sources.  To accomplish this, a relative
flux scaling solution is derived for each grism/tilt configuration using repeat
observations of bright stars in the self-calibration field (see
Sect.~\ref{sec:cal-rfx} for details).

The relative flux scaling is applied to the extracted spectra based on the
grism/tilt configuration and the location of the spectrum in the \ac{FP}.  This
correction ensures that a consistent instrumental flux (prior to absolute
scaling) will be reported for a given source no matter where it lands on the
detector/\ac{FP}, which grism/tilt combination was used, or the epoch of the
observation.  At present, the relative flux scaling solution appears to be
stable with time and hardly chromatic.

\subsubsection{Absolute flux scaling} % (SCI-AFX)
\label{sec:sci-afx}

% \qr{Phil}

After the relative flux scaling has been applied, each extracted spectrum is
divided by a grism- and tilt-specific sensitivity function produced by a
dedicated calibration pipeline (Sect.~\ref{sec:cal-afx}).  This converts the
instrumental flux units into physical units chosen to be
\si{erg.cm^{-2}.s^{-1}.\angstrom^{-1}}.  In this way, each individual spectrum
is flux calibrated in an absolute sense, making them intrinsically comparable.
The flags associated with the sensitivity product are carried forward in the
spectrum bitmask.

\subsubsection{Spectra combination} % (SCI-COM)
\label{sec:sci-com}

% \qr{YCo}

Multiple independent realisations -- from different detectors and \ac{ROS}
dithers, potentially from different pointings at their intersections -- of the
intrinsic spectrum of a given source are combined by the `spectra combination'
\ac{PE} to produce a consolidated -- both from a statistic and systematic point
of view -- estimate of the flux-calibrated source spectrum.  This operation is
run for every source on a MER tile basis, and combines all spectra for the
given source available to date, from pointings covering this tile.

Given the potential issues still affecting individual spectra (e.g.,
decontamination residuals, unmasked bad pixels, bright 0th-order diffraction
spikes, ghosts, etc.), a plain average of the single-dither spectra is not
robust enough.  On the other hand, given the small number of single-dither
spectra to be combined -- typically $N \approx 4$ corresponding to the four
dithers in an \ac{ROS} --, a plain median is not statistically efficient.

An outlier-detection scheme, similar to Grubbs' bilateral test
\citep{Grubbs1969} but using the `pull' in place of the \emph{z}-score, is
therefore run first at the pixel-level among the $N$ flux realisations
$f_{i} \pm \sigma_{i}$.  For each measurement $i$ of the $N$-sample, its pull
$p_{i}$ is defined as
\begin{equation}
  \label{eq:pull}
  p_i = \frac{f_{i} - \bar{f}_{i}}{%
    \sqrt{\sigma_{i}^{2} + \bar{\sigma}_{i}^{2} + \sigma_{0}^{2}}}
\end{equation}
where $\bar{f}_{i}$ (resp. $\bar{\sigma}_{i}$) is the inverse-variance weighted
average (resp. its associated error) of the sample \emph{without}
measurement~$i$, and $\sigma_{0}$ is an estimate of the \emph{intrinsic}
(beyond statistical) dispersion among the single-dither spectra (accounting
e.g., for flux calibration errors).  In practice, outlying flux realisations
are defined as $|p_{i}| > p_{\max} = 4$.  This procedure iteratively identifies
and masks out significantly discordant pixels still affected by yet-unflagged
bad pixels, decontamination residuals, bright 0th-order diffraction spikes,
ghosts, etc.

A standard inverse-variance weighted average is then performed over the
$n \leq N$ unclipped values, to compute the combined signal, its associated
variance, and set the following bitmask values (see Table~\ref{tab:bitmask}):
\begin{description}
\item[\path{NOT_USE},] if $n < 2$ or $n/N < 50\%$ (i.e., more than 50\% of the
  flux realisations were clipped out), or if $z > 5$, where $z$ is the
  $z$-score (statistical significance) of the final
  $\chi^{2} = \sum_{i=1}^{n} p_{i}^{2}$ (i.e., the distribution of selected
  flux realisations is not compatible with flux errors and intrinsic
  dispersion);
\item[\path{LOW_SNR},] if $n/N < 70\%$ (more than 30\% of the flux realisations
  were clipped out);
\item[\path{EXT_PBR},] if $z > 3$ (the distribution of selected flux
  realisations is barely compatible with flux errors and intrinsic dispersion);
\item[\path{HIGH},] if any $p_{i} > +3$, where $p_{i}$ is the pull of the
  \emph{selected} flux realisations.
\item[\path{LOW},] similar to \path{HIGH}, but if any $p_{i} < -3$.
\end{description}

\begin{table}
  \caption{Description of the bit used in the resampled spectrogram and
    combined spectrum bitmask.  All bits but \protect\path{NOT_USE} are
    warnings of suspicious behaviour.}
    \label{tab:bitmask}
    \centering
    \begin{tabular}{ccl}
      \hline\hline
      No & Description \\
      \hline
      0 & \path{NOT_USE}  & Could not be computed \\
      1 & \path{LOW_SNR}  & Low \ac{S/N}               \\
      2 & \path{EXT_PBR}  & Suspicious spectrum extraction \\
      3 & \path{HIGH}     & Suspected high \\
      4 & \path{LOW}      & Suspected low  \\
      5 & \path{REL_FLUX} & Suspicious relative flux scaling \\
      6 & \path{ABS_FLUX} & Suspicious absolute flux scaling \\
      \hline
    \end{tabular}
\end{table}

\begin{figure*}
  % \centering
  % \includegraphics[width=\linewidth]{EUC_SIR_W-COMBSPEC_2684805874647806467}
  \sidecaption
  \includegraphics[width=12cm]{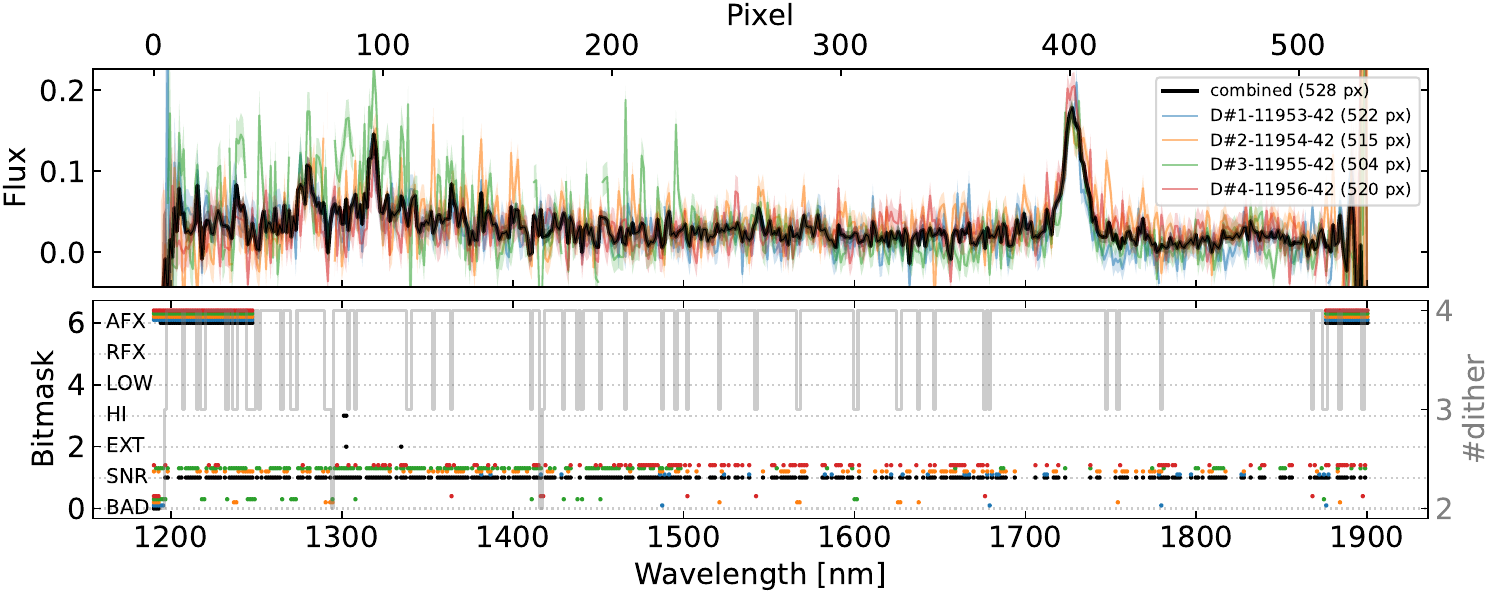}
  \caption{\emph{Top:} the four single-dither extracted spectra for object ID
    2684805874647806467 (coloured), as well as the combined spectrum (black);
    fluxes are in units of \SI{e-17}{erg.cm^{-2}.s^{-1}.\AA^{-1}}.
    \emph{Bottom:} individual bit flags (see Table~\ref{tab:bitmask}) for all
    the 531~pixels of the single-dither spectra (coloured) and combined
    spectrum (black), and effective number of pixels that entered the
    combination (grey).}
  \label{fig:combspec}
\end{figure*}

Furthermore, the standard deviation of the effective \ac{LSF} of the combined
spectrum is computed as the \ac{RMS} of the standard deviation of input
single-dither \acsp{LSF} (see Sect.~\ref{sec:sci-ext}).
Figure~\ref{fig:combspec} shows examples of single-dither and combined spectra
(after relative and absolute flux calibrations).

\subsection{Calibration processing elements}
\label{sec:calibration-pipeline}

The preprocessing calibration \acsp{PE} are presented in \cite{Q1-TP003}, and
we describe the relevant information there.

\subsubsection{Spectra location calibration} % (CAL-LOC)
\label{sec:cal-loc}

% \qr{MFu}

\paragraph{Astrometric modelling (OPT calibration).} % CAL-LOC-AST
This calibration \ac{PE} aims at deriving an astrometric model, i.e., the
mapping between the sky coordinates (RA, Dec) of a source, as identified in the
MER catalogue, and the corresponding 1st-order reference position in the
\ac{FP}, first in R-MOSAIC coordinates \citep[in~mm,][]{EuclidSkyNISP} and
ultimately in detector pixel coordinates (through the use of the metrologic
layout of the \ac{FP}): reference positions $(x_0, y_0)$ for 0th-order, and
$(x_1, y_1)$ for the 1st-order, at reference wavelength
$\lambda_{1} = \SI{1504}{nm}$ of the stellar spectral \ion{Mg}{I} (blended)
feature.

This calibration is derived from astrometric calibration pointings, in which
numerous bright point sources are observed simultaneously, by mapping sky
coordinates to measured 1st-order star absorption positions in the registered
exposure.  It uses a preliminary mapping specifically derived and validated
during the \ac{PV} phase, as well as the \ac{FP} metrology to convert \ac{FP}
coordinates (in~mm) to detector coordinates (in~pixels).  We note that the
ground-based metrology, derived from measurements at room temperature, was not
precise enough; an ad hoc effective metrology was developed -- only including
translation terms in the Q1 release -- to insure spectral continuity between
adjacent detectors.

\paragraph{Spectroscopic modelling (CRV and IDS calibrations).} % CAL-LOC-SPE
The spectroscopic model (sky position to spectrogram mapping), including
spectral distortions and wavelength solution, slightly varies over the~\ac{FP}.
The SIR reduction pipeline describes these changes using global spectroscopic
models, calibrated using the following two sets of observations.
\begin{description}
  % PSF not used in Q1, but used for validation of resolving power
\item[Curvature model (CRV):] the same astrometric calibration pointings, in
  which many point (stellar) sources are observed simultaneously, by measuring
  the cross-dispersion offset % and width
  of the spectral trace as a function of position along the dispersion axis, to
  accurately describe the geometrical shape of the spectrogram (spectral
  distortions).
\item[Wavelength solution (IDS):] dedicated observations during the \ac{PV}
  phase of a bright \acl{PN}\acused{PN} \cite[\acs{PN}
  SMC-SMP-20,][]{Paterson-EP32}, whereby the \ac{PN} is observed at
  $16 \times 5 = 80$ different positions in the NISP \ac{FP}, and the bright
  emission lines in the \ac{PN} spectrum are used to derive the mapping
  $\lambda(D)$ -- namely the \ac{IDS} -- between tabulated wavelengths
  $\lambda$ and measured positions $D$ along the spectral trace.  See Figure~15
  of \cite{EuclidSkyNISP} for examples of spectrograms, spectra, and reference
  wavelengths of \ac{PN} SMC-SMP-20 used in this procedure.
\end{description}
By constructing the spectroscopic model over the full \ac{FP}, the calibration
procedure allows us to predict for each source, given its coordinates in the
sky (but fully independently of its intensity), the geometric and chromatic
description of the 0th- and 1st-order spectrograms.  As a consequence, SIR
\ac{PF} can handle spectra for any source from the MER catalogue,
notwithstanding its magnitude.

\subsubsection{Detector scaling calibration} % (CAL-FF)
\label{sec:cal-ff}

% \qr{Phil, Ranga: awaiting final check by Ranga}

As mentioned in Sect.~\ref{sec:sci-ff}, the scaling acts as both small- and
large-scale flat fields -- correcting for detector-related \ac{QE} fluctuations
-- but also includes the effective \ac{QE} conversion factor for each pixel.
The pixel-level \ac{QE} was measured on the ground for all 16 detectors at
40~wavelengths between \num{600} and \SI{2550}{nm}, and shows a weak
percent-level stochastic dependence on wavelength (Euclid Collaboration: Kubik
et al., in prep.).  The detector-scaling calibration product (`master flat') is
a set of $2040\times 2040$ maps (one per detector) representing the
\emph{effective} \ac{QE} averaged over the \RGE passband (see below).

Except for the pixels illuminated by the brightest sources, most of the signal
in the pixels in the grism spectroscopic data arises from the zodiacal light,
which accounts for about 1000~electrons in a nominal 550~s exposure.  The
zodiacal light at these wavelengths is assumed to have an intensity power-law
spectral density $I_{\nu}\simeq\nu^{-0.8}$ per unit of frequency~$\nu$
\citep{Kelsall98, GWC00}, which, after conversion to electrons per spectral
pixel using the sensitivity curve (Sect.~\ref{sec:cal-afx}), is used to compute
the weighted average of the intrinsic \ac{QE} values for each
pixel. %, i.e., the effective \ac{QE} map.
Although uniform weighting by the zodiacal spectrum for all pixels in the field
of view is a crude approximation of the complex illumination scene, the \ac{QE}
spatial fluctuations are not significantly chromatic -- i.e.,
$\delta\text{QE}(i, j, \lambda) \approx \delta_{ij} \times
\overline{\text{QE}}(\lambda)$ with $\delta_{ij}$ the gray fluctuation of pixel
$(i,j)$ -- and this approach is therefore well justified.  The master flat is
the same for all red grism/tilt configurations, but since \ac{QE} varies with
wavelength, it still depends on the grism passband.

Local variations in the \ac{QE} maps averaged over several hundreds of pixels
are typically correcting the input signal at the percent level, whereas the
shot-noise in the detectors is typically 3 to 4\% in a blank region of the
detector for standard exposure times.  This means that the application of the
flat for a well-behaved part of the image is relatively benign.  However, the
detector scaling does have a significant net-positive effect, because it also
corrects for discontinuities at the boundaries of well identified detector
artefacts (e.g., the `fish'-shaped region in DET21, or the `duck'-shaped
structure in DET11, see Figure~A.1 of \citealt{EuclidSkyNISP}) showing abrupt
changes in \ac{QE} (at the 3 to 5\% level).  These effects are efficiently
handled by the detector scaling product, and lead to a significant flattening
of the images after application (see an illustration on the `duck' in
Fig.~\ref{fig:duckgone}).

In the current implementation, the detector scaling calibration does not use
the multi-chromatic flat exposures from the internal calibration unit
\citep{EuclidSkyNISP, EuclidSkyNISPCU}, but only relies on ground-based
multi-wavelength measurements.  It is a foreseen development of SIR \ac{PF} to
estimate and correct for potential time evolution of QE maps from in-flight
observations.

\subsubsection{Relative flux calibration} % (CAL-RFX)
\label{sec:cal-rfx}

% \qr{Dan}

The relative flux calibration module computes the relative transmission
variations of the instrument as a function of position on the \ac{FP},
wavelength, time, and the instrumental configuration (grism/tilt) used.  These
transmission variations are corrected for at the level of a single-dither
1D~extracted spectrum in the relative flux scaling \ac{PE}
(Sect.~\ref{sec:sci-rfx}), prior to combination of all the spectra.  It is
crucial that this module correctly estimates the transmission variations in
order to ensure consistent flux measurements for the mission.

The spectral flux measured for the same source observed at different locations
on the \ac{FP} will vary due to transmission (spatial) fluctuations of the
instrument, arising from a combination of optical and detector effects.  For
example, vignetting on the order of 10\% at one edge of the focal plane is
expected for acquisitions at the $\pm \ang{4}$ tilted-grism positions
\citep{EuclidSkyOverview}.  Moreover, the large-scale flat pattern may be
chromatic, differing at the blue and red end of the grism spectra.

To measure and correct for this effect, we use repeat observations of bright
($16 < \HE < 18$) stars at random positions in the self-calibration field
\citep{Q1-TP001}.  Such a large-scale retrospective relative spectrophotometric
self-calibration procedure has been described and tested in
\cite{2017MNRAS.467.3677M}, and a working version of it has also been
implemented for NISP-P \citep{Q1-TP003}.  The dithering pattern of the \Euclid
self-calibration observations ensures that sources will illuminate different
parts of the same detector, different detectors in the focal plane, at
different epochs, and with different grism/tilt configurations, thus providing
the necessary constraints to map variations in the large-scale response of the
instrument.  The extracted spectra used for calibration have low levels of
contamination (or have undergone successful decontamination).  By sampling the
same sources at different positions on the focal plane, we build up statistical
constraints on the large-scale flat pattern that needs to be corrected to
ensure consistent flux measurements regardless of source position.

\begin{figure}
  \centering
  \includegraphics[width=\linewidth]{%
    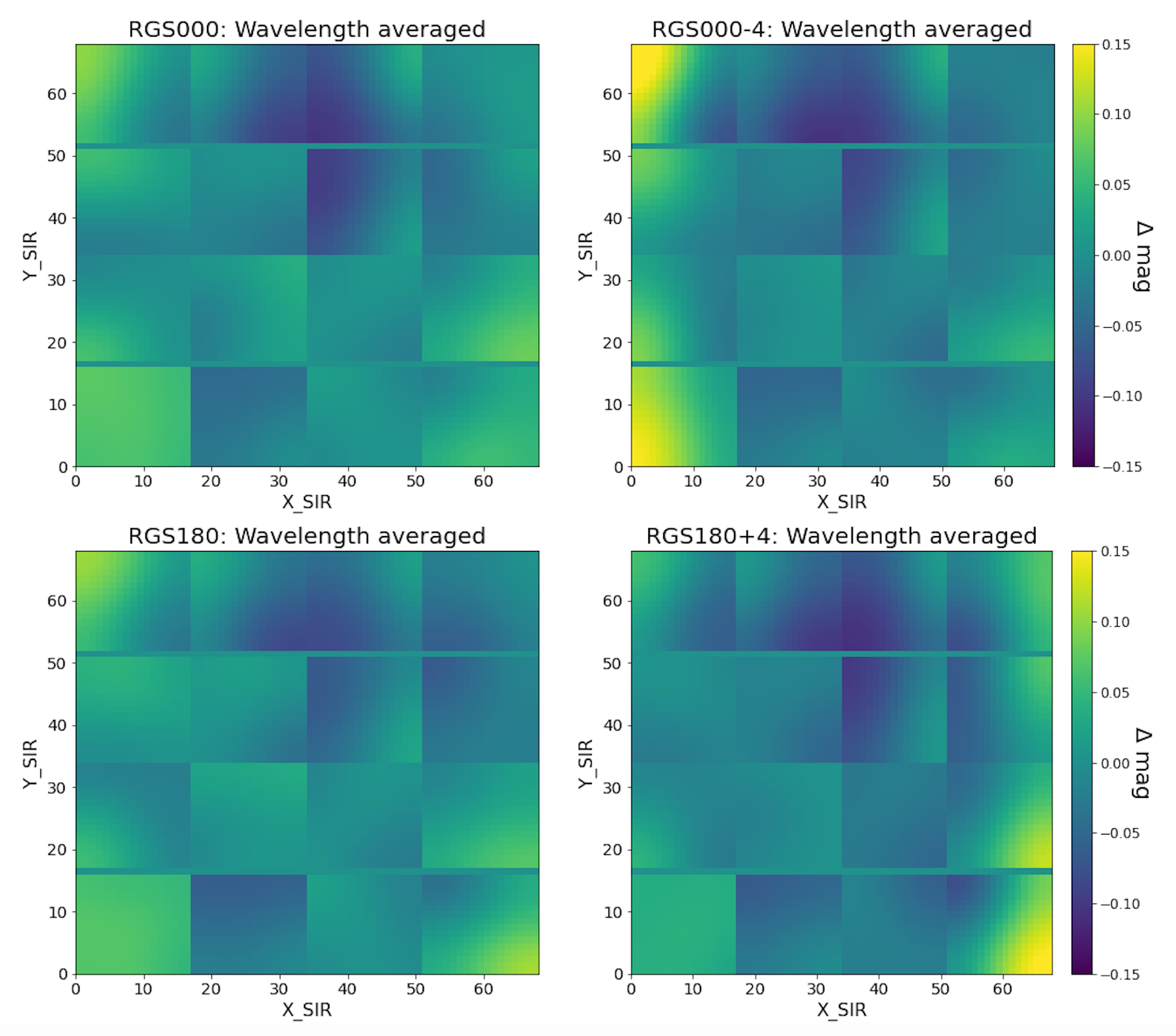}
  \caption{Relative flux solution in SIR coordinates derived for the four
    grism/tilt configurations by the relative flux calibration.  The value
    $\Delta$mag indicates the correction in magnitude that should be applied to
    the portion of a spectrum landing at the given focal plane position.  The
    vignetting at the focal plane sides in tilted configurations
    RGS000\texttt{-}4 (top right) and RGS180\texttt{+}4 (bottom right) is
    apparent, as are large-scale features in common with the imaging flux
    solution (so-called `flat field', \citealt{Q1-TP003}).  This solution is
    found to be mostly achromatic, and is therefore averaged over wavelength.}
  \label{fig:relflux}
\end{figure}

In practice for the Q1 release, the large-scale response is found to be nearly
achromatic, and therefore, to maximise the \ac{S/N} of the solution, an
achromatic solution is derived for all wavelengths that depends only on the
position of the observation on the focal plane (see Fig.~\ref{fig:relflux}).
This solution shows similarities to the NISP-P large-scale flat
\citep{Q1-TP003} and displays the vignetting pattern in the tilted grism
configurations expected based on optical simulations of the instrument.  It is
also shown to correct repeated spectra of bright sources such that they are in
agreement.  At present, the solution appears to be stable with time, and will
be monitored for evolution as the mission progresses.

\subsubsection{Absolute flux calibration} % (CAL-AFX)
\label{sec:cal-afx}

% \qr{Phil \& Jeff}

The absolute flux calibration pipeline is designed to create a sensitivity
function that is used to convert instrumental signal units (electrons per
sampling element) into astrophysical flux units
(\si{erg.s^{-1}.cm^{-2}.\AA^{-1}}, see Sect.~\ref{sec:sci-afx}).  The
sensitivity function was created by first averaging repeat observations of the
flux calibration star GRW+70\,5824, a DA2.4 white dwarf
\citep{2011ApJ...743..138G} acquired during the \ac{PV} phase in a pattern of
five points on each detector \citep{EuclidSkyNISP}, independently for each of
the four red grism/tilt combinations (RGS000\texttt{+}0/\texttt{-}4,
RGS180\texttt{+}0/\texttt{+}4).  For the Q1 release, a separate sensitivity
function is created for each grism/tilt combination from the pull-clipped
average of $16 \times 5 = 80$~single-dither spectra (after relative flux
scaling) of the standard star.

\begin{figure}
  \centering
  \includegraphics[width=.8\linewidth]{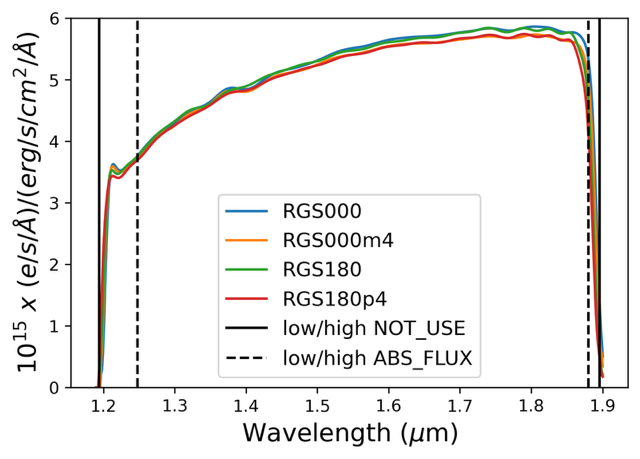}
  \caption{Sensitivity functions for the four red grisms
    (RGS000\texttt{+}0/\texttt{-}4, RGS180\texttt{+}0/\texttt{+}4), derived
    from \ac{PV} observations of the white dwarf GRW+70.  The vertical black
    solid and dashed lines represent the boundaries of the
    \protect\path{NOT_USE} and \protect\path{ABS_FLUX} (suspicious) flags.  In
    this figure, we show the limits for RGS180\texttt{+}0, but each grism/tilt
    configuration has slightly different limits.}
  \label{fig:AFX_sens}
\end{figure}

After the single-dither spectra of the reference star have been suitably
extracted and averaged, we convert the instrumental flux units into units of
\si{e.s^{-1}.\AA^{-1}}, and then divide by a suitably-matched reference
spectrum.  The reference spectrum used was a model spectrum from
CALSPEC\footnote{%
  \url{https://archive.stsci.edu/hlsps/reference-atlases/cdbs/current_calspec/grw_70d5824_mod_001.fits}}
\citep{2020AJ....160...21B}, resampled to \SI{0.1}{nm}, convolved to an
effective spectral resolution of \SI{3.3}{nm} (close to the spectral resolution
of the red grism for a point source), and resampled onto the SIR wavelengths.
The Q1 sensitivity curves for the four grism/tilt configurations are shown in
Fig.~\ref{fig:AFX_sens}.

The bitmask layer is used to flag the sensitivity function, wavelength by
wavelength: for the Q1 release, the \path{NOT_USE} flag is set on domains where
the sensitivity function deviates by more than 5\% from pre-launch
expectations; and the \path{ABS_FLUX} (suspicious) flag is intended to
emphasise spectral domains where the throughput is less than 80\% of the
maximum throughput at the band edges.  We note that, during the Q1 production,
a software error led to some excess flagging of \path{ABS_FLUX} pixels on the
blue-side of the spectral domain (see Figs.~\ref{fig:combspec} and
\ref{fig:AFX_sens}); this problem has been resolved for future releases.

\subsection{Q1 validation and data quality control}
\label{sec:dqc-validation}

\subsubsection{Validation pipeline}
\label{sec:validation}

The current validation process for the SIR \ac{PF} encompasses more than 20 test
cases, designed to ascertain the conformity of the pipeline and its data
products with the established requirements.  The principal objective of the
validation tests is to evaluate in detail the performance of each \ac{PE} of the
scientific pipeline, from the spectra location to combination (see
Sect.~\ref{sec:scientific-pipeline}).  The quality of the data is instead
assessed on a statistical basis through the Data Quality Control procedure (see
Sect.~\ref{sec:dqc}).  Validation tests are typically conducted on designated
reference fields to assess the impact of modifications and improvements
introduced in each pipeline release.

To validate the SIR pipeline used for the Q1 release, we selected four dithered
pointings in an \ac{ROS} over the COSMOS field \citep{2007ApJS..172....1S}, one
of the \Euclid ancillary fields including a multitude of additional data and
redshifts employed for validation
purposes. % \citep{Q1-TP009} % not part of Q1 splash
We present here the results of two of the most significant validation tests,
namely those that evaluate the accuracy of wavelength and flux calibration,
respectively.  These can provide an overall assessment of the performance of
the entire pipeline.

\begin{figure}
  \centering
  \includegraphics[width=.8\linewidth]{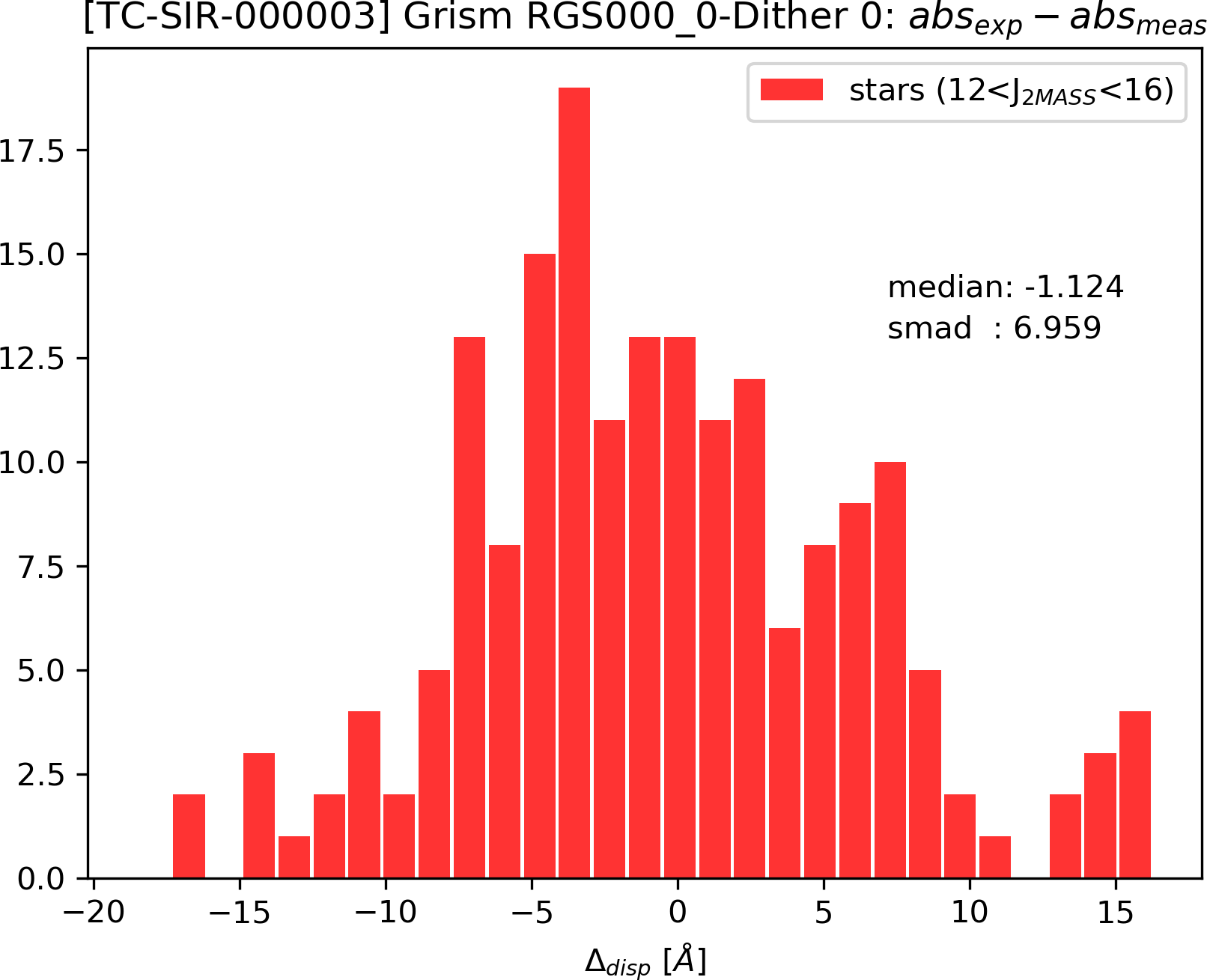}
  \caption{Distribution of the differences along the dispersion axis (in~\AA)
    between expected and measured positions of the \ion{Mg}{I} absorption
    feature used as reference in the 1D~extracted spectra for a sample of
    bright stars.  Results are presented for grism RGS000\texttt{+}0 in a
    single pointing.}
  \label{fig:validation-wave}
\end{figure}

Figure~\ref{fig:validation-wave} illustrates the outcome of the test on the
accuracy of the wavelength solution.  The histogram shows the difference along
the dispersion axis between the nominal position of the blended \ion{Mg}{I}
absorption line $\lambda\SI{1504}{nm}$ and the position measured on the spectra
in a single pointing for bright stars pre-selected in the 2MASS catalogue
($12 \leq \JMASS \leq 16$).  The distribution has a \ac{sMAD} of approximately
0.5~pixel, marginally higher than the requisite 0.4~pixel (\SI{0.54}{nm}).  It
appeared a posteriori that this slight non-conformity in the Q1 production was
mostly due to errors in automatic identification and measurement of the
\ion{Mg}{I} stellar feature; the procedure has been improved for future
releases.

\begin{figure}
  \centering
  \includegraphics[width=.8\linewidth]{%
    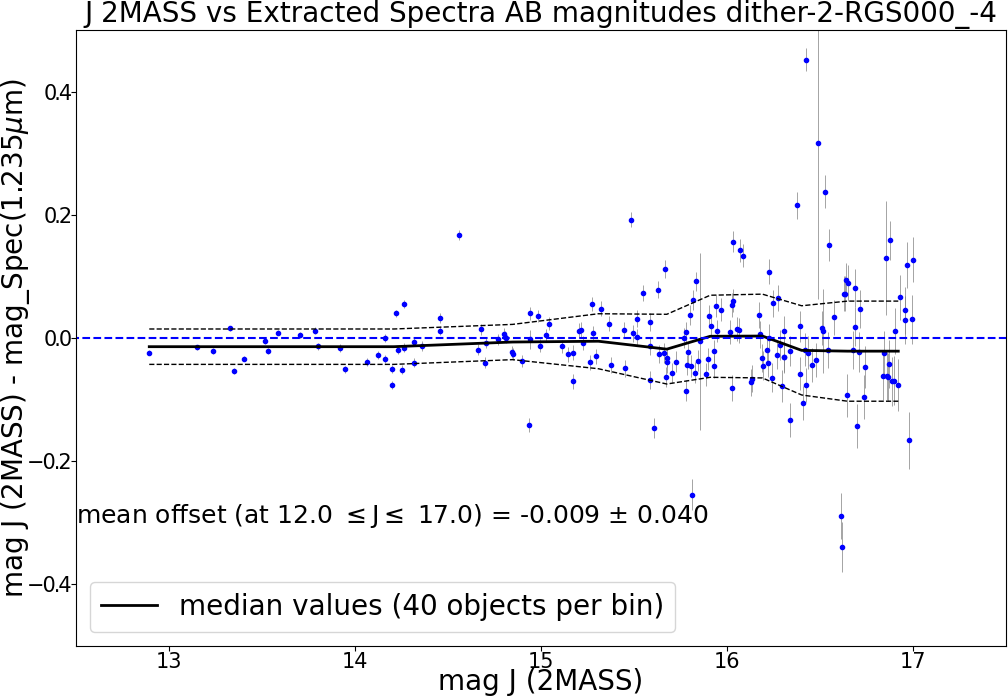}
  \caption{Comparison between the \JMASS magnitudes and magnitudes estimated
    from single-dither spectra in a \SI{50}{nm}-wide region around \JMASS
    effective wavelength (\SI{1235}{nm}) for bright stars in an
    RGS000\texttt{-}4 pointing. The black solid and dashed lines represent the
    40-point running median and \ac{sMAD}, respectively.}
  \label{fig:validation-mag}
\end{figure}

The accuracy of the flux calibration is shown in Fig.~\ref{fig:validation-mag}.
This plot displays the difference between the \JMASS magnitudes and the
magnitudes measured on the 1D spectra in a \SI{50}{nm} domain around the \JMASS
effective wavelength (\SI{1235}{nm}) in a single-dither pointing, for bright
stars ($12 \leq \JMASS \leq 17$).  The distribution is centred around a median
offset of $-0.01$, with an \ac{sMAD} of $0.04$ consistent with flux calibration
objectives.

\subsubsection{Data quality control}
\label{sec:dqc}

The SIR \ac{PF} includes the calculation of a number of \ac{DQC} parameters at
each data-processing step.  These parameters are critical in assessing the
quality of incoming data and identifying potential calibration or reduction
problems.  The \ac{DQC} parameters are statistical quantities, calculated on
each archived data product using pre-selected sources (e.g., bright stars) or
regions of the sky (e.g., excluding bad or contaminated pixels).  They are
computed on the fly within each \ac{PE}, in both the calibration and scientific
pipelines, and then stored in the \Euclid archive system in the XML metadata
associated with each data product (e.g.,
\path{DpdSirScienceFrame}, %\path{DpdSirLocationTable},
%\path{DpdSirExtracted}\-\path{SpectraCollection},
\path{DpdSirCombined}\-\path{SpectraCollection}, etc.).  By collecting all
\acsp{DQC} on hundreds of observations, we can get an overview of the general
trend of each parameter and thus the average quality of the data.  It is beyond
the scope of this paper to provide an exhaustive and complete description of
all SIR \ac{PF} \ac{DQC} parameters.  Instead, the following discussion will
focus on some of the key parameters derived from the scientific pipeline for
the Q1 data processing.  This will illustrate the method used for data
validation and the quality of the Q1 release.

\subsubsection{Quality control for Q1 data set}

The SIR data released in Q1 include 117~observations with red grisms, each
\ac{ROS} consisting of four dithers obtained with the different grism/tilt
configurations (RGS000\texttt{+}0/\texttt{-}4, RGS180\texttt{+}0/\texttt{+}4).
In this section, we present the results for the Q1 release in relation to the
following quantities, derived at each run of the SIR scientific pipeline for
each detector on sub-samples of point sources (typically 10 to 30 per
detector).

\begin{figure}
  \begin{subfigure}{\linewidth}
    \centering
    \includegraphics[width=0.8\linewidth]{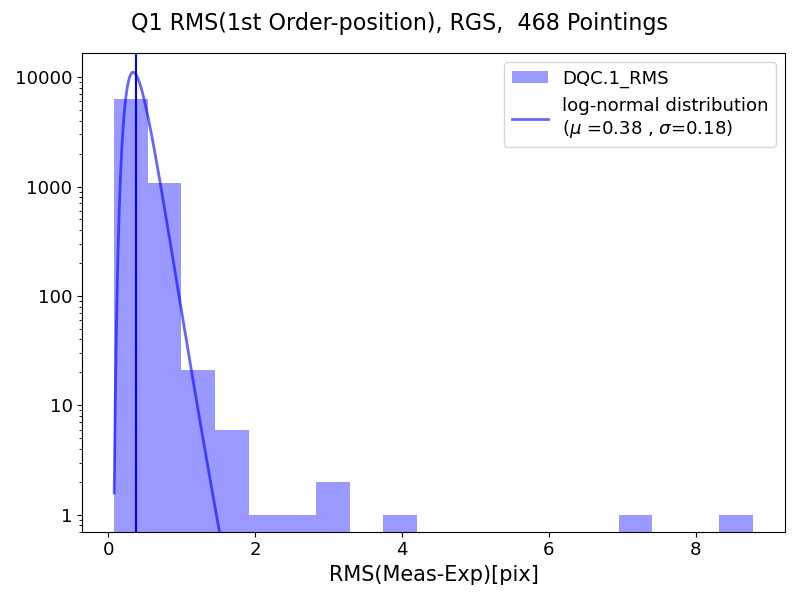}
    % \caption{DQC.1}
    \label{fig:dqc-1storders}
  \end{subfigure}
  \begin{subfigure}{\linewidth}
    \centering
    \includegraphics[width=0.8\linewidth]{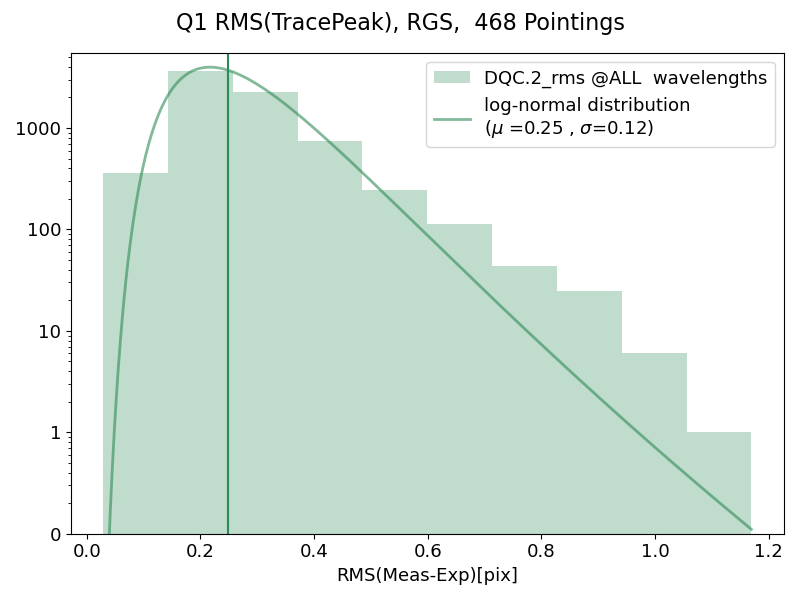}
    % \caption{DQC.2}
    \label{fig:dqc-trace}
  \end{subfigure}
  \begin{subfigure}{\linewidth}
    \centering
    \includegraphics[width=0.8\linewidth]{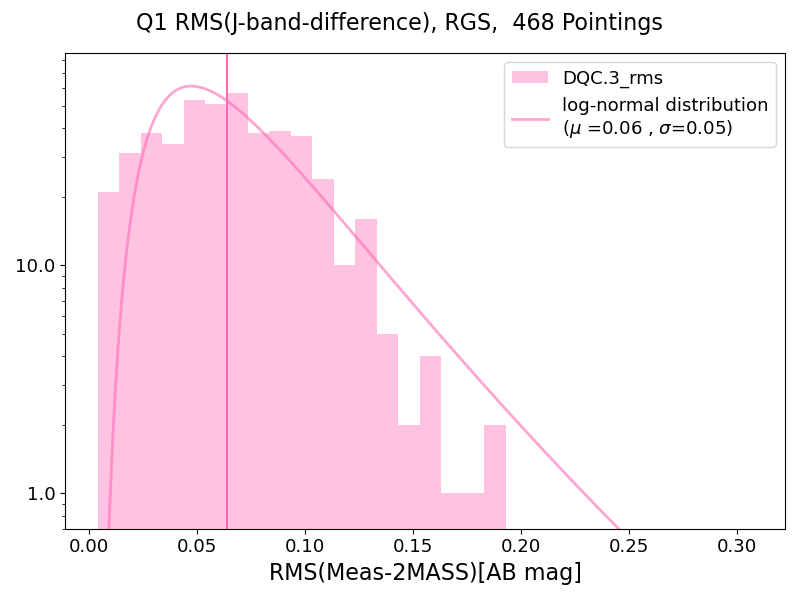}
    % \caption{DQC.3}
    \label{fig:dqc-magrms}
  \end{subfigure}
  \caption{\emph{Top:} distribution of the robust \ac{RMS} of the spectrogram reference
    position offsets (DQC.1) for the Q1 release.  The vertical solid line is
    the median of the distribution and the solid curve is the log-normal PDF.
    \emph{Middle:} same as above but for the cross-dispersion peak-position
    offsets (DQC.2); since the results are not significantly chromatic, we show
    the robust \ac{RMS} averaged over the seven wavelengths.
    \emph{Bottom:} same as above but for $J$-band magnitude offsets (DQC.3).
    \label{fig:dqc}
  }
\end{figure}

\begin{description}
\item[DQC.1:] difference (in~pixels) between the measured and expected position
  of the 1504-nm absorption feature on the 1st-order spectrograms along the
  dispersion axis for bright point sources ($12 < \JMASS < 16$).  This is
  related to the accuracy of the spectra location (Sect.~\ref{sec:sci-loc}),
  and in particular to the optical model and zero-point of the wavelength
  solution (Sect.~\ref{sec:cal-loc});
\item[DQC.2:] difference (in~pixels) between the measured and predicted peak
  position (of the cross-dispersion profile) along the spectrograms at seven
  different wavelengths for bright point sources ($12 \leq \JMASS \leq 16$).
  This is also probing the validity of the spectra location, in relation to the
  curvature model (Sect.~\ref{sec:cal-loc}).
\item[DQC.3:] difference between $J$-band magnitudes measured on the spectra
  and those from the 2MASS catalogue for point sources with
  $16 \leq \JMASS \leq 18$, probing the reliability of the overall flux
  calibration (Sects.~\ref{sec:sci-afx} and~\ref{sec:cal-afx}).
\end{description}
The median and \ac{sMAD} per detector of each \ac{DQC} parameter are stored in
the metadata of the SIR products.  Given their derivation from the difference
between a predicted and a measured quantity, a positive outcome of the \ac{DQC}
parameters is associated with a median value close to zero and an \ac{sMAD}
within a specified threshold.  In order to evaluate a single \ac{DQC} criterion
for each quantity, extending König--Huygens formula, the sum in quadrature of
the median and \ac{sMAD} is used as a robust estimate of the~\ac{RMS}:
\begin{equation}
  \label{eq:rms}
  \rm rRMS = \sqrt{\rm median^2 + sMAD^2}.
\end{equation}
This is illustrated in Fig.~\ref{fig:dqc}, showing the distribution of the
robust \ac{RMS} obtained for the Q1 release for all three \ac{DQC} parameters;
in each case, the distribution is reasonably well represented by a log-normal
PDF.  A specific pointing may exhibit anomalies if any DQC parameter yields
outliers for more than three detectors in these distributions.  Similarly, a
standard \ac{ROS} observation may be problematic if more than two grisms fail
the DQC threshold.  Some outlying detectors were identified in
Fig.~\ref{fig:dqc} (i.e., $\gtrsim 5\sigma$ away from the median \ac{RMS}
value), but in most cases, these are from different pointings.  In cases where
more than three detectors per pointing exhibited outliers, an investigation was
carried out.

While not all the Q1 spectroscopic data release is strictly within the
specified requirements, no invalidating systematic errors were identified.  In
conclusion, this first Q1 release is considered to be of reasonable accuracy
level, in line with the initial performance of the SIR~pipeline, and there are
reasons to be confident that it will further improve in future releases.

\section{Validation of spectroscopic requirements}
\label{sc:perfs}

% \qr{YCo}

In this section, we briefly assess the spectroscopic performance of the NISP
instrument and SIR pipeline from on-orbit observations, in regard to \Euclid's
top-level mission requirements \citep{EuclidSkyOverview}.  Since the Q1 release
only covers red-grism observations from the \ac{EWS}, we do not address here
the specificities of the blue grism and the deep survey.

We note that flux requirements have not been evaluated at the Q1 stage, and are
therefore not addressed here.  Preliminary analyses show that flux performance
(relative flux accuracy and flux limits) are globally on par with requirements,
but we postpone in-depth (red/blue, wide/deep) analyses and validations to
later SIR \ac{PF} publications.

\subsection{Spectral resolution}
\label{sec:specres-reqs}

% See EUCL-IPN-TN-8-005

\Euclid's top-level requirement on spectral resolution (actually resolving
power) of red-grism observations states:
\begin{quote}
  % \item[\texttt{MRD-GC-004}]
  ``The NISP spectrometric channel spectral resolution considering a reference
  \ang{;;0.5}-diameter source shall be
  $\mathcal{R} = \lambda/\Delta\lambda > 380$ over the \SIrange{1250}{1850}{nm}
  spectral range.

  Note: resolution element ($\Delta\lambda$) is defined as the minimum
  wavelength separation at which two spectral lines produced by a \ang{;;0.5}
  object and with the same equivalent width can still be separated.''
\end{quote}
This requirement specifically refers to the \emph{Sparrow
  criterion} %, i.e., nullity of the first and second derivatives at mid-point
\citep{1916ApJ....44...76S, 1995ASPC...77..503J}, for which the physical
resolution element (in~pixels) is $r(\lambda) = 2\sigma(\lambda)$ in the
Gaussian approximation.  Given the native spectral sampling
$s(\lambda) \defeq \d\lambda/\d D$ (in \si{nm.pix^{-1}}, before any spectral
resampling), one defines the resolving power as
\begin{equation}
  \label{eq:respow}
  \mathcal{R} \defeq \frac{\lambda}{\Delta\lambda} =
  \frac{\lambda}{2\sigma(\lambda)\,s(\lambda)}.
\end{equation}

The native spectral sampling $s(\lambda)$ is estimated from measured positions,
directly in R-MOSAIC coordinates, of significant emission lines in the \ac{PN}
SMC-SMP-20 spectrograms acquired all over the \ac{FP} during the \ac{PV} phase
\citep[see Figure~15 of][for an illustration]{EuclidSkyNISP}.  Overall, the
sampling is not found to be significantly dependent on positions in the
\ac{FP}, on wavelength, or on red grism/tilt combination, and hence we adopt
the following constant value:
\begin{equation}
  \label{eq:srgs}
  s = \SI{1.368 \pm 0.025}{nm.pix^{-1}} \quad(\text{median} \pm \text{sMAD}).
\end{equation}
This value is, as it should be, slightly larger than the adopted spectral bin
after wavelength resampling ($\delta\lambda = \SI{1.34}{nm}$, see
Sect.~\ref{sec:sci-ext}).

\begin{figure}
  \centering
  \includegraphics[width=.8\linewidth]{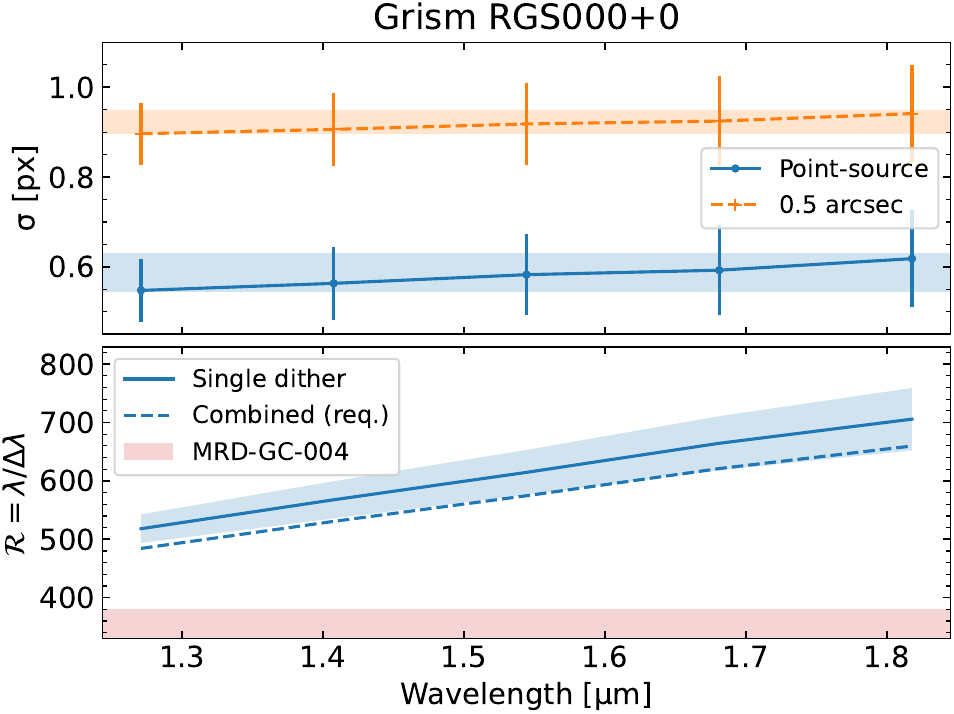}
  \caption{\emph{Top:} distribution (median $\pm$ \ac{sMAD}) of intrinsic
    spectral resolution $\sigma$ (in blue) and effective \ang{;;0.5}-source
    $\sigma_{e}$ (in orange) for RGS000\texttt{+}0 (pointing ID 266626).
    \emph{Bottom:} corresponding resolving power $\mathcal{R}$, for a
    single-dither spectrum (continuous line) and a 4-dither combined spectrum
    (dashed line), assuming an as-required wavelength accuracy (38\% of a
    resolution element, see text); the red zone corresponds to the requirement
    from top-level mission document.}
  \label{fig:specres-RGS000+0}
\end{figure}

Under the assumption of an axisymmetric \ac{PSF}, the \emph{intrinsic}
resolution $\sigma$ of NISP-S is evaluated during the CRV calibration
(Sect.~\ref{sec:cal-loc}) from Gaussian error-function fits to the
cross-dispersion profiles of 1st-order spectrograms of bright-yet-unsaturated
point sources (approximately 20 per detector).  The collection of measurements
-- for all stars and wavelengths -- is then robustly combined into five
spectral bins over the spectral extent (see Fig.~\ref{fig:specres-RGS000+0}),
without noticeable variations over the \ac{FP}.  As expected from instrumental
design, the spectral resolution does not show any significant differences
between grism/tilt configurations.
% We note also that at this stage, the precise wavelength solution is not yet
% available, so the conversion between position along dispersion (in~pix) and
% wavelength is only approximate, with no practical consequence.

In dispersed imaging, the resolution element is directly degraded by the source
extent projected onto the dispersion direction (see Sect.~\ref{sec:sci-ext}).
The requirement refers to a fiducial \ang{;;0.5} source, understood as the full
width at half maximum of an axi-symmetric Gaussian source.  With a nominal NISP
pixel scale of \ang{;;0.3}\,\si{pixel^{-1}}, this corresponds to a
self-contamination contribution to the resolution of
$\sigma_{\text{c}} = \SI{0.710}{pixel}$, to be added in quadrature to
\emph{intrinsic} resolution $\sigma$ estimated from point sources (see
Fig.~\ref{fig:specres-RGS000+0}).  We note that, while the cross-dispersion
profile is significantly under-sampled for point sources
($\sigma \approx \SI{0.6}{pixel}$), it becomes reasonably sampled for
\ang{;;0.5} distant galaxies, \Euclid's primary targets.

Ultimately, the resulting resolving power $\mathcal{R}$ is computed from the
\emph{effective} resolution $\sqrt{\sigma^{2} + \sigma_{\text{c}}^{2}}$, and
shown in Fig.~\ref{fig:specres-RGS000+0}, along with \Euclid's top-level
requirement.

The spectral resolution can be computed either from individual `single-dither'
spectra or on `combined' (multi-dither) spectra.  The later estimate should
therefore include a contribution from the residual wavelength solution errors,
since the wavelengths may not be exactly aligned and a spectral feature is
artificially broadened by the \ac{IDS} inaccuracies.  If, in the
worst-yet-acceptable-case scenario, the wavelength accuracy only marginally
meets requirement (38\% of a resolution element, see below), its effective
contribution to the resolution element is a net increase by approximately
$\sqrt{1 + 0.38^{2}} - 1 = 7\%$ (see Fig.~\ref{fig:specres-RGS000+0}).

In conclusion, the resolving power for a reference \ang{;;0.5}-diameter source
is compatible with $\mathcal{R} \approx \text{\numrange{500}{700}}$, well above
\Euclid's top-level requirement, $\mathcal{R} > 380$ for the red grisms.  It
does not show significant dependence on \ac{FP} position and grism/tilt
configuration.  This seemingly high resolving power is a direct indication of
the superb quality of the NISP-S optics; we note, however, that it is also
dependent on the specific adopted definition of the resolution element.

\subsection{Wavelength accuracy}
\label{sec:wcal-req}

% See EUCL-IPN-TN-8-006

Regarding wavelength accuracy, \Euclid's top-level requirement reads:
\begin{quote}
  % \item[\texttt{MRD-GC-005}]
  ``After calibration, the maximum error in the measured position of a spectral
  feature in the NISP red spectrometric channel (\SIrange{1250}{1850}{nm})
  shall be $< 38\%$ of one resolution element.''
\end{quote}

An analysis of the wavelength accuracy is performed for grism RGS000\texttt{+}0
from intermediate quantities obtained during the \ac{IDS} calibration
(Sect.~\ref{sec:cal-loc}) applied to spectrograms of \ac{PN} SMC-SMP-20.  The
wavelength accuracy is estimated from the robust \ac{RMS} error of the expected
(calibrated) wavelength position compared to the observed one (the emission
line position).  The resolution element, presented in
Sect.~\ref{sec:specres-reqs} for a point source, has been converted to account
for the $\sigma_{\text{PN}} = \SI{0.37}{pixel}$ extent of \ac{PN} SMC-SMP-20
\citep[80\%-energy radius $r_{80\%} = \ang{;;0.20}$,][]{Paterson-EP32}.

\begin{figure}
  \centering
  \includegraphics[width=0.8\linewidth]{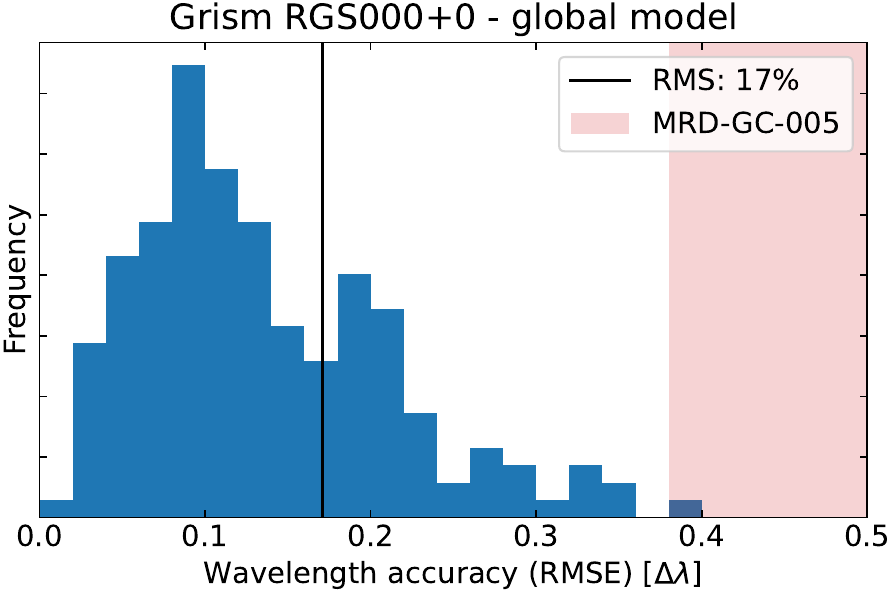}
  \caption{Overall wavelength accuracy distribution for the RGS000\texttt{+}0
    \ac{IDS} (\ac{RMS} error in units of resolution element~$\Delta\lambda$,
    marginalized over all wavelengths and detectors).  The red zone corresponds
    to the maximum error as stipulated in top-level mission requirements
    document.}
  \label{fig:ids-rgs000-rrms}
\end{figure}

Finally, the wavelength accuracy, i.e., the wavelength RMS error in units of
resolution element, is computed per detector and reference line.  Its
distribution does not show strong chromatic or spatial variations (as a
function of wavelength and detector in the \ac{FP}), and the marginalised
distribution is shown in Fig.~\ref{fig:ids-rgs000-rrms}, along with \Euclid's
top-level requirement.

Overall, the \ac{IDS} delivers an estimated mean wavelength accuracy of
\SI{0.23}{pix}, corresponding to only 17\% of the effective resolution element
for SMC-SMP-20 (averaged over wavelengths and detectors in the \ac{FP}), well
below the requirement of 38\%.  Even though a similar analysis has not been
performed on other grism/tilt configurations (namely RGS180\texttt{+}4,
RGS000\texttt{-}4, and RGS180\texttt{+}0) at the time of the Q1 release, it is
expected that a similar wavelength accuracy would be obtained with these
analogous optical configurations.

We note, however, that this analysis is a lower limit, since the wavelength
accuracy has been evaluated on wavelength reference \ac{PN} SMC-SMP-20 itself,
a bright ($J \simeq 15.9$) and compact ($r_{80\%} = \ang{;;0.20}$) source not
exactly representative of the \hbox{\ang{;;0.5}} galaxies that will constitute
the core of the \Euclid sample.  Yet, the overall wavelength accuracy is
confirmed by more general pipeline validation analyses (see
Sect.~\ref{sec:validation}), even though it is limited to bright stars and the
reference spectral feature \ion{Mg}{I} wavelength (\SI{1504}{nm}).  It has been
repeatedly found that the robust \ac{RMS} of the wavelength scatter is around
\SI{0.7}{nm} (see Fig.~\ref{fig:validation-wave}), which corresponds to an
error of less than 30\% of the effective resolution element for a \ang{;;0.5}
source (see Fig.~\ref{fig:specres-RGS000+0}).  Ultimately, a consolidated
assessment of the wavelength accuracy will come from redshift measurements of
reference galaxies performed by SPE \ac{PF} \citep{Q1-TP007}.

\section{Conclusions and planned future work}
\label{sc:conclusion}

In this paper, we have detailed the status of the SIR slitless spectroscopy
science, calibration, and validation pipelines, as well as its interfaces and
principal data products, at the time of the Q1 release \citep{Q1cite}.  The SIR
\ac{PF}, with a codebase exceeding \num{150000} lines -- primarily written in
Python and C\texttt{++} -- is designed to address the complexities of slitless
spectroscopy data from the NISP instrument on board \Euclid.

\begin{figure}
  \centering
  \includegraphics[width=\linewidth]{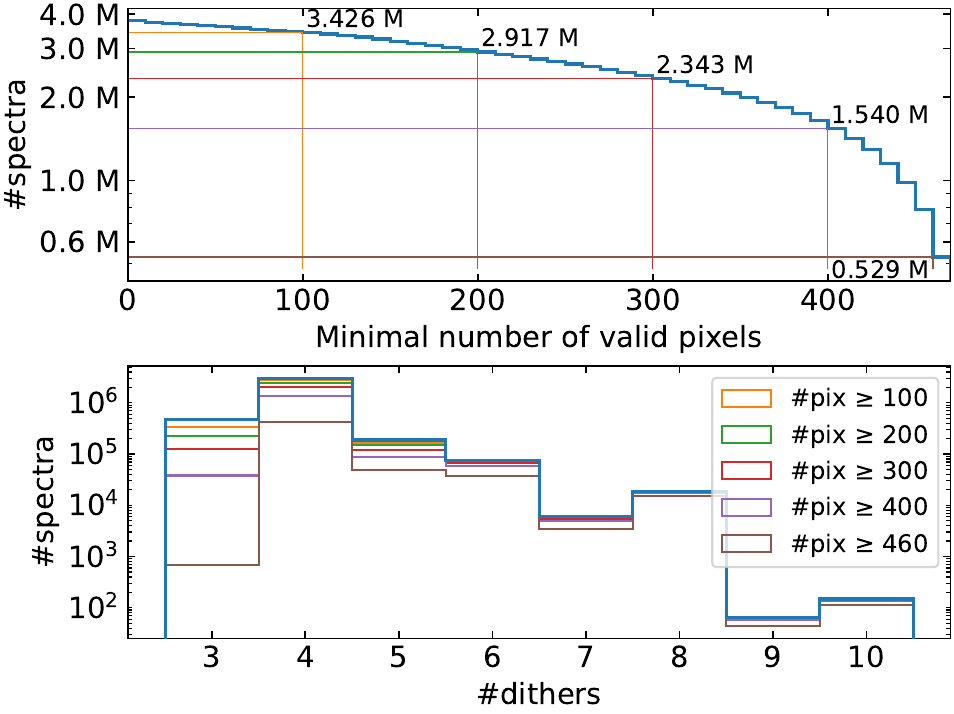}
  \caption{\emph{Top:} distribution, in the Q1 release, of the number (in
    millions) of combined spectra originating from at least two dithers with a
    minimal number of `valid' pixels (see text).  \emph{Bottom:} distribution
    of the number of dithers included in the spectra combination, as a function
    of the minimal number of valid pixels.}
  \label{fig:q1-sample}
\end{figure}

The final Q1 spectroscopic sample includes 4.314~million entries out of the
5.134~million sources with $\HE \leq 22.5$ catalogued by the MER \ac{PF} over
an area of \SI{63.2}{deg^{2}} \citep{Q1-TP001}.  Considering only combined
spectra originating from at least two dithers (to make the outlier clipping
meaningful during combination), 3.778~million spectra have at least one `valid'
pixel (Fig.~\ref{fig:q1-sample}), defined as pixels not flagged as
\path{NOT_USE} or \path{ABS_FLUX} (see Fig.~\ref{fig:AFX_sens}), and
2.343~million with 300~valid pixels or more (up to 468~pixels).  As expected
from the \ac{ROS}, a vast majority (92\%) of the spectra results of the
combination of three or four dithers, but a substantial number of spectra are
computed from eight dithers or more (17733 with at least 400~valid pixels).

We also reported spectroscopic performance in favorable agreement with
top-level mission requirements, in particular regarding the resolving power
$\mathcal{R} \approx \text{\numrange{500}{700}}$, thanks to the exquisite
optical quality of the instrument.

We are fully aware of the current limitations and shortcomings of the pipeline,
and we urge end-users of SIR spectra to validate them thoroughly before drawing
any scientific conclusion \citep[see][]{Q1-TP007}.  However, the modular and
flexible structure of the SIR pipelines ensures that it can continuously
evolve, with ongoing improvements and refinements at each step of the data
calibration and reduction process.  The ability to integrate new methodologies,
enhance existing algorithms, and incorporate feedback from the scientific
community will guarantee that the pipeline remains robust and adaptable to
future requirements.

For the forthcoming first data release (DR1, in prep.), several major
improvements are being implemented, notably:
\begin{itemize}
\item improved \ac{FP} metrology, accounting for translation and rotation of
  the detectors in the \ac{FP};
\item updated Gaussian error-function-based curvature and NISP-S chromatic
  \ac{PSF} models;
\item incorporation of an optimal extraction \citep{robertson_optimal_1986,
    horne_optimal_1986} based on cross-dispersion profiles derived from NISP-S
  PSF-matched MER thumbnails;
\item integration with the spectroscopic survey visibility mask estimation
  process, interfacing with both the SIM and LE3 \acp{PF}, respectively in
  charge of \Euclid data simulations and cosmology analyses.
\end{itemize}

Looking beyond DR1, the pipeline will continue to evolve, with several
promising improvements addressing further technical issues: persistence
correction on spectroscopic exposures \citep{2024SPIE13103E..15K}; masking
and/or subtraction of ghosts and stray light (building on Euclid Collaboration:
Paterson et al., in prep.); improved background subtraction methods
\citep[e.g.,][]{2015ApJS..220....1A}; the introduction of a dedicated model for
0th-order masking/subtraction, accounting for its complex shape and
position-dependent variations across the \ac{FP}.  The pipeline shall also
address the challenges posed by dispersion direction jitter (due to
$\lesssim\ang{0.1}$-RMS fluctuations in the \ac{GWA} position), and extend its
decontamination capabilities to spectrograms from $-1$ and $+2$ dispersion
orders.  To ensure large-scale flux accuracy requirements, it should also
implement an übercal flux-calibration scheme \citep{2008ApJ...674.1217P,
  2017MNRAS.467.3677M}.

Ultimately, the SIR \ac{PF} could move to more advanced dispersed imaging
methods, like advanced decontamination strategies
\citep[e.g.,][]{bella_decontamination_2022}, and other forward-modelling
techniques \citep{2018PASP..130c4501R, 2021PASP..133f4001R,
  2024A&A...684A..21N} to further enhance the reliability and precision of the
extracted spectra.  However, one has to keep in mind that these improved
algorithms have to match the constraints on memory and computing time from the
\ac{SGS}; it is therefore foreseen that these developments would need to be
restricted to a fraction of selected targets of interest among the
approximately \num{50 000} sources of a typical NISP-S exposure.

In conclusion, the SIR \acl{PF} represents a significant achievement in the
reduction of slitless spectroscopic data for the \Euclid mission.  As the
survey progresses, along with our knowledge of the NISP-S instrument, the SIR
\ac{PF} ongoing development, coupled with its modular design, ensures that it
will remain a key tool for advancing cosmological and astrophysical research.

%
% Add the acknowledgement using the achnowledgements environment.
% Do not use \acknowledgement{....} as this affects the formatting
% of the references.
%

\begin{acknowledgements}
  Funded by the European Union – Next Generation EU, Mission 4 Component 1
  Large Scale Lab (LaScaLa), CUP C53D23001390006. % Georg Herzog

  \AckQone                      % Q1 release
  {}
  \AckEC                        % Standard Euclid

  This publication makes use of data products from the Two Micron All Sky
  Survey, which is a joint project of the University of Massachusetts and the
  Infrared Processing and Analysis Center/California Institute of Technology,
  funded by the National Aeronautics and Space Administration and the National
  Science Foundation.

  In the development of our pipeline, we acknowledge use of the Python
  libraries Numpy %\footnote{\url{http://www.numpy.org}}
  \citep{2020Natur.585..357H}, Scipy %\footnote{\url{http://www.scipy.org}}
  \citep{2020NatMe..17..261V},
  Matplotlib %\footnote{\url{http://www.matplotlib.org}}
  \citep{2007CSE.....9...90H}, Astropy %\footnote{\url{http://www.astropy.org}}
  \citep{astropy_2013, astropy_2018, astropy_2022} and
  Pandas %\footnote{\url{http://pandas.pydata.org}}
  \citep{pandas-devpandas_2024}.
\end{acknowledgements}

%
% Here comes the reference list, generated via bibtex from
% your bibfile my.bib and Euclid.bib. Please make sure that
% the same paper is not referenced twice, one from your my.bib
% file, and once from Euclid.bib.
%

\bibliography{Euclid, Q1, sir}

% Now you can add appendices.
% In this example, the appendices are in one column mode.
% If that is not requires, comment out \onecolumn
% Note that appendices in A\&A come {\it after\/} the references.

% \begin{appendix}
%   \onecolumn %If you don't want single column for the Appendix, please
%              %   comment this out
%
% \section{Acronyms}
% \label{sec:acronyms}

\begin{acronym}
  \acro{ADU}{analog-to-digital unit} %
  \acro{DQC}{data quality control} %
  \acro{EWS}{Euclid Wide Survey} %
  \acro{FP}{focal plane} %
  \acro{IDS}{inverse dispersion solution} %
  \acro{GWA}{grism-wheel assembly} %
  \acro{LSF}{line-spread function} %
  \acro{PE}{processing element} %
  \acro{PF}{processing function} %
  \acro{PN}{planetary nebula} %
  \acro{PV}{performance-verification} %
  \acro{PSF}{point spread function} %
  \acro{QE}{quantum efficiency} %
  \acro{QF}{quality factor} %
  \acro{sMAD}{normally-scaled median absolute deviation} %
  \acro{RMS}{root mean square} %
  \acro{ROS}{reference observing sequence} %
  \acro{SGS}{science ground segment} %
  \acro{S/N}{signal-to-noise ratio} %
\end{acronym}

% \end{appendix}

\label{LastPage}
\end{document}

%% file: authors.tex
%%%% Version Friday 14th of March 2025 03:14:24 PM UT
%%%% Please do not edit the author list -- contact ECEB Bureau for changes
% \newcommand{\orcid}[1]{} %% if already defined in aa.cls: comment, or use renewcommand
\author{Euclid Collaboration: Y.~Copin\orcid{0000-0002-5317-7518}\thanks{\email{y.copin@ipnl.in2p3.fr}}\inst{\ref{aff1}}
\and M.~Fumana\orcid{0000-0001-6787-5950}\inst{\ref{aff2}}
\and C.~Mancini\orcid{0000-0002-4297-0561}\inst{\ref{aff2}}
\and P.~N.~Appleton\orcid{0000-0002-7607-8766}\inst{\ref{aff3},\ref{aff4}}
\and R.~Chary\orcid{0000-0001-7583-0621}\inst{\ref{aff4},\ref{aff5}}
\and S.~Conseil\orcid{0000-0002-3657-4191}\inst{\ref{aff1}}
\and A.~L.~Faisst\orcid{0000-0002-9382-9832}\inst{\ref{aff3}}
\and S.~Hemmati\orcid{0000-0003-2226-5395}\inst{\ref{aff3}}
\and D.~C.~Masters\orcid{0000-0001-5382-6138}\inst{\ref{aff4}}
\and C.~Scarlata\orcid{0000-0002-9136-8876}\inst{\ref{aff6}}
\and M.~Scodeggio\inst{\ref{aff2}}
\and A.~Alavi\orcid{0000-0002-8630-6435}\inst{\ref{aff3}}
\and A.~Carle\inst{\ref{aff1}}
\and P.~Casenove\orcid{0009-0006-6736-1670}\inst{\ref{aff7}}
\and T.~Contini\orcid{0000-0003-0275-938X}\inst{\ref{aff8}}
\and I.~Das\orcid{0009-0007-7088-2044}\inst{\ref{aff3}}
\and W.~Gillard\orcid{0000-0003-4744-9748}\inst{\ref{aff9}}
\and G.~Herzog\orcid{0000-0001-5909-4054}\inst{\ref{aff2}}
\and J.~Jacobson\inst{\ref{aff3}}
\and V.~Le~Brun\orcid{0000-0002-5027-1939}\inst{\ref{aff10}}
\and D.~Maino\inst{\ref{aff11},\ref{aff2},\ref{aff12}}
\and G.~Setnikar\orcid{0009-0000-0136-3397}\inst{\ref{aff1}}
\and N.~R.~Stickley\orcid{0000-0003-0987-5738}\inst{\ref{aff13}}
\and D.~Tavagnacco\orcid{0000-0001-7475-9894}\inst{\ref{aff14}}
\and Q.~Xie\inst{\ref{aff3}}
\and N.~Aghanim\orcid{0000-0002-6688-8992}\inst{\ref{aff15}}
\and B.~Altieri\orcid{0000-0003-3936-0284}\inst{\ref{aff16}}
\and A.~Amara\inst{\ref{aff17}}
\and S.~Andreon\orcid{0000-0002-2041-8784}\inst{\ref{aff18}}
\and N.~Auricchio\orcid{0000-0003-4444-8651}\inst{\ref{aff19}}
\and H.~Aussel\orcid{0000-0002-1371-5705}\inst{\ref{aff20}}
\and C.~Baccigalupi\orcid{0000-0002-8211-1630}\inst{\ref{aff21},\ref{aff14},\ref{aff22},\ref{aff23}}
\and M.~Baldi\orcid{0000-0003-4145-1943}\inst{\ref{aff24},\ref{aff19},\ref{aff25}}
\and A.~Balestra\orcid{0000-0002-6967-261X}\inst{\ref{aff26}}
\and S.~Bardelli\orcid{0000-0002-8900-0298}\inst{\ref{aff19}}
\and A.~Basset\inst{\ref{aff7}}
\and P.~Battaglia\orcid{0000-0002-7337-5909}\inst{\ref{aff19}}
\and A.~N.~Belikov\inst{\ref{aff27},\ref{aff28}}
\and A.~Biviano\orcid{0000-0002-0857-0732}\inst{\ref{aff14},\ref{aff21}}
\and A.~Bonchi\orcid{0000-0002-2667-5482}\inst{\ref{aff29}}
\and E.~Branchini\orcid{0000-0002-0808-6908}\inst{\ref{aff30},\ref{aff31},\ref{aff18}}
\and M.~Brescia\orcid{0000-0001-9506-5680}\inst{\ref{aff32},\ref{aff33}}
\and J.~Brinchmann\orcid{0000-0003-4359-8797}\inst{\ref{aff34},\ref{aff35}}
\and S.~Camera\orcid{0000-0003-3399-3574}\inst{\ref{aff36},\ref{aff37},\ref{aff38}}
\and G.~Ca\~nas-Herrera\orcid{0000-0003-2796-2149}\inst{\ref{aff39},\ref{aff40},\ref{aff41}}
\and V.~Capobianco\orcid{0000-0002-3309-7692}\inst{\ref{aff38}}
\and C.~Carbone\orcid{0000-0003-0125-3563}\inst{\ref{aff2}}
\and J.~Carretero\orcid{0000-0002-3130-0204}\inst{\ref{aff42},\ref{aff43}}
\and S.~Casas\orcid{0000-0002-4751-5138}\inst{\ref{aff44}}
\and F.~J.~Castander\orcid{0000-0001-7316-4573}\inst{\ref{aff45},\ref{aff46}}
\and M.~Castellano\orcid{0000-0001-9875-8263}\inst{\ref{aff47}}
\and G.~Castignani\orcid{0000-0001-6831-0687}\inst{\ref{aff19}}
\and S.~Cavuoti\orcid{0000-0002-3787-4196}\inst{\ref{aff33},\ref{aff48}}
\and K.~C.~Chambers\orcid{0000-0001-6965-7789}\inst{\ref{aff49}}
\and A.~Cimatti\inst{\ref{aff50}}
\and C.~Colodro-Conde\inst{\ref{aff51}}
\and G.~Congedo\orcid{0000-0003-2508-0046}\inst{\ref{aff52}}
\and C.~J.~Conselice\orcid{0000-0003-1949-7638}\inst{\ref{aff53}}
\and L.~Conversi\orcid{0000-0002-6710-8476}\inst{\ref{aff54},\ref{aff16}}
\and F.~Courbin\orcid{0000-0003-0758-6510}\inst{\ref{aff55},\ref{aff56}}
\and H.~M.~Courtois\orcid{0000-0003-0509-1776}\inst{\ref{aff57}}
\and A.~Da~Silva\orcid{0000-0002-6385-1609}\inst{\ref{aff58},\ref{aff59}}
\and R.~da~Silva\orcid{0000-0003-4788-677X}\inst{\ref{aff47},\ref{aff29}}
\and H.~Degaudenzi\orcid{0000-0002-5887-6799}\inst{\ref{aff60}}
\and S.~de~la~Torre\inst{\ref{aff10}}
\and G.~De~Lucia\orcid{0000-0002-6220-9104}\inst{\ref{aff14}}
\and A.~M.~Di~Giorgio\orcid{0000-0002-4767-2360}\inst{\ref{aff61}}
\and H.~Dole\orcid{0000-0002-9767-3839}\inst{\ref{aff15}}
\and F.~Dubath\orcid{0000-0002-6533-2810}\inst{\ref{aff60}}
\and X.~Dupac\inst{\ref{aff16}}
\and S.~Dusini\orcid{0000-0002-1128-0664}\inst{\ref{aff62}}
\and A.~Ealet\orcid{0000-0003-3070-014X}\inst{\ref{aff1}}
\and S.~Escoffier\orcid{0000-0002-2847-7498}\inst{\ref{aff9}}
\and M.~Farina\orcid{0000-0002-3089-7846}\inst{\ref{aff61}}
\and R.~Farinelli\inst{\ref{aff19}}
\and S.~Ferriol\inst{\ref{aff1}}
\and F.~Finelli\orcid{0000-0002-6694-3269}\inst{\ref{aff19},\ref{aff63}}
\and S.~Fotopoulou\orcid{0000-0002-9686-254X}\inst{\ref{aff64}}
\and N.~Fourmanoit\orcid{0009-0005-6816-6925}\inst{\ref{aff9}}
\and M.~Frailis\orcid{0000-0002-7400-2135}\inst{\ref{aff14}}
\and E.~Franceschi\orcid{0000-0002-0585-6591}\inst{\ref{aff19}}
\and P.~Franzetti\inst{\ref{aff2}}
\and S.~Galeotta\orcid{0000-0002-3748-5115}\inst{\ref{aff14}}
\and K.~George\orcid{0000-0002-1734-8455}\inst{\ref{aff65}}
\and B.~Gillis\orcid{0000-0002-4478-1270}\inst{\ref{aff52}}
\and C.~Giocoli\orcid{0000-0002-9590-7961}\inst{\ref{aff19},\ref{aff25}}
\and J.~Gracia-Carpio\inst{\ref{aff66}}
\and B.~R.~Granett\orcid{0000-0003-2694-9284}\inst{\ref{aff18}}
\and A.~Grazian\orcid{0000-0002-5688-0663}\inst{\ref{aff26}}
\and F.~Grupp\inst{\ref{aff66},\ref{aff65}}
\and L.~Guzzo\orcid{0000-0001-8264-5192}\inst{\ref{aff11},\ref{aff18},\ref{aff12}}
\and S.~V.~H.~Haugan\orcid{0000-0001-9648-7260}\inst{\ref{aff67}}
\and J.~Hoar\inst{\ref{aff16}}
\and H.~Hoekstra\orcid{0000-0002-0641-3231}\inst{\ref{aff41}}
\and W.~Holmes\inst{\ref{aff68}}
\and I.~M.~Hook\orcid{0000-0002-2960-978X}\inst{\ref{aff69}}
\and F.~Hormuth\inst{\ref{aff70}}
\and A.~Hornstrup\orcid{0000-0002-3363-0936}\inst{\ref{aff71},\ref{aff72}}
\and P.~Hudelot\inst{\ref{aff73}}
\and K.~Jahnke\orcid{0000-0003-3804-2137}\inst{\ref{aff74}}
\and M.~Jhabvala\inst{\ref{aff75}}
\and B.~Joachimi\orcid{0000-0001-7494-1303}\inst{\ref{aff76}}
\and E.~Keih\"anen\orcid{0000-0003-1804-7715}\inst{\ref{aff77}}
\and S.~Kermiche\orcid{0000-0002-0302-5735}\inst{\ref{aff9}}
\and A.~Kiessling\orcid{0000-0002-2590-1273}\inst{\ref{aff68}}
\and B.~Kubik\orcid{0009-0006-5823-4880}\inst{\ref{aff1}}
\and K.~Kuijken\orcid{0000-0002-3827-0175}\inst{\ref{aff41}}
\and M.~K\"ummel\orcid{0000-0003-2791-2117}\inst{\ref{aff65}}
\and M.~Kunz\orcid{0000-0002-3052-7394}\inst{\ref{aff78}}
\and H.~Kurki-Suonio\orcid{0000-0002-4618-3063}\inst{\ref{aff79},\ref{aff80}}
\and Q.~Le~Boulc'h\inst{\ref{aff81}}
\and A.~M.~C.~Le~Brun\orcid{0000-0002-0936-4594}\inst{\ref{aff82}}
\and D.~Le~Mignant\orcid{0000-0002-5339-5515}\inst{\ref{aff10}}
\and S.~Ligori\orcid{0000-0003-4172-4606}\inst{\ref{aff38}}
\and P.~B.~Lilje\orcid{0000-0003-4324-7794}\inst{\ref{aff67}}
\and V.~Lindholm\orcid{0000-0003-2317-5471}\inst{\ref{aff79},\ref{aff80}}
\and I.~Lloro\orcid{0000-0001-5966-1434}\inst{\ref{aff83}}
\and G.~Mainetti\orcid{0000-0003-2384-2377}\inst{\ref{aff81}}
\and E.~Maiorano\orcid{0000-0003-2593-4355}\inst{\ref{aff19}}
\and O.~Mansutti\orcid{0000-0001-5758-4658}\inst{\ref{aff14}}
\and S.~Marcin\inst{\ref{aff84}}
\and O.~Marggraf\orcid{0000-0001-7242-3852}\inst{\ref{aff85}}
\and K.~Markovic\orcid{0000-0001-6764-073X}\inst{\ref{aff68}}
\and M.~Martinelli\orcid{0000-0002-6943-7732}\inst{\ref{aff47},\ref{aff86}}
\and N.~Martinet\orcid{0000-0003-2786-7790}\inst{\ref{aff10}}
\and F.~Marulli\orcid{0000-0002-8850-0303}\inst{\ref{aff87},\ref{aff19},\ref{aff25}}
\and R.~Massey\orcid{0000-0002-6085-3780}\inst{\ref{aff88}}
\and S.~Maurogordato\inst{\ref{aff89}}
\and E.~Medinaceli\orcid{0000-0002-4040-7783}\inst{\ref{aff19}}
\and S.~Mei\orcid{0000-0002-2849-559X}\inst{\ref{aff90},\ref{aff91}}
\and M.~Melchior\inst{\ref{aff92}}
\and Y.~Mellier\inst{\ref{aff93},\ref{aff73}}
\and M.~Meneghetti\orcid{0000-0003-1225-7084}\inst{\ref{aff19},\ref{aff25}}
\and E.~Merlin\orcid{0000-0001-6870-8900}\inst{\ref{aff47}}
\and G.~Meylan\inst{\ref{aff94}}
\and A.~Mora\orcid{0000-0002-1922-8529}\inst{\ref{aff95}}
\and M.~Moresco\orcid{0000-0002-7616-7136}\inst{\ref{aff87},\ref{aff19}}
\and P.~W.~Morris\orcid{0000-0002-5186-4381}\inst{\ref{aff13}}
\and L.~Moscardini\orcid{0000-0002-3473-6716}\inst{\ref{aff87},\ref{aff19},\ref{aff25}}
\and R.~Nakajima\orcid{0009-0009-1213-7040}\inst{\ref{aff85}}
\and C.~Neissner\orcid{0000-0001-8524-4968}\inst{\ref{aff96},\ref{aff43}}
\and R.~C.~Nichol\orcid{0000-0003-0939-6518}\inst{\ref{aff17}}
\and S.-M.~Niemi\inst{\ref{aff39}}
\and J.~W.~Nightingale\orcid{0000-0002-8987-7401}\inst{\ref{aff97}}
\and C.~Padilla\orcid{0000-0001-7951-0166}\inst{\ref{aff96}}
\and S.~Paltani\orcid{0000-0002-8108-9179}\inst{\ref{aff60}}
\and F.~Pasian\orcid{0000-0002-4869-3227}\inst{\ref{aff14}}
\and K.~Pedersen\inst{\ref{aff98}}
\and W.~J.~Percival\orcid{0000-0002-0644-5727}\inst{\ref{aff99},\ref{aff100},\ref{aff101}}
\and V.~Pettorino\inst{\ref{aff39}}
\and S.~Pires\orcid{0000-0002-0249-2104}\inst{\ref{aff20}}
\and G.~Polenta\orcid{0000-0003-4067-9196}\inst{\ref{aff29}}
\and M.~Poncet\inst{\ref{aff7}}
\and L.~A.~Popa\inst{\ref{aff102}}
\and L.~Pozzetti\orcid{0000-0001-7085-0412}\inst{\ref{aff19}}
\and G.~D.~Racca\inst{\ref{aff39},\ref{aff41}}
\and F.~Raison\orcid{0000-0002-7819-6918}\inst{\ref{aff66}}
\and R.~Rebolo\orcid{0000-0003-3767-7085}\inst{\ref{aff51},\ref{aff103},\ref{aff104}}
\and A.~Renzi\orcid{0000-0001-9856-1970}\inst{\ref{aff105},\ref{aff62}}
\and J.~Rhodes\orcid{0000-0002-4485-8549}\inst{\ref{aff68}}
\and G.~Riccio\inst{\ref{aff33}}
\and E.~Romelli\orcid{0000-0003-3069-9222}\inst{\ref{aff14}}
\and M.~Roncarelli\orcid{0000-0001-9587-7822}\inst{\ref{aff19}}
\and E.~Rossetti\orcid{0000-0003-0238-4047}\inst{\ref{aff24}}
\and R.~Saglia\orcid{0000-0003-0378-7032}\inst{\ref{aff65},\ref{aff66}}
\and Z.~Sakr\orcid{0000-0002-4823-3757}\inst{\ref{aff106},\ref{aff8},\ref{aff107}}
\and A.~G.~S\'anchez\orcid{0000-0003-1198-831X}\inst{\ref{aff66}}
\and D.~Sapone\orcid{0000-0001-7089-4503}\inst{\ref{aff108}}
\and B.~Sartoris\orcid{0000-0003-1337-5269}\inst{\ref{aff65},\ref{aff14}}
\and J.~A.~Schewtschenko\orcid{0000-0002-4913-6393}\inst{\ref{aff52}}
\and M.~Schirmer\orcid{0000-0003-2568-9994}\inst{\ref{aff74}}
\and P.~Schneider\orcid{0000-0001-8561-2679}\inst{\ref{aff85}}
\and T.~Schrabback\orcid{0000-0002-6987-7834}\inst{\ref{aff109}}
\and A.~Secroun\orcid{0000-0003-0505-3710}\inst{\ref{aff9}}
\and E.~Sefusatti\orcid{0000-0003-0473-1567}\inst{\ref{aff14},\ref{aff21},\ref{aff22}}
\and G.~Seidel\orcid{0000-0003-2907-353X}\inst{\ref{aff74}}
\and S.~Serrano\orcid{0000-0002-0211-2861}\inst{\ref{aff46},\ref{aff110},\ref{aff45}}
\and P.~Simon\inst{\ref{aff85}}
\and C.~Sirignano\orcid{0000-0002-0995-7146}\inst{\ref{aff105},\ref{aff62}}
\and G.~Sirri\orcid{0000-0003-2626-2853}\inst{\ref{aff25}}
\and A.~Spurio~Mancini\orcid{0000-0001-5698-0990}\inst{\ref{aff111}}
\and L.~Stanco\orcid{0000-0002-9706-5104}\inst{\ref{aff62}}
\and J.~Steinwagner\orcid{0000-0001-7443-1047}\inst{\ref{aff66}}
\and P.~Tallada-Cresp\'{i}\orcid{0000-0002-1336-8328}\inst{\ref{aff42},\ref{aff43}}
\and A.~N.~Taylor\inst{\ref{aff52}}
\and H.~I.~Teplitz\orcid{0000-0002-7064-5424}\inst{\ref{aff4}}
\and I.~Tereno\inst{\ref{aff58},\ref{aff112}}
\and N.~Tessore\orcid{0000-0002-9696-7931}\inst{\ref{aff76}}
\and S.~Toft\orcid{0000-0003-3631-7176}\inst{\ref{aff113},\ref{aff114}}
\and R.~Toledo-Moreo\orcid{0000-0002-2997-4859}\inst{\ref{aff115}}
\and F.~Torradeflot\orcid{0000-0003-1160-1517}\inst{\ref{aff43},\ref{aff42}}
\and I.~Tutusaus\orcid{0000-0002-3199-0399}\inst{\ref{aff8}}
\and L.~Valenziano\orcid{0000-0002-1170-0104}\inst{\ref{aff19},\ref{aff63}}
\and J.~Valiviita\orcid{0000-0001-6225-3693}\inst{\ref{aff79},\ref{aff80}}
\and T.~Vassallo\orcid{0000-0001-6512-6358}\inst{\ref{aff65},\ref{aff14}}
\and G.~Verdoes~Kleijn\orcid{0000-0001-5803-2580}\inst{\ref{aff27}}
\and A.~Veropalumbo\orcid{0000-0003-2387-1194}\inst{\ref{aff18},\ref{aff31},\ref{aff30}}
\and Y.~Wang\orcid{0000-0002-4749-2984}\inst{\ref{aff4}}
\and J.~Weller\orcid{0000-0002-8282-2010}\inst{\ref{aff65},\ref{aff66}}
\and A.~Zacchei\orcid{0000-0003-0396-1192}\inst{\ref{aff14},\ref{aff21}}
\and G.~Zamorani\orcid{0000-0002-2318-301X}\inst{\ref{aff19}}
\and F.~M.~Zerbi\inst{\ref{aff18}}
\and I.~A.~Zinchenko\orcid{0000-0002-2944-2449}\inst{\ref{aff65}}
\and E.~Zucca\orcid{0000-0002-5845-8132}\inst{\ref{aff19}}
\and V.~Allevato\orcid{0000-0001-7232-5152}\inst{\ref{aff33}}
\and M.~Ballardini\orcid{0000-0003-4481-3559}\inst{\ref{aff116},\ref{aff117},\ref{aff19}}
\and M.~Bolzonella\orcid{0000-0003-3278-4607}\inst{\ref{aff19}}
\and E.~Bozzo\orcid{0000-0002-8201-1525}\inst{\ref{aff60}}
\and C.~Burigana\orcid{0000-0002-3005-5796}\inst{\ref{aff118},\ref{aff63}}
\and R.~Cabanac\orcid{0000-0001-6679-2600}\inst{\ref{aff8}}
\and A.~Cappi\inst{\ref{aff19},\ref{aff89}}
\and D.~Di~Ferdinando\inst{\ref{aff25}}
\and J.~A.~Escartin~Vigo\inst{\ref{aff66}}
\and L.~Gabarra\orcid{0000-0002-8486-8856}\inst{\ref{aff119}}
\and M.~Huertas-Company\orcid{0000-0002-1416-8483}\inst{\ref{aff51},\ref{aff120},\ref{aff121},\ref{aff122}}
\and J.~Mart\'{i}n-Fleitas\orcid{0000-0002-8594-569X}\inst{\ref{aff95}}
\and S.~Matthew\orcid{0000-0001-8448-1697}\inst{\ref{aff52}}
\and N.~Mauri\orcid{0000-0001-8196-1548}\inst{\ref{aff50},\ref{aff25}}
\and R.~B.~Metcalf\orcid{0000-0003-3167-2574}\inst{\ref{aff87},\ref{aff19}}
\and A.~Pezzotta\orcid{0000-0003-0726-2268}\inst{\ref{aff123},\ref{aff66}}
\and M.~P\"ontinen\orcid{0000-0001-5442-2530}\inst{\ref{aff79}}
\and C.~Porciani\orcid{0000-0002-7797-2508}\inst{\ref{aff85}}
\and I.~Risso\orcid{0000-0003-2525-7761}\inst{\ref{aff124}}
\and V.~Scottez\inst{\ref{aff93},\ref{aff125}}
\and M.~Sereno\orcid{0000-0003-0302-0325}\inst{\ref{aff19},\ref{aff25}}
\and M.~Tenti\orcid{0000-0002-4254-5901}\inst{\ref{aff25}}
\and M.~Viel\orcid{0000-0002-2642-5707}\inst{\ref{aff21},\ref{aff14},\ref{aff23},\ref{aff22},\ref{aff126}}
\and M.~Wiesmann\orcid{0009-0000-8199-5860}\inst{\ref{aff67}}
\and Y.~Akrami\orcid{0000-0002-2407-7956}\inst{\ref{aff127},\ref{aff128}}
\and S.~Alvi\orcid{0000-0001-5779-8568}\inst{\ref{aff116}}
\and I.~T.~Andika\orcid{0000-0001-6102-9526}\inst{\ref{aff129},\ref{aff130}}
\and S.~Anselmi\orcid{0000-0002-3579-9583}\inst{\ref{aff62},\ref{aff105},\ref{aff131}}
\and M.~Archidiacono\orcid{0000-0003-4952-9012}\inst{\ref{aff11},\ref{aff12}}
\and F.~Atrio-Barandela\orcid{0000-0002-2130-2513}\inst{\ref{aff132}}
\and C.~Benoist\inst{\ref{aff89}}
\and K.~Benson\inst{\ref{aff133}}
\and P.~Bergamini\orcid{0000-0003-1383-9414}\inst{\ref{aff11},\ref{aff19}}
\and D.~Bertacca\orcid{0000-0002-2490-7139}\inst{\ref{aff105},\ref{aff26},\ref{aff62}}
\and M.~Bethermin\orcid{0000-0002-3915-2015}\inst{\ref{aff134}}
\and L.~Bisigello\orcid{0000-0003-0492-4924}\inst{\ref{aff26}}
\and A.~Blanchard\orcid{0000-0001-8555-9003}\inst{\ref{aff8}}
\and L.~Blot\orcid{0000-0002-9622-7167}\inst{\ref{aff135},\ref{aff131}}
\and M.~L.~Brown\orcid{0000-0002-0370-8077}\inst{\ref{aff53}}
\and S.~Bruton\orcid{0000-0002-6503-5218}\inst{\ref{aff13}}
\and A.~Calabro\orcid{0000-0003-2536-1614}\inst{\ref{aff47}}
\and B.~Camacho~Quevedo\orcid{0000-0002-8789-4232}\inst{\ref{aff46},\ref{aff45}}
\and F.~Caro\inst{\ref{aff47}}
\and T.~Castro\orcid{0000-0002-6292-3228}\inst{\ref{aff14},\ref{aff22},\ref{aff21},\ref{aff126}}
\and F.~Cogato\orcid{0000-0003-4632-6113}\inst{\ref{aff87},\ref{aff19}}
\and A.~R.~Cooray\orcid{0000-0002-3892-0190}\inst{\ref{aff136}}
\and O.~Cucciati\orcid{0000-0002-9336-7551}\inst{\ref{aff19}}
\and S.~Davini\orcid{0000-0003-3269-1718}\inst{\ref{aff31}}
\and F.~De~Paolis\orcid{0000-0001-6460-7563}\inst{\ref{aff137},\ref{aff138},\ref{aff139}}
\and G.~Desprez\orcid{0000-0001-8325-1742}\inst{\ref{aff27}}
\and A.~D\'iaz-S\'anchez\orcid{0000-0003-0748-4768}\inst{\ref{aff140}}
\and J.~J.~Diaz\inst{\ref{aff51}}
\and S.~Di~Domizio\orcid{0000-0003-2863-5895}\inst{\ref{aff30},\ref{aff31}}
\and J.~M.~Diego\orcid{0000-0001-9065-3926}\inst{\ref{aff141}}
\and P.-A.~Duc\orcid{0000-0003-3343-6284}\inst{\ref{aff134}}
\and A.~Enia\orcid{0000-0002-0200-2857}\inst{\ref{aff24},\ref{aff19}}
\and Y.~Fang\inst{\ref{aff65}}
\and A.~M.~N.~Ferguson\inst{\ref{aff52}}
\and A.~G.~Ferrari\orcid{0009-0005-5266-4110}\inst{\ref{aff25}}
\and A.~Finoguenov\orcid{0000-0002-4606-5403}\inst{\ref{aff79}}
\and A.~Fontana\orcid{0000-0003-3820-2823}\inst{\ref{aff47}}
\and A.~Franco\orcid{0000-0002-4761-366X}\inst{\ref{aff138},\ref{aff137},\ref{aff139}}
\and K.~Ganga\orcid{0000-0001-8159-8208}\inst{\ref{aff90}}
\and J.~Garc\'ia-Bellido\orcid{0000-0002-9370-8360}\inst{\ref{aff127}}
\and T.~Gasparetto\orcid{0000-0002-7913-4866}\inst{\ref{aff14}}
\and V.~Gautard\inst{\ref{aff142}}
\and E.~Gaztanaga\orcid{0000-0001-9632-0815}\inst{\ref{aff45},\ref{aff46},\ref{aff143}}
\and F.~Giacomini\orcid{0000-0002-3129-2814}\inst{\ref{aff25}}
\and F.~Gianotti\orcid{0000-0003-4666-119X}\inst{\ref{aff19}}
\and G.~Gozaliasl\orcid{0000-0002-0236-919X}\inst{\ref{aff144},\ref{aff79}}
\and A.~Gregorio\orcid{0000-0003-4028-8785}\inst{\ref{aff145},\ref{aff14},\ref{aff22}}
\and M.~Guidi\orcid{0000-0001-9408-1101}\inst{\ref{aff24},\ref{aff19}}
\and C.~M.~Gutierrez\orcid{0000-0001-7854-783X}\inst{\ref{aff146}}
\and A.~Hall\orcid{0000-0002-3139-8651}\inst{\ref{aff52}}
\and W.~G.~Hartley\inst{\ref{aff60}}
\and C.~Hern\'andez-Monteagudo\orcid{0000-0001-5471-9166}\inst{\ref{aff104},\ref{aff51}}
\and H.~Hildebrandt\orcid{0000-0002-9814-3338}\inst{\ref{aff147}}
\and J.~Hjorth\orcid{0000-0002-4571-2306}\inst{\ref{aff98}}
\and S.~Hosseini\inst{\ref{aff8}}
\and J.~J.~E.~Kajava\orcid{0000-0002-3010-8333}\inst{\ref{aff148},\ref{aff149}}
\and Y.~Kang\orcid{0009-0000-8588-7250}\inst{\ref{aff60}}
\and V.~Kansal\orcid{0000-0002-4008-6078}\inst{\ref{aff150},\ref{aff151}}
\and D.~Karagiannis\orcid{0000-0002-4927-0816}\inst{\ref{aff116},\ref{aff152}}
\and K.~Kiiveri\inst{\ref{aff77}}
\and C.~C.~Kirkpatrick\inst{\ref{aff77}}
\and S.~Kruk\orcid{0000-0001-8010-8879}\inst{\ref{aff16}}
\and J.~Le~Graet\orcid{0000-0001-6523-7971}\inst{\ref{aff9}}
\and L.~Legrand\orcid{0000-0003-0610-5252}\inst{\ref{aff153},\ref{aff154}}
\and M.~Lembo\orcid{0000-0002-5271-5070}\inst{\ref{aff116},\ref{aff117}}
\and F.~Lepori\orcid{0009-0000-5061-7138}\inst{\ref{aff155}}
\and G.~Leroy\orcid{0009-0004-2523-4425}\inst{\ref{aff156},\ref{aff88}}
\and G.~F.~Lesci\orcid{0000-0002-4607-2830}\inst{\ref{aff87},\ref{aff19}}
\and J.~Lesgourgues\orcid{0000-0001-7627-353X}\inst{\ref{aff44}}
\and L.~Leuzzi\orcid{0009-0006-4479-7017}\inst{\ref{aff87},\ref{aff19}}
\and T.~I.~Liaudat\orcid{0000-0002-9104-314X}\inst{\ref{aff157}}
\and A.~Loureiro\orcid{0000-0002-4371-0876}\inst{\ref{aff158},\ref{aff159}}
\and J.~Macias-Perez\orcid{0000-0002-5385-2763}\inst{\ref{aff160}}
\and G.~Maggio\orcid{0000-0003-4020-4836}\inst{\ref{aff14}}
\and M.~Magliocchetti\orcid{0000-0001-9158-4838}\inst{\ref{aff61}}
\and F.~Mannucci\orcid{0000-0002-4803-2381}\inst{\ref{aff161}}
\and R.~Maoli\orcid{0000-0002-6065-3025}\inst{\ref{aff162},\ref{aff47}}
\and C.~J.~A.~P.~Martins\orcid{0000-0002-4886-9261}\inst{\ref{aff163},\ref{aff34}}
\and L.~Maurin\orcid{0000-0002-8406-0857}\inst{\ref{aff15}}
\and C.~J.~R.~McPartland\orcid{0000-0003-0639-025X}\inst{\ref{aff72},\ref{aff114}}
\and M.~Miluzio\inst{\ref{aff16},\ref{aff164}}
\and P.~Monaco\orcid{0000-0003-2083-7564}\inst{\ref{aff145},\ref{aff14},\ref{aff22},\ref{aff21}}
\and A.~Montoro\orcid{0000-0003-4730-8590}\inst{\ref{aff45},\ref{aff46}}
\and C.~Moretti\orcid{0000-0003-3314-8936}\inst{\ref{aff23},\ref{aff126},\ref{aff14},\ref{aff21},\ref{aff22}}
\and G.~Morgante\inst{\ref{aff19}}
\and C.~Murray\inst{\ref{aff90}}
\and S.~Nadathur\orcid{0000-0001-9070-3102}\inst{\ref{aff143}}
\and K.~Naidoo\orcid{0000-0002-9182-1802}\inst{\ref{aff143}}
\and A.~Navarro-Alsina\orcid{0000-0002-3173-2592}\inst{\ref{aff85}}
\and S.~Nesseris\orcid{0000-0002-0567-0324}\inst{\ref{aff127}}
\and F.~Passalacqua\orcid{0000-0002-8606-4093}\inst{\ref{aff105},\ref{aff62}}
\and K.~Paterson\orcid{0000-0001-8340-3486}\inst{\ref{aff74}}
\and L.~Patrizii\inst{\ref{aff25}}
\and A.~Pisani\orcid{0000-0002-6146-4437}\inst{\ref{aff9},\ref{aff165}}
\and D.~Potter\orcid{0000-0002-0757-5195}\inst{\ref{aff155}}
\and S.~Quai\orcid{0000-0002-0449-8163}\inst{\ref{aff87},\ref{aff19}}
\and M.~Radovich\orcid{0000-0002-3585-866X}\inst{\ref{aff26}}
\and P.-F.~Rocci\inst{\ref{aff15}}
\and G.~Rodighiero\orcid{0000-0002-9415-2296}\inst{\ref{aff105},\ref{aff26}}
\and S.~Sacquegna\orcid{0000-0002-8433-6630}\inst{\ref{aff137},\ref{aff138},\ref{aff139}}
\and M.~Sahl\'en\orcid{0000-0003-0973-4804}\inst{\ref{aff166}}
\and D.~B.~Sanders\orcid{0000-0002-1233-9998}\inst{\ref{aff49}}
\and E.~Sarpa\orcid{0000-0002-1256-655X}\inst{\ref{aff23},\ref{aff126},\ref{aff22}}
\and A.~Schneider\orcid{0000-0001-7055-8104}\inst{\ref{aff155}}
\and D.~Sciotti\orcid{0009-0008-4519-2620}\inst{\ref{aff47},\ref{aff86}}
\and E.~Sellentin\inst{\ref{aff167},\ref{aff41}}
\and L.~C.~Smith\orcid{0000-0002-3259-2771}\inst{\ref{aff168}}
\and K.~Tanidis\orcid{0000-0001-9843-5130}\inst{\ref{aff119}}
\and C.~Tao\orcid{0000-0001-7961-8177}\inst{\ref{aff9}}
\and G.~Testera\inst{\ref{aff31}}
\and R.~Teyssier\orcid{0000-0001-7689-0933}\inst{\ref{aff165}}
\and S.~Tosi\orcid{0000-0002-7275-9193}\inst{\ref{aff30},\ref{aff31},\ref{aff18}}
\and A.~Troja\orcid{0000-0003-0239-4595}\inst{\ref{aff105},\ref{aff62}}
\and M.~Tucci\inst{\ref{aff60}}
\and C.~Valieri\inst{\ref{aff25}}
\and A.~Venhola\orcid{0000-0001-6071-4564}\inst{\ref{aff169}}
\and D.~Vergani\orcid{0000-0003-0898-2216}\inst{\ref{aff19}}
\and G.~Verza\orcid{0000-0002-1886-8348}\inst{\ref{aff170}}
\and P.~Vielzeuf\orcid{0000-0003-2035-9339}\inst{\ref{aff9}}
\and N.~A.~Walton\orcid{0000-0003-3983-8778}\inst{\ref{aff168}}
\and M.~Bella\orcid{0000-0002-6406-4789}\inst{\ref{aff8}}
\and D.~Scott\orcid{0000-0002-6878-9840}\inst{\ref{aff171}}}

%%%% please do not edit the affiliation list -- contact ECEB Bureau for changes
\institute{Universit\'e Claude Bernard Lyon 1, CNRS/IN2P3, IP2I Lyon, UMR 5822, Villeurbanne, F-69100, France\label{aff1}
\and
INAF-IASF Milano, Via Alfonso Corti 12, 20133 Milano, Italy\label{aff2}
\and
Caltech/IPAC, 1200 E. California Blvd., Pasadena, CA 91125, USA\label{aff3}
\and
Infrared Processing and Analysis Center, California Institute of Technology, Pasadena, CA 91125, USA\label{aff4}
\and
University of California, Los Angeles, CA 90095-1562, USA\label{aff5}
\and
Minnesota Institute for Astrophysics, University of Minnesota, 116 Church St SE, Minneapolis, MN 55455, USA\label{aff6}
\and
Centre National d'Etudes Spatiales -- Centre spatial de Toulouse, 18 avenue Edouard Belin, 31401 Toulouse Cedex 9, France\label{aff7}
\and
Institut de Recherche en Astrophysique et Plan\'etologie (IRAP), Universit\'e de Toulouse, CNRS, UPS, CNES, 14 Av. Edouard Belin, 31400 Toulouse, France\label{aff8}
\and
Aix-Marseille Universit\'e, CNRS/IN2P3, CPPM, Marseille, France\label{aff9}
\and
Aix-Marseille Universit\'e, CNRS, CNES, LAM, Marseille, France\label{aff10}
\and
Dipartimento di Fisica "Aldo Pontremoli", Universit\`a degli Studi di Milano, Via Celoria 16, 20133 Milano, Italy\label{aff11}
\and
INFN-Sezione di Milano, Via Celoria 16, 20133 Milano, Italy\label{aff12}
\and
California Institute of Technology, 1200 E California Blvd, Pasadena, CA 91125, USA\label{aff13}
\and
INAF-Osservatorio Astronomico di Trieste, Via G. B. Tiepolo 11, 34143 Trieste, Italy\label{aff14}
\and
Universit\'e Paris-Saclay, CNRS, Institut d'astrophysique spatiale, 91405, Orsay, France\label{aff15}
\and
ESAC/ESA, Camino Bajo del Castillo, s/n., Urb. Villafranca del Castillo, 28692 Villanueva de la Ca\~nada, Madrid, Spain\label{aff16}
\and
School of Mathematics and Physics, University of Surrey, Guildford, Surrey, GU2 7XH, UK\label{aff17}
\and
INAF-Osservatorio Astronomico di Brera, Via Brera 28, 20122 Milano, Italy\label{aff18}
\and
INAF-Osservatorio di Astrofisica e Scienza dello Spazio di Bologna, Via Piero Gobetti 93/3, 40129 Bologna, Italy\label{aff19}
\and
Universit\'e Paris-Saclay, Universit\'e Paris Cit\'e, CEA, CNRS, AIM, 91191, Gif-sur-Yvette, France\label{aff20}
\and
IFPU, Institute for Fundamental Physics of the Universe, via Beirut 2, 34151 Trieste, Italy\label{aff21}
\and
INFN, Sezione di Trieste, Via Valerio 2, 34127 Trieste TS, Italy\label{aff22}
\and
SISSA, International School for Advanced Studies, Via Bonomea 265, 34136 Trieste TS, Italy\label{aff23}
\and
Dipartimento di Fisica e Astronomia, Universit\`a di Bologna, Via Gobetti 93/2, 40129 Bologna, Italy\label{aff24}
\and
INFN-Sezione di Bologna, Viale Berti Pichat 6/2, 40127 Bologna, Italy\label{aff25}
\and
INAF-Osservatorio Astronomico di Padova, Via dell'Osservatorio 5, 35122 Padova, Italy\label{aff26}
\and
Kapteyn Astronomical Institute, University of Groningen, PO Box 800, 9700 AV Groningen, The Netherlands\label{aff27}
\and
ATG Europe BV, Huygensstraat 34, 2201 DK Noordwijk, The Netherlands\label{aff28}
\and
Space Science Data Center, Italian Space Agency, via del Politecnico snc, 00133 Roma, Italy\label{aff29}
\and
Dipartimento di Fisica, Universit\`a di Genova, Via Dodecaneso 33, 16146, Genova, Italy\label{aff30}
\and
INFN-Sezione di Genova, Via Dodecaneso 33, 16146, Genova, Italy\label{aff31}
\and
Department of Physics "E. Pancini", University Federico II, Via Cinthia 6, 80126, Napoli, Italy\label{aff32}
\and
INAF-Osservatorio Astronomico di Capodimonte, Via Moiariello 16, 80131 Napoli, Italy\label{aff33}
\and
Instituto de Astrof\'isica e Ci\^encias do Espa\c{c}o, Universidade do Porto, CAUP, Rua das Estrelas, PT4150-762 Porto, Portugal\label{aff34}
\and
Faculdade de Ci\^encias da Universidade do Porto, Rua do Campo de Alegre, 4150-007 Porto, Portugal\label{aff35}
\and
Dipartimento di Fisica, Universit\`a degli Studi di Torino, Via P. Giuria 1, 10125 Torino, Italy\label{aff36}
\and
INFN-Sezione di Torino, Via P. Giuria 1, 10125 Torino, Italy\label{aff37}
\and
INAF-Osservatorio Astrofisico di Torino, Via Osservatorio 20, 10025 Pino Torinese (TO), Italy\label{aff38}
\and
European Space Agency/ESTEC, Keplerlaan 1, 2201 AZ Noordwijk, The Netherlands\label{aff39}
\and
Institute Lorentz, Leiden University, Niels Bohrweg 2, 2333 CA Leiden, The Netherlands\label{aff40}
\and
Leiden Observatory, Leiden University, Einsteinweg 55, 2333 CC Leiden, The Netherlands\label{aff41}
\and
Centro de Investigaciones Energ\'eticas, Medioambientales y Tecnol\'ogicas (CIEMAT), Avenida Complutense 40, 28040 Madrid, Spain\label{aff42}
\and
Port d'Informaci\'{o} Cient\'{i}fica, Campus UAB, C. Albareda s/n, 08193 Bellaterra (Barcelona), Spain\label{aff43}
\and
Institute for Theoretical Particle Physics and Cosmology (TTK), RWTH Aachen University, 52056 Aachen, Germany\label{aff44}
\and
Institute of Space Sciences (ICE, CSIC), Campus UAB, Carrer de Can Magrans, s/n, 08193 Barcelona, Spain\label{aff45}
\and
Institut d'Estudis Espacials de Catalunya (IEEC),  Edifici RDIT, Campus UPC, 08860 Castelldefels, Barcelona, Spain\label{aff46}
\and
INAF-Osservatorio Astronomico di Roma, Via Frascati 33, 00078 Monteporzio Catone, Italy\label{aff47}
\and
INFN section of Naples, Via Cinthia 6, 80126, Napoli, Italy\label{aff48}
\and
Institute for Astronomy, University of Hawaii, 2680 Woodlawn Drive, Honolulu, HI 96822, USA\label{aff49}
\and
Dipartimento di Fisica e Astronomia "Augusto Righi" - Alma Mater Studiorum Universit\`a di Bologna, Viale Berti Pichat 6/2, 40127 Bologna, Italy\label{aff50}
\and
Instituto de Astrof\'{\i}sica de Canarias, V\'{\i}a L\'actea, 38205 La Laguna, Tenerife, Spain\label{aff51}
\and
Institute for Astronomy, University of Edinburgh, Royal Observatory, Blackford Hill, Edinburgh EH9 3HJ, UK\label{aff52}
\and
Jodrell Bank Centre for Astrophysics, Department of Physics and Astronomy, University of Manchester, Oxford Road, Manchester M13 9PL, UK\label{aff53}
\and
European Space Agency/ESRIN, Largo Galileo Galilei 1, 00044 Frascati, Roma, Italy\label{aff54}
\and
Institut de Ci\`{e}ncies del Cosmos (ICCUB), Universitat de Barcelona (IEEC-UB), Mart\'{i} i Franqu\`{e}s 1, 08028 Barcelona, Spain\label{aff55}
\and
Instituci\'o Catalana de Recerca i Estudis Avan\c{c}ats (ICREA), Passeig de Llu\'{\i}s Companys 23, 08010 Barcelona, Spain\label{aff56}
\and
UCB Lyon 1, CNRS/IN2P3, IUF, IP2I Lyon, 4 rue Enrico Fermi, 69622 Villeurbanne, France\label{aff57}
\and
Departamento de F\'isica, Faculdade de Ci\^encias, Universidade de Lisboa, Edif\'icio C8, Campo Grande, PT1749-016 Lisboa, Portugal\label{aff58}
\and
Instituto de Astrof\'isica e Ci\^encias do Espa\c{c}o, Faculdade de Ci\^encias, Universidade de Lisboa, Campo Grande, 1749-016 Lisboa, Portugal\label{aff59}
\and
Department of Astronomy, University of Geneva, ch. d'Ecogia 16, 1290 Versoix, Switzerland\label{aff60}
\and
INAF-Istituto di Astrofisica e Planetologia Spaziali, via del Fosso del Cavaliere, 100, 00100 Roma, Italy\label{aff61}
\and
INFN-Padova, Via Marzolo 8, 35131 Padova, Italy\label{aff62}
\and
INFN-Bologna, Via Irnerio 46, 40126 Bologna, Italy\label{aff63}
\and
School of Physics, HH Wills Physics Laboratory, University of Bristol, Tyndall Avenue, Bristol, BS8 1TL, UK\label{aff64}
\and
Universit\"ats-Sternwarte M\"unchen, Fakult\"at f\"ur Physik, Ludwig-Maximilians-Universit\"at M\"unchen, Scheinerstrasse 1, 81679 M\"unchen, Germany\label{aff65}
\and
Max Planck Institute for Extraterrestrial Physics, Giessenbachstr. 1, 85748 Garching, Germany\label{aff66}
\and
Institute of Theoretical Astrophysics, University of Oslo, P.O. Box 1029 Blindern, 0315 Oslo, Norway\label{aff67}
\and
Jet Propulsion Laboratory, California Institute of Technology, 4800 Oak Grove Drive, Pasadena, CA, 91109, USA\label{aff68}
\and
Department of Physics, Lancaster University, Lancaster, LA1 4YB, UK\label{aff69}
\and
Felix Hormuth Engineering, Goethestr. 17, 69181 Leimen, Germany\label{aff70}
\and
Technical University of Denmark, Elektrovej 327, 2800 Kgs. Lyngby, Denmark\label{aff71}
\and
Cosmic Dawn Center (DAWN), Denmark\label{aff72}
\and
Institut d'Astrophysique de Paris, UMR 7095, CNRS, and Sorbonne Universit\'e, 98 bis boulevard Arago, 75014 Paris, France\label{aff73}
\and
Max-Planck-Institut f\"ur Astronomie, K\"onigstuhl 17, 69117 Heidelberg, Germany\label{aff74}
\and
NASA Goddard Space Flight Center, Greenbelt, MD 20771, USA\label{aff75}
\and
Department of Physics and Astronomy, University College London, Gower Street, London WC1E 6BT, UK\label{aff76}
\and
Department of Physics and Helsinki Institute of Physics, Gustaf H\"allstr\"omin katu 2, 00014 University of Helsinki, Finland\label{aff77}
\and
Universit\'e de Gen\`eve, D\'epartement de Physique Th\'eorique and Centre for Astroparticle Physics, 24 quai Ernest-Ansermet, CH-1211 Gen\`eve 4, Switzerland\label{aff78}
\and
Department of Physics, P.O. Box 64, 00014 University of Helsinki, Finland\label{aff79}
\and
Helsinki Institute of Physics, Gustaf H{\"a}llstr{\"o}min katu 2, University of Helsinki, Helsinki, Finland\label{aff80}
\and
Centre de Calcul de l'IN2P3/CNRS, 21 avenue Pierre de Coubertin 69627 Villeurbanne Cedex, France\label{aff81}
\and
Laboratoire d'etude de l'Univers et des phenomenes eXtremes, Observatoire de Paris, Universit\'e PSL, Sorbonne Universit\'e, CNRS, 92190 Meudon, France\label{aff82}
\and
SKA Observatory, Jodrell Bank, Lower Withington, Macclesfield, Cheshire SK11 9FT, UK\label{aff83}
\and
University of Applied Sciences and Arts of Northwestern Switzerland, School of Computer Science, 5210 Windisch, Switzerland\label{aff84}
\and
Universit\"at Bonn, Argelander-Institut f\"ur Astronomie, Auf dem H\"ugel 71, 53121 Bonn, Germany\label{aff85}
\and
INFN-Sezione di Roma, Piazzale Aldo Moro, 2 - c/o Dipartimento di Fisica, Edificio G. Marconi, 00185 Roma, Italy\label{aff86}
\and
Dipartimento di Fisica e Astronomia "Augusto Righi" - Alma Mater Studiorum Universit\`a di Bologna, via Piero Gobetti 93/2, 40129 Bologna, Italy\label{aff87}
\and
Department of Physics, Institute for Computational Cosmology, Durham University, South Road, Durham, DH1 3LE, UK\label{aff88}
\and
Universit\'e C\^{o}te d'Azur, Observatoire de la C\^{o}te d'Azur, CNRS, Laboratoire Lagrange, Bd de l'Observatoire, CS 34229, 06304 Nice cedex 4, France\label{aff89}
\and
Universit\'e Paris Cit\'e, CNRS, Astroparticule et Cosmologie, 75013 Paris, France\label{aff90}
\and
CNRS-UCB International Research Laboratory, Centre Pierre Binetruy, IRL2007, CPB-IN2P3, Berkeley, USA\label{aff91}
\and
University of Applied Sciences and Arts of Northwestern Switzerland, School of Engineering, 5210 Windisch, Switzerland\label{aff92}
\and
Institut d'Astrophysique de Paris, 98bis Boulevard Arago, 75014, Paris, France\label{aff93}
\and
Institute of Physics, Laboratory of Astrophysics, Ecole Polytechnique F\'ed\'erale de Lausanne (EPFL), Observatoire de Sauverny, 1290 Versoix, Switzerland\label{aff94}
\and
Aurora Technology for European Space Agency (ESA), Camino bajo del Castillo, s/n, Urbanizacion Villafranca del Castillo, Villanueva de la Ca\~nada, 28692 Madrid, Spain\label{aff95}
\and
Institut de F\'{i}sica d'Altes Energies (IFAE), The Barcelona Institute of Science and Technology, Campus UAB, 08193 Bellaterra (Barcelona), Spain\label{aff96}
\and
School of Mathematics, Statistics and Physics, Newcastle University, Herschel Building, Newcastle-upon-Tyne, NE1 7RU, UK\label{aff97}
\and
DARK, Niels Bohr Institute, University of Copenhagen, Jagtvej 155, 2200 Copenhagen, Denmark\label{aff98}
\and
Waterloo Centre for Astrophysics, University of Waterloo, Waterloo, Ontario N2L 3G1, Canada\label{aff99}
\and
Department of Physics and Astronomy, University of Waterloo, Waterloo, Ontario N2L 3G1, Canada\label{aff100}
\and
Perimeter Institute for Theoretical Physics, Waterloo, Ontario N2L 2Y5, Canada\label{aff101}
\and
Institute of Space Science, Str. Atomistilor, nr. 409 M\u{a}gurele, Ilfov, 077125, Romania\label{aff102}
\and
Consejo Superior de Investigaciones Cientificas, Calle Serrano 117, 28006 Madrid, Spain\label{aff103}
\and
Universidad de La Laguna, Departamento de Astrof\'{\i}sica, 38206 La Laguna, Tenerife, Spain\label{aff104}
\and
Dipartimento di Fisica e Astronomia "G. Galilei", Universit\`a di Padova, Via Marzolo 8, 35131 Padova, Italy\label{aff105}
\and
Institut f\"ur Theoretische Physik, University of Heidelberg, Philosophenweg 16, 69120 Heidelberg, Germany\label{aff106}
\and
Universit\'e St Joseph; Faculty of Sciences, Beirut, Lebanon\label{aff107}
\and
Departamento de F\'isica, FCFM, Universidad de Chile, Blanco Encalada 2008, Santiago, Chile\label{aff108}
\and
Universit\"at Innsbruck, Institut f\"ur Astro- und Teilchenphysik, Technikerstr. 25/8, 6020 Innsbruck, Austria\label{aff109}
\and
Satlantis, University Science Park, Sede Bld 48940, Leioa-Bilbao, Spain\label{aff110}
\and
Department of Physics, Royal Holloway, University of London, TW20 0EX, UK\label{aff111}
\and
Instituto de Astrof\'isica e Ci\^encias do Espa\c{c}o, Faculdade de Ci\^encias, Universidade de Lisboa, Tapada da Ajuda, 1349-018 Lisboa, Portugal\label{aff112}
\and
Cosmic Dawn Center (DAWN)\label{aff113}
\and
Niels Bohr Institute, University of Copenhagen, Jagtvej 128, 2200 Copenhagen, Denmark\label{aff114}
\and
Universidad Polit\'ecnica de Cartagena, Departamento de Electr\'onica y Tecnolog\'ia de Computadoras,  Plaza del Hospital 1, 30202 Cartagena, Spain\label{aff115}
\and
Dipartimento di Fisica e Scienze della Terra, Universit\`a degli Studi di Ferrara, Via Giuseppe Saragat 1, 44122 Ferrara, Italy\label{aff116}
\and
Istituto Nazionale di Fisica Nucleare, Sezione di Ferrara, Via Giuseppe Saragat 1, 44122 Ferrara, Italy\label{aff117}
\and
INAF, Istituto di Radioastronomia, Via Piero Gobetti 101, 40129 Bologna, Italy\label{aff118}
\and
Department of Physics, Oxford University, Keble Road, Oxford OX1 3RH, UK\label{aff119}
\and
Instituto de Astrof\'isica de Canarias (IAC); Departamento de Astrof\'isica, Universidad de La Laguna (ULL), 38200, La Laguna, Tenerife, Spain\label{aff120}
\and
Universit\'e PSL, Observatoire de Paris, Sorbonne Universit\'e, CNRS, LERMA, 75014, Paris, France\label{aff121}
\and
Universit\'e Paris-Cit\'e, 5 Rue Thomas Mann, 75013, Paris, France\label{aff122}
\and
INAF - Osservatorio Astronomico di Brera, via Emilio Bianchi 46, 23807 Merate, Italy\label{aff123}
\and
INAF-Osservatorio Astronomico di Brera, Via Brera 28, 20122 Milano, Italy, and INFN-Sezione di Genova, Via Dodecaneso 33, 16146, Genova, Italy\label{aff124}
\and
ICL, Junia, Universit\'e Catholique de Lille, LITL, 59000 Lille, France\label{aff125}
\and
ICSC - Centro Nazionale di Ricerca in High Performance Computing, Big Data e Quantum Computing, Via Magnanelli 2, Bologna, Italy\label{aff126}
\and
Instituto de F\'isica Te\'orica UAM-CSIC, Campus de Cantoblanco, 28049 Madrid, Spain\label{aff127}
\and
CERCA/ISO, Department of Physics, Case Western Reserve University, 10900 Euclid Avenue, Cleveland, OH 44106, USA\label{aff128}
\and
Technical University of Munich, TUM School of Natural Sciences, Physics Department, James-Franck-Str.~1, 85748 Garching, Germany\label{aff129}
\and
Max-Planck-Institut f\"ur Astrophysik, Karl-Schwarzschild-Str.~1, 85748 Garching, Germany\label{aff130}
\and
Laboratoire Univers et Th\'eorie, Observatoire de Paris, Universit\'e PSL, Universit\'e Paris Cit\'e, CNRS, 92190 Meudon, France\label{aff131}
\and
Departamento de F{\'\i}sica Fundamental. Universidad de Salamanca. Plaza de la Merced s/n. 37008 Salamanca, Spain\label{aff132}
\and
Mullard Space Science Laboratory, University College London, Holmbury St Mary, Dorking, Surrey RH5 6NT, UK\label{aff133}
\and
Universit\'e de Strasbourg, CNRS, Observatoire astronomique de Strasbourg, UMR 7550, 67000 Strasbourg, France\label{aff134}
\and
Center for Data-Driven Discovery, Kavli IPMU (WPI), UTIAS, The University of Tokyo, Kashiwa, Chiba 277-8583, Japan\label{aff135}
\and
Department of Physics \& Astronomy, University of California Irvine, Irvine CA 92697, USA\label{aff136}
\and
Department of Mathematics and Physics E. De Giorgi, University of Salento, Via per Arnesano, CP-I93, 73100, Lecce, Italy\label{aff137}
\and
INFN, Sezione di Lecce, Via per Arnesano, CP-193, 73100, Lecce, Italy\label{aff138}
\and
INAF-Sezione di Lecce, c/o Dipartimento Matematica e Fisica, Via per Arnesano, 73100, Lecce, Italy\label{aff139}
\and
Departamento F\'isica Aplicada, Universidad Polit\'ecnica de Cartagena, Campus Muralla del Mar, 30202 Cartagena, Murcia, Spain\label{aff140}
\and
Instituto de F\'isica de Cantabria, Edificio Juan Jord\'a, Avenida de los Castros, 39005 Santander, Spain\label{aff141}
\and
CEA Saclay, DFR/IRFU, Service d'Astrophysique, Bat. 709, 91191 Gif-sur-Yvette, France\label{aff142}
\and
Institute of Cosmology and Gravitation, University of Portsmouth, Portsmouth PO1 3FX, UK\label{aff143}
\and
Department of Computer Science, Aalto University, PO Box 15400, Espoo, FI-00 076, Finland\label{aff144}
\and
Dipartimento di Fisica - Sezione di Astronomia, Universit\`a di Trieste, Via Tiepolo 11, 34131 Trieste, Italy\label{aff145}
\and
Instituto de Astrof\'\i sica de Canarias, c/ Via Lactea s/n, La Laguna 38200, Spain. Departamento de Astrof\'\i sica de la Universidad de La Laguna, Avda. Francisco Sanchez, La Laguna, 38200, Spain\label{aff146}
\and
Ruhr University Bochum, Faculty of Physics and Astronomy, Astronomical Institute (AIRUB), German Centre for Cosmological Lensing (GCCL), 44780 Bochum, Germany\label{aff147}
\and
Department of Physics and Astronomy, Vesilinnantie 5, 20014 University of Turku, Finland\label{aff148}
\and
Serco for European Space Agency (ESA), Camino bajo del Castillo, s/n, Urbanizacion Villafranca del Castillo, Villanueva de la Ca\~nada, 28692 Madrid, Spain\label{aff149}
\and
ARC Centre of Excellence for Dark Matter Particle Physics, Melbourne, Australia\label{aff150}
\and
Centre for Astrophysics \& Supercomputing, Swinburne University of Technology,  Hawthorn, Victoria 3122, Australia\label{aff151}
\and
Department of Physics and Astronomy, University of the Western Cape, Bellville, Cape Town, 7535, South Africa\label{aff152}
\and
DAMTP, Centre for Mathematical Sciences, Wilberforce Road, Cambridge CB3 0WA, UK\label{aff153}
\and
Kavli Institute for Cosmology Cambridge, Madingley Road, Cambridge, CB3 0HA, UK\label{aff154}
\and
Department of Astrophysics, University of Zurich, Winterthurerstrasse 190, 8057 Zurich, Switzerland\label{aff155}
\and
Department of Physics, Centre for Extragalactic Astronomy, Durham University, South Road, Durham, DH1 3LE, UK\label{aff156}
\and
IRFU, CEA, Universit\'e Paris-Saclay 91191 Gif-sur-Yvette Cedex, France\label{aff157}
\and
Oskar Klein Centre for Cosmoparticle Physics, Department of Physics, Stockholm University, Stockholm, SE-106 91, Sweden\label{aff158}
\and
Astrophysics Group, Blackett Laboratory, Imperial College London, London SW7 2AZ, UK\label{aff159}
\and
Univ. Grenoble Alpes, CNRS, Grenoble INP, LPSC-IN2P3, 53, Avenue des Martyrs, 38000, Grenoble, France\label{aff160}
\and
INAF-Osservatorio Astrofisico di Arcetri, Largo E. Fermi 5, 50125, Firenze, Italy\label{aff161}
\and
Dipartimento di Fisica, Sapienza Universit\`a di Roma, Piazzale Aldo Moro 2, 00185 Roma, Italy\label{aff162}
\and
Centro de Astrof\'{\i}sica da Universidade do Porto, Rua das Estrelas, 4150-762 Porto, Portugal\label{aff163}
\and
HE Space for European Space Agency (ESA), Camino bajo del Castillo, s/n, Urbanizacion Villafranca del Castillo, Villanueva de la Ca\~nada, 28692 Madrid, Spain\label{aff164}
\and
Department of Astrophysical Sciences, Peyton Hall, Princeton University, Princeton, NJ 08544, USA\label{aff165}
\and
Theoretical astrophysics, Department of Physics and Astronomy, Uppsala University, Box 515, 751 20 Uppsala, Sweden\label{aff166}
\and
Mathematical Institute, University of Leiden, Einsteinweg 55, 2333 CA Leiden, The Netherlands\label{aff167}
\and
Institute of Astronomy, University of Cambridge, Madingley Road, Cambridge CB3 0HA, UK\label{aff168}
\and
Space physics and astronomy research unit, University of Oulu, Pentti Kaiteran katu 1, FI-90014 Oulu, Finland\label{aff169}
\and
Center for Computational Astrophysics, Flatiron Institute, 162 5th Avenue, 10010, New York, NY, USA\label{aff170}
\and
Department of Physics and Astronomy, University of British Columbia, Vancouver, BC V6T 1Z1, Canada\label{aff171}}

%% file: main.bbl
\begin{thebibliography}{46}
\expandafter\ifx\csname natexlab\endcsname\relax\def\natexlab#1{#1}\fi

\bibitem[{{Akhlaghi} \& {Ichikawa}(2015)}]{2015ApJS..220....1A}
{Akhlaghi}, M. \& {Ichikawa}, T. 2015, \apjs, 220, 1

\bibitem[{{Astropy Collaboration} {et~al.}(2022){Astropy Collaboration},
  {Price-Whelan}, Lian~Lim, Earl, Starkman, Bradley, Shupe, Patil, Corrales,
  Brasseur, N{\"o}the, Donath, Tollerud, Morris, Ginsburg, Vaher, Weaver,
  Tocknell, Jamieson, {van Kerkwijk}, Robitaille, Merry, Bachetti, G{\"u}nther,
  Aldcroft, {Alvarado-Montes}, Archibald, B{\'o}di, Bapat, Barentsen,
  Baz{\'a}n, Biswas, Boquien, Burke, Cara, Cara, E~Conroy, Conseil, Craig,
  Cross, Cruz, D'Eugenio, Dencheva, Devillepoix, Dietrich, Davis~Eigenbrot,
  Erben, Ferreira, {Foreman-Mackey}, Fox, Freij, Garg, Geda, Glattly,
  Gondhalekar, Gordon, Grant, Greenfield, Groener, Guest, Gurovich, Handberg,
  Hart, {Hatfield-Dodds}, Homeier, Hosseinzadeh, Jenness, Jones, Joseph,
  Bryce~Kalmbach, Karamehmetoglu, Ka{\l}uszy{\'n}ski, Kelley, Kern, Kerzendorf,
  Koch, Kulumani, Lee, Ly, Ma, MacBride, Maljaars, Muna, Murphy, Norman,
  O'Steen, Oman, Pacifici, Pascual, {Pascual-Granado}, Patil, Perren,
  Pickering, Rastogi, Roulston, Ryan, Rykoff, Sabater, Sakurikar, Salgado,
  Sanghi, Saunders, Savchenko, Schwardt, {Seifert-Eckert}, Shih, Shrey~Jain,
  Shukla, Sick, Simpson, Singanamalla, Singer, Singhal, Sinha, Sip{\H o}cz,
  Spitler, Stansby, Streicher, {\v S}umak, Swinbank, Taranu, Tewary, Tremblay,
  {de Val-Borro}, Van~Kooten, Vasovi{\'c}, Verma, Cardoso, Williams, Wilson,
  Winkel, {Wood-Vasey}, Xue, Yoachim, ZHANG, \& Zonca}]{astropy_2022}
{Astropy Collaboration}, {Price-Whelan}, A.~M., Lian~Lim, P., {et~al.} 2022,
  The {{Astropy Project}}: {{Sustaining}} and {{Growing}} a
  {{Community-oriented Open-source Project}} and the {{Latest Major Release}}
  (v5.0) of the {{Core Package}}

\bibitem[{{Astropy Collaboration} {et~al.}(2018){Astropy Collaboration},
  {Price-Whelan}, Sip{\H o}cz, G{\"u}nther, Lim, Crawford, Conseil, Shupe,
  Craig, Dencheva, Ginsburg, VanderPlas, Bradley, {P{\'e}rez-Su{\'a}rez}, {de
  Val-Borro}, Aldcroft, Cruz, Robitaille, Tollerud, Ardelean, Babej, Bach,
  Bachetti, Bakanov, Bamford, Barentsen, Barmby, Baumbach, Berry, Biscani,
  Boquien, Bostroem, Bouma, Brammer, Bray, Breytenbach, Buddelmeijer, Burke,
  Calderone, Cano~Rodr{\'i}guez, Cara, Cardoso, Cheedella, Copin, Corrales,
  Crichton, D'Avella, Deil, Depagne, Dietrich, Donath, Droettboom, Earl, Erben,
  Fabbro, Ferreira, Finethy, Fox, Garrison, Gibbons, Goldstein, Gommers, Greco,
  Greenfield, Groener, Grollier, Hagen, Hirst, Homeier, Horton, Hosseinzadeh,
  Hu, Hunkeler, Ivezi{\'c}, Jain, Jenness, Kanarek, Kendrew, Kern, Kerzendorf,
  Khvalko, King, Kirkby, Kulkarni, Kumar, Lee, Lenz, Littlefair, Ma, Macleod,
  Mastropietro, McCully, Montagnac, Morris, Mueller, Mumford, Muna, Murphy,
  Nelson, Nguyen, Ninan, N{\"o}the, Ogaz, Oh, Parejko, Parley, Pascual, Patil,
  Patil, Plunkett, Prochaska, Rastogi, Reddy~Janga, Sabater, Sakurikar,
  Seifert, Sherbert, {Sherwood-Taylor}, Shih, Sick, Silbiger, Singanamalla,
  Singer, Sladen, Sooley, Sornarajah, Streicher, Teuben, Thomas, Tremblay,
  Turner, Terr{\'o}n, {van Kerkwijk}, {de la Vega}, Watkins, Weaver, Whitmore,
  Woillez, Zabalza, \& {Astropy Contributors}}]{astropy_2018}
{Astropy Collaboration}, {Price-Whelan}, A.~M., Sip{\H o}cz, B.~M., {et~al.}
  2018, \aj, 156, 123

\bibitem[{{Astropy Collaboration} {et~al.}(2013){Astropy Collaboration},
  Robitaille, Tollerud, Greenfield, Droettboom, Bray, Aldcroft, Davis,
  Ginsburg, {Price-Whelan}, Kerzendorf, Conley, Crighton, Barbary, Muna,
  Ferguson, Grollier, Parikh, Nair, Unther, Deil, Woillez, Conseil, Kramer,
  Turner, Singer, Fox, Weaver, Zabalza, Edwards, Azalee~Bostroem, Burke, Casey,
  Crawford, Dencheva, Ely, Jenness, Labrie, Lim, Pierfederici, Pontzen, Ptak,
  Refsdal, Servillat, \& Streicher}]{astropy_2013}
{Astropy Collaboration}, Robitaille, T.~P., Tollerud, E.~J., {et~al.} 2013,
  \aap, 558, A33

\bibitem[{Bella {et~al.}(2022)Bella, Hosseini, Saylani, Contini, Gregoire,
  Guerrero, \& Deville}]{bella_decontamination_2022}
Bella, M., Hosseini, S., Saylani, H., {et~al.} 2022, in 30th {{European Signal
  Processing Conference}} ({{EUSIPCO}}), Belgrade, 5

\bibitem[{{Bohlin} {et~al.}(2020){Bohlin}, {Hubeny}, \&
  {Rauch}}]{2020AJ....160...21B}
{Bohlin}, R.~C., {Hubeny}, I., \& {Rauch}, T. 2020, \aj, 160, 21

\bibitem[{{Casertano} {et~al.}(2000){Casertano}, {de Mello}, {Dickinson},
  {Ferguson}, {Fruchter}, {Gonzalez-Lopezlira}, {Heyer}, {Hook}, {Levay},
  {Lucas}, {Mack}, {Makidon}, {Mutchler}, {Smith}, {Stiavelli}, {Wiggs}, \&
  {Williams}}]{2000AJ....120.2747C}
{Casertano}, S., {de Mello}, D., {Dickinson}, M., {et~al.} 2000, \aj, 120, 2747

\bibitem[{{Chambers} {et~al.}(2016){Chambers}, {Magnier}, {Metcalfe},
  {Flewelling}, {Huber}, {Waters}, {Denneau}, {Draper}, {Farrow}, {Finkbeiner},
  {Holmberg}, {Koppenhoefer}, {Price}, {Rest}, {Saglia}, {Schlafly}, {Smartt},
  {Sweeney}, {Wainscoat}, {Burgett}, {Chastel}, {Grav}, {Heasley}, {Hodapp},
  {Jedicke}, {Kaiser}, {Kudritzki}, {Luppino}, {Lupton}, {Monet}, {Morgan},
  {Onaka}, {Shiao}, {Stubbs}, {Tonry}, {White}, {Ba{\~n}ados}, {Bell},
  {Bender}, {Bernard}, {Boegner}, {Boffi}, {Botticella}, {Calamida},
  {Casertano}, {Chen}, {Chen}, {Cole}, {Deacon}, {Frenk}, {Fitzsimmons},
  {Gezari}, {Gibbs}, {Goessl}, {Goggia}, {Gourgue}, {Goldman}, {Grant},
  {Grebel}, {Hambly}, {Hasinger}, {Heavens}, {Heckman}, {Henderson}, {Henning},
  {Holman}, {Hopp}, {Ip}, {Isani}, {Jackson}, {Keyes}, {Koekemoer}, {Kotak},
  {Le}, {Liska}, {Long}, {Lucey}, {Liu}, {Martin}, {Masci}, {McLean}, {Mindel},
  {Misra}, {Morganson}, {Murphy}, {Obaika}, {Narayan}, {Nieto-Santisteban},
  {Norberg}, {Peacock}, {Pier}, {Postman}, {Primak}, {Rae}, {Rai}, {Riess},
  {Riffeser}, {Rix}, {R{\"o}ser}, {Russel}, {Rutz}, {Schilbach}, {Schultz},
  {Scolnic}, {Strolger}, {Szalay}, {Seitz}, {Small}, {Smith}, {Soderblom},
  {Taylor}, {Thomson}, {Taylor}, {Thakar}, {Thiel}, {Thilker}, {Unger},
  {Urata}, {Valenti}, {Wagner}, {Walder}, {Walter}, {Watters}, {Werner},
  {Wood-Vasey}, \& {Wyse}}]{2016arXiv161205560C}
{Chambers}, K.~C., {Magnier}, E.~A., {Metcalfe}, N., {et~al.} 2016, arXiv
  e-prints, arXiv:1612.05560

\bibitem[{{Dey} {et~al.}(2019){Dey}, {Schlegel}, {Lang}, {Blum}, {Burleigh},
  {Fan}, {Findlay}, {Finkbeiner}, {Herrera}, {Juneau}, {Landriau}, {Levi},
  {McGreer}, {Meisner}, {Myers}, {Moustakas}, {Nugent}, {Patej}, {Schlafly},
  {Walker}, {Valdes}, {Weaver}, {Y{\`e}che}, {Zou}, {Zhou}, {Abareshi},
  {Abbott}, {Abolfathi}, {Aguilera}, {Alam}, {Allen}, {Alvarez}, {Annis},
  {Ansarinejad}, {Aubert}, {Beechert}, {Bell}, {BenZvi}, {Beutler}, {Bielby},
  {Bolton}, {Brice{\~n}o}, {Buckley-Geer}, {Butler}, {Calamida}, {Carlberg},
  {Carter}, {Casas}, {Castander}, {Choi}, {Comparat}, {Cukanovaite}, {Delubac},
  {DeVries}, {Dey}, {Dhungana}, {Dickinson}, {Ding}, {Donaldson}, {Duan},
  {Duckworth}, {Eftekharzadeh}, {Eisenstein}, {Etourneau}, {Fagrelius},
  {Farihi}, {Fitzpatrick}, {Font-Ribera}, {Fulmer}, {G{\"a}nsicke},
  {Gaztanaga}, {George}, {Gerdes}, {Gontcho}, {Gorgoni}, {Green}, {Guy},
  {Harmer}, {Hernandez}, {Honscheid}, {Huang}, {James}, {Jannuzi}, {Jiang},
  {Joyce}, {Karcher}, {Karkar}, {Kehoe}, {Kneib}, {Kueter-Young}, {Lan},
  {Lauer}, {Le Guillou}, {Le Van Suu}, {Lee}, {Lesser}, {Perreault Levasseur},
  {Li}, {Mann}, {Marshall}, {Mart{\'\i}nez-V{\'a}zquez}, {Martini}, {du Mas des
  Bourboux}, {McManus}, {Meier}, {M{\'e}nard}, {Metcalfe},
  {Mu{\~n}oz-Guti{\'e}rrez}, {Najita}, {Napier}, {Narayan}, {Newman}, {Nie},
  {Nord}, {Norman}, {Olsen}, {Paat}, {Palanque-Delabrouille}, {Peng},
  {Poppett}, {Poremba}, {Prakash}, {Rabinowitz}, {Raichoor}, {Rezaie},
  {Robertson}, {Roe}, {Ross}, {Ross}, {Rudnick}, {Safonova}, {Saha},
  {S{\'a}nchez}, {Savary}, {Schweiker}, {Scott}, {Seo}, {Shan}, {Silva},
  {Slepian}, {Soto}, {Sprayberry}, {Staten}, {Stillman}, {Stupak}, {Summers},
  {Sien Tie}, {Tirado}, {Vargas-Maga{\~n}a}, {Vivas}, {Wechsler}, {Williams},
  {Yang}, {Yang}, {Yapici}, {Zaritsky}, {Zenteno}, {Zhang}, {Zhang}, {Zhou}, \&
  {Zhou}}]{2019AJ....157..168D}
{Dey}, A., {Schlegel}, D.~J., {Lang}, D., {et~al.} 2019, \aj, 157, 168

\bibitem[{{Euclid Collaboration: Aussel} {et~al.}(2025){Euclid Collaboration:
  Aussel}, {Tereno}, {Schirmer}, {et~al.}}]{Q1-TP001}
{Euclid Collaboration: Aussel}, H., {Tereno}, I., {Schirmer}, M., {et~al.}
  2025, \aap, submitted

\bibitem[{{Euclid Collaboration: Cropper} {et~al.}(2024){Euclid Collaboration:
  Cropper}, {Al Bahlawan}, {Amiaux}, {et~al.}}]{EuclidSkyVIS}
{Euclid Collaboration: Cropper}, M., {Al Bahlawan}, A., {Amiaux}, J., {et~al.}
  2024, \aap, accepted, arXiv:2405.13492

\bibitem[{{Euclid Collaboration: Gabarra} {et~al.}(2023){Euclid Collaboration:
  Gabarra}, {Mancini}, {Rodriguez Mu{\~n}oz}, {et~al.}}]{Gabarra-EP31}
{Euclid Collaboration: Gabarra}, L., {Mancini}, C., {Rodriguez Mu{\~n}oz}, L.,
  {et~al.} 2023, \aap, 676, A34

\bibitem[{{Euclid Collaboration: Hormuth} {et~al.}(2024){Euclid Collaboration:
  Hormuth}, {Jahnke}, {Schirmer}, {et~al.}}]{EuclidSkyNISPCU}
{Euclid Collaboration: Hormuth}, F., {Jahnke}, K., {Schirmer}, M., {et~al.}
  2024, \aap, accepted, arXiv:2405.13494

\bibitem[{{Euclid Collaboration: Jahnke} {et~al.}(2024){Euclid Collaboration:
  Jahnke}, {Gillard}, {Schirmer}, {et~al.}}]{EuclidSkyNISP}
{Euclid Collaboration: Jahnke}, K., {Gillard}, W., {Schirmer}, M., {et~al.}
  2024, \aap, accepted, arXiv:2405.13493

\bibitem[{{Euclid Collaboration: Le Brun} {et~al.}(2025){Euclid Collaboration:
  Le Brun}, {Bethermin}, {et~al.}}]{Q1-TP007}
{Euclid Collaboration: Le Brun}, V., {Bethermin}, B., {et~al.} 2025, \aap,
  submitted

\bibitem[{{Euclid Collaboration: McCracken} {et~al.}(2025){Euclid
  Collaboration: McCracken}, {Benson}, {et~al.}}]{Q1-TP002}
{Euclid Collaboration: McCracken}, H., {Benson}, K., {et~al.} 2025, \aap,
  submitted

\bibitem[{{Euclid Collaboration: Mellier} {et~al.}(2024){Euclid Collaboration:
  Mellier}, {Abdurro'uf}, {Acevedo~Barroso}, {et~al.}}]{EuclidSkyOverview}
{Euclid Collaboration: Mellier}, Y., {Abdurro'uf}, {Acevedo~Barroso}, J.,
  {et~al.} 2024, \aap, accepted, arXiv:2405.13491

\bibitem[{{Euclid Collaboration: Paterson} {et~al.}(2023){Euclid Collaboration:
  Paterson}, {Schirmer}, {Copin}, {et~al.}}]{Paterson-EP32}
{Euclid Collaboration: Paterson}, K., {Schirmer}, M., {Copin}, Y., {et~al.}
  2023, \aap, 674, A172

\bibitem[{{Euclid Collaboration: Polenta} {et~al.}(2025){Euclid Collaboration:
  Polenta}, {Frailis}, {Alavi}, {et~al.}}]{Q1-TP003}
{Euclid Collaboration: Polenta}, G., {Frailis}, M., {Alavi}, A., {et~al.} 2025,
  \aap, submitted

\bibitem[{{Euclid Collaboration: Romelli} {et~al.}(2025){Euclid Collaboration:
  Romelli}, {K\"ummel}, {Dole}, {et~al.}}]{Q1-TP004}
{Euclid Collaboration: Romelli}, E., {K\"ummel}, M., {Dole}, H., {et~al.} 2025,
  \aap, submitted

\bibitem[{{Euclid Collaboration: Scaramella} {et~al.}(2022){Euclid
  Collaboration: Scaramella}, {Amiaux}, {Mellier}, {et~al.}}]{Scaramella-EP1}
{Euclid Collaboration: Scaramella}, R., {Amiaux}, J., {Mellier}, Y., {et~al.}
  2022, \aap, 662, A112

\bibitem[{{Euclid Collaboration: Schirmer} {et~al.}(2022){Euclid Collaboration:
  Schirmer}, {Jahnke}, {Seidel}, {et~al.}}]{Schirmer-EP18}
{Euclid Collaboration: Schirmer}, M., {Jahnke}, K., {Seidel}, G., {et~al.}
  2022, \aap, 662, A92

\bibitem[{{Euclid Quick Release Q1}(2025)}]{Q1cite}
{Euclid Quick Release Q1}. 2025, \url{https://doi.org/10.57780/esa-2853f3b}

\bibitem[{{Gianninas} {et~al.}(2011){Gianninas}, {Bergeron}, \&
  {Ruiz}}]{2011ApJ...743..138G}
{Gianninas}, A., {Bergeron}, P., \& {Ruiz}, M.~T. 2011, \apj, 743, 138

\bibitem[{{Gorjian} {et~al.}(2000){Gorjian}, {Wright}, \& {Chary}}]{GWC00}
{Gorjian}, V., {Wright}, E.~L., \& {Chary}, R.~R. 2000, \apj, 536, 550

\bibitem[{Grubbs(1969)}]{Grubbs1969}
Grubbs, F.~E. 1969, Technometrics, 11, 1

\bibitem[{{Harris} {et~al.}(2020){Harris}, {Millman}, {van der Walt},
  {Gommers}, {Virtanen}, {Cournapeau}, {Wieser}, {Taylor}, {Berg}, {Smith},
  {Kern}, {Picus}, {Hoyer}, {van Kerkwijk}, {Brett}, {Haldane}, {del R{\'\i}o},
  {Wiebe}, {Peterson}, {G{\'e}rard-Marchant}, {Sheppard}, {Reddy}, {Weckesser},
  {Abbasi}, {Gohlke}, \& {Oliphant}}]{2020Natur.585..357H}
{Harris}, C.~R., {Millman}, K.~J., {van der Walt}, S.~J., {et~al.} 2020, \nat,
  585, 357

\bibitem[{Horne(1986)}]{horne_optimal_1986}
Horne, K. 1986, \pasp, 98, 609

\bibitem[{{Hunter}(2007)}]{2007CSE.....9...90H}
{Hunter}, J.~D. 2007, Computing in Science and Engineering, 9, 90

\bibitem[{{Jones} {et~al.}(1995){Jones}, {Bland-Hawthorn}, \&
  {Shopbell}}]{1995ASPC...77..503J}
{Jones}, A.~W., {Bland-Hawthorn}, J., \& {Shopbell}, P.~L. 1995, in ASP Conf.
  Ser., Vol.~77, Astronomical Data Analysis Software and Systems IV, ed. R.~A.
  {Shaw}, H.~E. {Payne}, \& J.~J.~E. {Hayes}, 503

\bibitem[{{Kelsall} {et~al.}(1998){Kelsall}, {Weiland}, {Franz}, {Reach},
  {Arendt}, {Dwek}, {Freudenreich}, {Hauser}, {Moseley}, {Odegard},
  {Silverberg}, \& {Wright}}]{Kelsall98}
{Kelsall}, T., {Weiland}, J.~L., {Franz}, B.~A., {et~al.} 1998, \apj, 508, 44

\bibitem[{{Kubik} {et~al.}(2016){Kubik}, {Barbier}, {Chabanat}, {Chapon},
  {Clemens}, {Ealet}, {Ferriol}, {Gillard}, {Secroun}, {Serra}, {Smadja}, \&
  {Tilquin}}]{2016PASP..128j4504K}
{Kubik}, B., {Barbier}, R., {Chabanat}, E., {et~al.} 2016, \pasp, 128, 104504

\bibitem[{{Kubik} {et~al.}(2024){Kubik}, {Barbier}, {Smadja}, {Ferriol},
  {Conseil}, {Copin}, {Gillard}, {Dusini}, {Jahnke}, {Prieto}, {Auricchio},
  {Balbi}, {Balestra}, {Battaglia}, {Capobianco}, {Chary}, {Corcione},
  {Cogato}, {Delucchi}, {Franceschi}, {Gabarra}, {Gianotti}, {Grupp},
  {Lentini}, {Ligori}, {Medinaceli}, {Morgante}, {Paterson}, {Romelli},
  {Sauniere}, {Schirmer}, {Sirignano}, {Testera}, {Trifoglio}, {Troja},
  {Valenziano}, {Frailis}, {Scodeggio}, {Barriere}, {Berthe}, {Bodendorf},
  {Caillat}, {Carle}, {Casas}, {Cho}, {Costille}, {Ducret}, {Garilli},
  {Holmes}, {Hormuth}, {Hornstrup}, {Jhabvala}, {Kohley}, {Le Mignant},
  {Lilje}, {Lloro}, {Padilla}, {Polenta}, {Salvignol}, {Seidel}, {Serra},
  {Secroun}, {Stanco}, {Toledo-Moreo}, {Anselmi}, {Borsato}, {Caillat},
  {Colodro-Conde}, {Conforti}, {Davies}, {Renzi}, {Dal Corso}, {Davini},
  {Derosa}, {Diaz}, {Di Domizio}, {Di Ferdinando}, {Farinelli}, {Ferrari},
  {Fornari}, {Giacomini}, {Krause}, {Laudisio}, {Macias-Perez}, {Marpaud},
  {Mauri}, {da Silva}, {Niclas}, {Passalacqua}, {Risso}, {Lagier}, {Sorensen},
  {Stassi}, {Steinwagner}, {Tenti}, {Thizy}, {Tosi}, {Travaglini}, {Tubio},
  {Valieri}, {Ventura}, {Vescovi}, \& {Zoubian}}]{2024SPIE13103E..15K}
{Kubik}, B., {Barbier}, R., {Smadja}, G., {et~al.} 2024, in SPIE Conf. Ser.,
  Vol. 13103, X-Ray, Optical, and Infrared Detectors for Astronomy XI, ed.
  A.~D. {Holland} \& K.~{Minoglou}, 1310315

\bibitem[{{Laureijs} {et~al.}(2011){Laureijs}, {Amiaux}, {Arduini},
  {Augu{\`e}res}, {Brinchmann}, {Cole}, {Cropper}, {Dabin}, {Duvet}, {Ealet},
  {Garilli}, {Gondoin}, {Guzzo}, {Hoar}, {Hoekstra}, {Holmes}, {Kitching},
  {Maciaszek}, {Mellier}, {Pasian}, {Percival}, {Rhodes}, {Saavedra Criado},
  {Sauvage}, {Scaramella}, {Valenziano}, {Warren}, {Bender}, {Castander},
  {Cimatti}, {Le F{\`e}vre}, {Kurki-Suonio}, {Levi}, {Lilje}, {Meylan},
  {Nichol}, {Pedersen}, {Popa}, {Rebolo Lopez}, {Rix}, {Rottgering},
  {Zeilinger}, {Grupp}, {Hudelot}, {Massey}, {Meneghetti}, {Miller}, {Paltani},
  {Paulin-Henriksson}, {Pires}, {Saxton}, {Schrabback}, {Seidel}, {Walsh},
  {Aghanim}, {Amendola}, {Bartlett}, {Baccigalupi}, {Beaulieu}, {Benabed},
  {Cuby}, {Elbaz}, {Fosalba}, {Gavazzi}, {Helmi}, {Hook}, {Irwin}, {Kneib},
  {Kunz}, {Mannucci}, {Moscardini}, {Tao}, {Teyssier}, {Weller}, {Zamorani},
  {Zapatero Osorio}, {Boulade}, {Foumond}, {Di Giorgio}, {Guttridge}, {James},
  {Kemp}, {Martignac}, {Spencer}, {Walton}, {Bl{\"u}mchen}, {Bonoli},
  {Bortoletto}, {Cerna}, {Corcione}, {Fabron}, {Jahnke}, {Ligori}, {Madrid},
  {Martin}, {Morgante}, {Pamplona}, {Prieto}, {Riva}, {Toledo}, {Trifoglio},
  {Zerbi}, {Abdalla}, {Douspis}, {Grenet}, {Borgani}, {Bouwens}, {Courbin},
  {Delouis}, {Dubath}, {Fontana}, {Frailis}, {Grazian}, {Koppenh{\"o}fer},
  {Mansutti}, {Melchior}, {Mignoli}, {Mohr}, {Neissner}, {Noddle}, {Poncet},
  {Scodeggio}, {Serrano}, {Shane}, {Starck}, {Surace}, {Taylor},
  {Verdoes-Kleijn}, {Vuerli}, {Williams}, {Zacchei}, {Altieri}, {Escudero
  Sanz}, {Kohley}, {Oosterbroek}, {Astier}, {Bacon}, {Bardelli}, {Baugh},
  {Bellagamba}, {Benoist}, {Bianchi}, {Biviano}, {Branchini}, {Carbone},
  {Cardone}, {Clements}, {Colombi}, {Conselice}, {Cresci}, {Deacon}, {Dunlop},
  {Fedeli}, {Fontanot}, {Franzetti}, {Giocoli}, {Garcia-Bellido}, {Gow},
  {Heavens}, {Hewett}, {Heymans}, {Holland}, {Huang}, {Ilbert}, {Joachimi},
  {Jennins}, {Kerins}, {Kiessling}, {Kirk}, {Kotak}, {Krause}, {Lahav}, {van
  Leeuwen}, {Lesgourgues}, {Lombardi}, {Magliocchetti}, {Maguire}, {Majerotto},
  {Maoli}, {Marulli}, {Maurogordato}, {McCracken}, {McLure}, {Melchiorri},
  {Merson}, {Moresco}, {Nonino}, {Norberg}, {Peacock}, {Pello}, {Penny},
  {Pettorino}, {Di Porto}, {Pozzetti}, {Quercellini}, {Radovich}, {Rassat},
  {Roche}, {Ronayette}, {Rossetti}, {Sartoris}, {Schneider}, {Semboloni},
  {Serjeant}, {Simpson}, {Skordis}, {Smadja}, {Smartt}, {Spano}, {Spiro},
  {Sullivan}, {Tilquin}, {Trotta}, {Verde}, {Wang}, {Williger}, {Zhao},
  {Zoubian}, \& {Zucca}}]{Laureijs11}
{Laureijs}, R., {Amiaux}, J., {Arduini}, S., {et~al.} 2011, ESA/SRE(2011)12,
  arXiv:1110.3193

\bibitem[{{Markovi{\v{c}}} {et~al.}(2017){Markovi{\v{c}}}, {Percival},
  {Scodeggio}, {Ealet}, {Wachter}, {Garilli}, {Guzzo}, {Scaramella},
  {Maiorano}, \& {Amiaux}}]{2017MNRAS.467.3677M}
{Markovi{\v{c}}}, K., {Percival}, W.~J., {Scodeggio}, M., {et~al.} 2017,
  \mnras, 467, 3677

\bibitem[{{Neveu} {et~al.}(2024){Neveu}, {Br{\'e}maud}, {Antilogus}, {Barret},
  {Bongard}, {Copin}, {Dagoret-Campagne}, {Juramy}, {Le Guillou}, {Moniez},
  {Sepulveda}, \& {LSST Dark Energy Science
  Collaboration}}]{2024A&A...684A..21N}
{Neveu}, J., {Br{\'e}maud}, V., {Antilogus}, P., {et~al.} 2024, \aap, 684, A21

\bibitem[{{Outini} \& {Copin}(2020)}]{2020A&A...633A..43O}
{Outini}, M. \& {Copin}, Y. 2020, \aap, 633, A43

\bibitem[{{Padmanabhan} {et~al.}(2008){Padmanabhan}, {Schlegel}, {Finkbeiner},
  {Barentine}, {Blanton}, {Brewington}, {Gunn}, {Harvanek}, {Hogg},
  {Ivezi{\'c}}, {Johnston}, {Kent}, {Kleinman}, {Knapp}, {Krzesinski}, {Long},
  {Neilsen}, {Nitta}, {Loomis}, {Lupton}, {Roweis}, {Snedden}, {Strauss}, \&
  {Tucker}}]{2008ApJ...674.1217P}
{Padmanabhan}, N., {Schlegel}, D.~J., {Finkbeiner}, D.~P., {et~al.} 2008, \apj,
  674, 1217

\bibitem[{Robertson(1986)}]{robertson_optimal_1986}
Robertson, J.~G. 1986, \pasp, 98, 1220

\bibitem[{{Rubin} {et~al.}(2021){Rubin}, {Cikota}, {Aldering}, {Fruchter},
  {Perlmutter}, \& {Sako}}]{2021PASP..133f4001R}
{Rubin}, D., {Cikota}, A., {Aldering}, G., {et~al.} 2021, \pasp, 133, 064001

\bibitem[{{Ryan} {et~al.}(2018){Ryan}, {Casertano}, \&
  {Pirzkal}}]{2018PASP..130c4501R}
{Ryan}, Jr., R.~E., {Casertano}, S., \& {Pirzkal}, N. 2018, \pasp, 130, 034501

\bibitem[{{Scoville} {et~al.}(2007){Scoville}, {Aussel}, {Brusa}, {Capak},
  {Carollo}, {Elvis}, {Giavalisco}, {Guzzo}, {Hasinger}, {Impey}, {Kneib},
  {LeFevre}, {Lilly}, {Mobasher}, {Renzini}, {Rich}, {Sanders}, {Schinnerer},
  {Schminovich}, {Shopbell}, {Taniguchi}, \& {Tyson}}]{2007ApJS..172....1S}
{Scoville}, N., {Aussel}, H., {Brusa}, M., {et~al.} 2007, \apjs, 172, 1

\bibitem[{{Skrutskie} {et~al.}(2006){Skrutskie}, {Cutri}, {Stiening},
  {Weinberg}, {Schneider}, {Carpenter}, {Beichman}, {Capps}, {Chester},
  {Elias}, {Huchra}, {Liebert}, {Lonsdale}, {Monet}, {Price}, {Seitzer},
  {Jarrett}, {Kirkpatrick}, {Gizis}, {Howard}, {Evans}, {Fowler}, {Fullmer},
  {Hurt}, {Light}, {Kopan}, {Marsh}, {McCallon}, {Tam}, {Van Dyk}, \&
  {Wheelock}}]{2006AJ....131.1163S}
{Skrutskie}, M.~F., {Cutri}, R.~M., {Stiening}, R., {et~al.} 2006, \aj, 131,
  1163

\bibitem[{{Sparrow}(1916)}]{1916ApJ....44...76S}
{Sparrow}, C.~M. 1916, \apj, 44, 76

\bibitem[{{The Pandas development team}(2024)}]{pandas-devpandas_2024}
{The Pandas development team}. 2024, Pandas-Dev/Pandas: {{Pandas}}

\bibitem[{{Virtanen} {et~al.}(2020){Virtanen}, {Gommers}, {Oliphant},
  {Haberland}, {Reddy}, {Cournapeau}, {Burovski}, {Peterson}, {Weckesser},
  {Bright}, {van der Walt}, {Brett}, {Wilson}, {Millman}, {Mayorov}, {Nelson},
  {Jones}, {Kern}, {Larson}, {Carey}, {Polat}, {Feng}, {Moore}, {VanderPlas},
  {Laxalde}, {Perktold}, {Cimrman}, {Henriksen}, {Quintero}, {Harris},
  {Archibald}, {Ribeiro}, {Pedregosa}, {van Mulbregt}, \& {SciPy 1. 0
  Contributors}}]{2020NatMe..17..261V}
{Virtanen}, P., {Gommers}, R., {Oliphant}, T.~E., {et~al.} 2020, Nature
  Methods, 17, 261

\end{thebibliography}
